\documentclass[
    aps,
    prx,
    superscriptaddress,
    reprint,
    floatfix,
    longbibliography]{revtex4-2}

\usepackage{amsmath}
\usepackage{amsfonts}
\usepackage{newtxtext}
\usepackage{newtxmath}
\usepackage{mathtools}
\usepackage{bbm}
\usepackage[mathcal]{eucal}
\usepackage{bm}

\usepackage{amsbsy}

\usepackage{pict2e}

\usepackage[version=4]{mhchem}
\usepackage{comment}

\usepackage[table,dvipsnames]{xcolor}

\usepackage[export]{adjustbox}

\usepackage{tabularray}
\usepackage[percent]{overpic}
\UseTblrLibrary{booktabs}
\usepackage{array}
\newcolumntype{P}[1]{>{\centering\arraybackslash}p{#1}}
\newcolumntype{M}[1]{>{\centering\arraybackslash}m{#1}}
\definecolor{lgray}{rgb}{.93,.93,.93}
\usepackage{tabularx}
\usepackage{multirow}

\usepackage{enumitem}

\usepackage[bookmarksnumbered,pdfpagelabels=true,plainpages=false,colorlinks=true,linkcolor=blue,citecolor=magenta,urlcolor=blue]{hyperref}
\usepackage[capitalize]{cleveref}
\definecolor{cblue}{RGB}{55,126,184}
\hypersetup{colorlinks=true,linkcolor=cblue,citecolor=cblue,urlcolor=cblue}

\usepackage{tikz}

\usepackage{setspace}

\newcounter{mysubsection}[section] 

\newcommand{\mysubsection}[1]{%
  \refstepcounter{mysubsection}%
  \begin{center}
    \textbf{\small \themysubsection. #1}
  \end{center}
}

\newcommand{\cf}{\textit{cf}.}

\newcommand{\s}{\mathscr{s}}

\renewcommand{\o}{\mathrm{o}}
\newcommand{\q}{\mathscr{q}}
\newcommand{\D}{\mathscr{D}}

\newcommand{\N}{\mathcal{N}}
\renewcommand{\c}{\mathsf{c}}

\newcommand{\moment}{d}

\newcommand{\bq}{\bm{q}}
\DeclareMathOperator{\Tr}{Tr}

\newcommand{\llangle}{\langle \! \langle}
\newcommand{\rrangle}{\rangle \! \rangle}

\newcommand{\A}{\mathsf{A}}
\newcommand{\B}{\mathsf{B}}

\newcommand{\fragmented}{\boxplus}
\newcommand{\mixed}{\oplus}

\newcommand{\TCuni}{\smash{T_c^{\text{uni}}}}
\newcommand{\TCbi}{\smash{T_c^{\text{bi}}}}

\begin{document}

\title{
    Fracton Spin Liquid and Exotic Frustrated Phases in Ising-like Octochlore Magnets
}

\author{Matthew Stern}
\altaffiliation[]
{These authors contributed equally to the project.}
\affiliation{Department of Physics and Astronomy, University of California, Irvine, California 92697, USA}

\author{Michael D. Burke}
\altaffiliation[]
{These authors contributed equally to the project.}
\affiliation{Department of Physics and Astronomy, University of Waterloo, Waterloo, Ontario, N2L 3G1, Canada}

\author{Michel J. P. Gingras}
\affiliation{Department of Physics and Astronomy, University of Waterloo, Waterloo, Ontario, N2L 3G1, Canada}
\affiliation{Waterloo Institute for Nanotechnology, University of Waterloo, Waterloo, Ontario, N2L 3G1, Canada}

\author{Judit Romh\'{a}nyi}
\affiliation{Department of Physics and Astronomy, University of California, Irvine, California 92697, USA}

\author{Kristian Tyn Kai Chung}
\altaffiliation[]{Corresponding author: ktchung.phys@gmail.com}
\affiliation{Max Planck Institute for the Physics of Complex Systems, N\"othnitzer Strasse 38, 01187 Dresden, Germany}
\affiliation{Department of Physics and Astronomy, Rice University, Houston, Texas 77005, USA}

\begin{abstract}
For nearly three decades, research on frustrated magnetism in three dimensions (3D) has centered on the pyrochlore lattice of corner-sharing tetrahedra and the classical spin liquid (CSL) known as spin ice.
In this work, we propose that a lattice of corner-sharing octahedra---aptly dubbed the octochlore lattice---may provide a next-generation platform for the study of 3D frustrated magnetism, with realizations in anti-perovskite and 
certain alkali-rare-earth fluoride compounds. 
We study the phase diagram of Ising spins on the octochlore lattice with first- and second-neighbor interactions within each octahedron, finding that it displays a variety of frustrated phases.
These include CSLs with extensive ground state degeneracies, as well as phases with subextensive ground state degeneracies intermediate between spin liquids and long-range order.
Utilizing the framework of cluster multipole moments, we present a unified treatment of the variety of frustrated behaviors found in the phase diagram.
In addition to a spin ice CSL, we identify a novel fracton CSL with subdimensional excitations restricted to move along one-dimensional (1D) lines, realizing a classical U(1) equivalent of the paradigmatic X-cube model harboring fracton topological order.
These ``lineon'' quasiparticles carry magnetic quadrupole moments, contrasting the famous magnetic monopoles of spin ice.
These two CSLs lie at the boundaries of a parent ``frustrated chains'' phase with subextensive degeneracy which exhibits a dimensional crossover from 3D paramagnet to disordered 1D chains. 
Each CSL corresponds to condensation of distinct bound states of chain-based ferro-spinons, giving rise to quasi-criticality near the ends of the phase associated to avoided Kasteleyn-like transitions.
We also find a spin nematic phase whose ground states may be viewed as fracton crystals.
Upon cooling, the system first enters a uniaxial nematic phase, then transitions to a biaxial nematic phase.
The latter transition is driven by fluctuations arising from deconfinement of 1D antiferro-spinons owing to accidental symmetries of the subextensive ground state manifold, giving rise to dimensional reduction.
This work paves the way for the realization of fracton CSLs and the exploration of other exotic states in underexplored 3D frustrated octochlore magnetic materials.
\end{abstract}

\date{\today}

\maketitle

\tableofcontents


\section{Introduction}
    
For the past thirty years, frustrated magnets have been a leading platform for the study of exotic condensed matter physics phenomena, with a rich interplay of theoretical and experimental developments~\cite{lacroixIntroductionFrustratedMagnetism2011,gardnerMagneticPyrochloreOxides2010,rauFrustratedQuantumRareEarth2019,udagawaSpinIce2021,starykhUnusualOrderedPhases2015,vojtaFrustrationQuantumCriticality2018,henleyCoulombPhaseFrustrated2010,balentsSpinLiquidsFrustrated2010, castelnovoSpinIceFractionalization2012,gingrasQuantumSpinIce2014,savaryQuantumSpinLiquids2016,knolleFieldGuideSpin2019}.
Frustration, arising from competing spin-spin interactions, can significantly suppress the development of magnetic order, leaving systems paramagnetic well below their Curie-Weiss temperature.
In this regime, strong correlations and quantum fluctuations can give rise to complex behaviors.
In particular, they may stabilize a spin liquid state that hosts emergent deconfined gauge fields rather than conventional symmetry-breaking long-range order~\cite{wenQuantumOrdersSymmetric2002,henleyCoulombPhaseFrustrated2010,balentsSpinLiquidsFrustrated2010,castelnovoSpinIceFractionalization2012,gingrasQuantumSpinIce2014,savaryQuantumSpinLiquids2016,zhouQuantumSpinLiquid2017,knolleFieldGuideSpin2019,broholmQuantumSpinLiquids2020}.

In three dimensions (3D), pyrochlore rare-earth magnets have become standard bearers of the field, which are naturally frustrated by tetrahedral coordination of magnetic ions, with the tetrahedra forming a corner-sharing network~\cite{gardnerMagneticPyrochloreOxides2010,hallasExperimentalInsightsGroundState2018,rauFrustratedQuantumRareEarth2019,smithExperimentalInsightsQuantum2025}.
The interplay of strong spin-orbit coupling and crystalline electric field effects in these systems often results in Ising-like moments~\cite{gardnerMagneticPyrochloreOxides2010,rauFrustratedQuantumRareEarth2019,rauMagnitudeQuantumEffects2015}, with the most famous examples being the compounds
\ce{Ho2Ti2O7}~\cite{harrisGeometricalFrustrationFerromagnetic1997,bramwellSpinCorrelationsHo2Ti2O72001,fennellMagneticCoulombPhase2009} and
\ce{Dy2Ti2O7}~\cite{ramirezZeropointEntropySpin1999,yavorskiiDy2Ti2O7SpinIce2008,morrisDiracStringsMagnetic2009,heneliusRefrustrationCompetingOrders2016}.
At low temperatures, these materials realize a classical spin liquid (CSL) called spin ice~\cite{harrisGeometricalFrustrationFerromagnetic1997,bramwellSpinIceState2001,bramwellHistorySpinIce2020}. 
These systems are governed at low temperatures by a local 2-in--2-out ``ice rule'' constraint of the four Ising moments on each tetrahedron~\footnote{
    Thanks to the magnetostatic dipolar interactions, these materials should, in principle, order at very low temperature~\cite{heneliusRefrustrationCompetingOrders2016,melkoLongRangeOrderLow2001,melkoMonteCarloStudies2004}, but such order has never been observed experimentally.
}, an emergent zero-divergence Gauss constraint, giving rise to fractionalized quasiparticle excitations behaving as magnetic monopoles~\cite{udagawaSpinIce2021,castelnovoMagneticMonopolesSpin2008,castelnovoSpinIceFractionalization2012}.

\begin{figure*}[t]
    \centering
    \phantom{a}
    \hfill
    \begin{overpic}[width=0.48\textwidth]{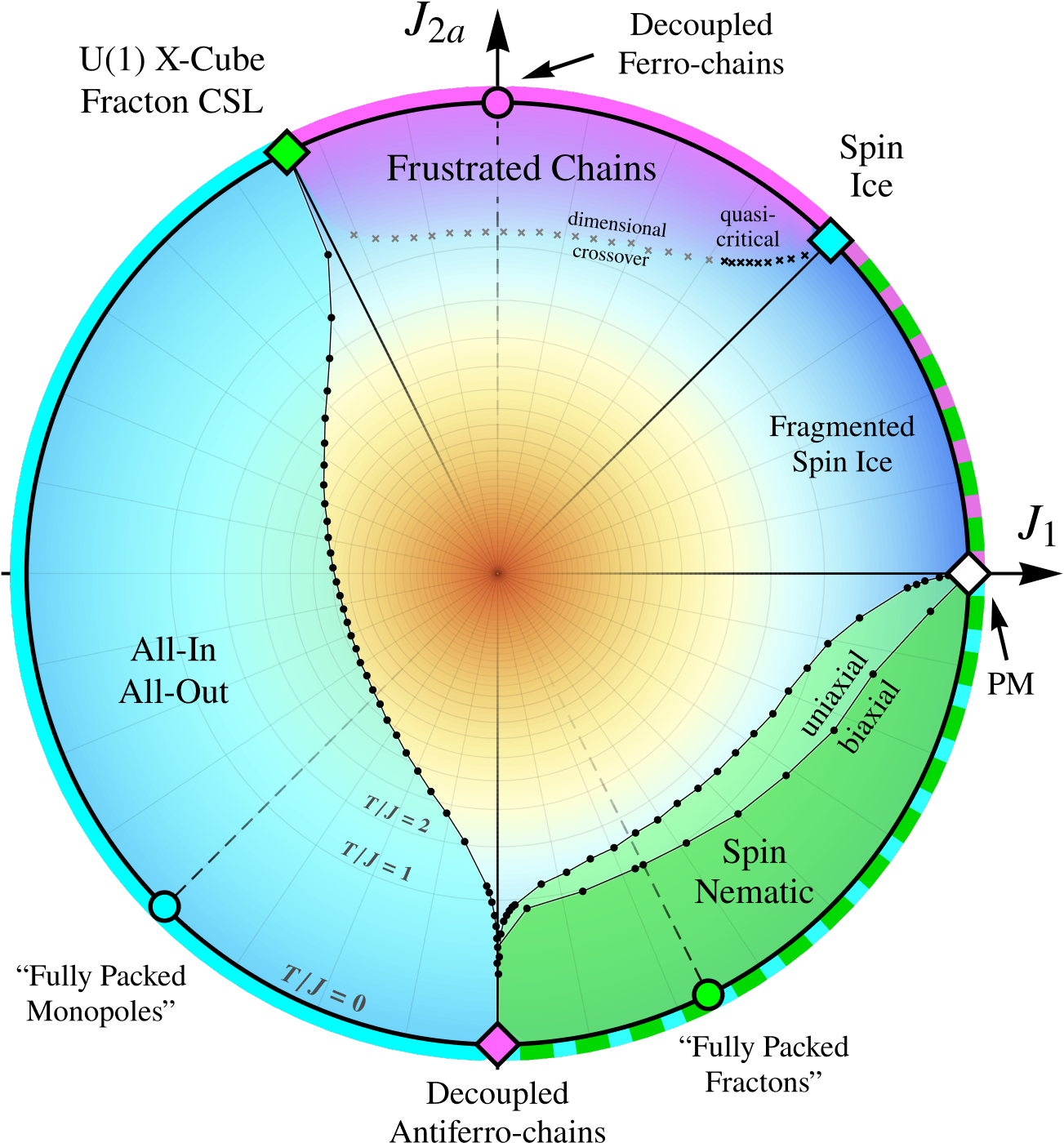}
        \put(-99,33){
            \begin{overpic}[width=.32\linewidth]{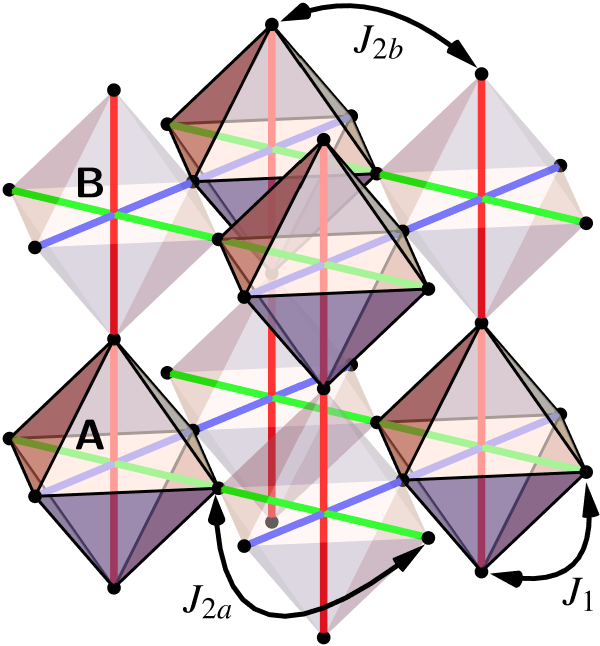}
            \end{overpic}
        }
        \put(-27,-1){
            \includegraphics[width=.121
            \linewidth]{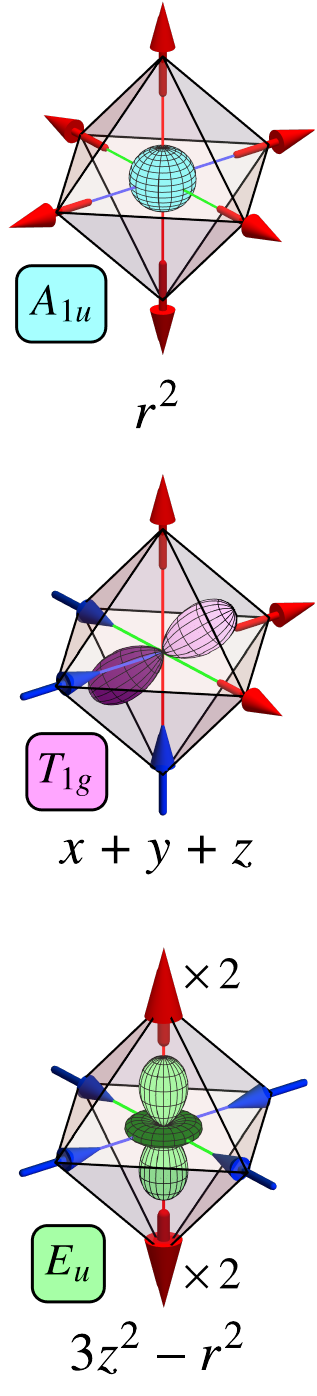}
        }
        \put(-104,-1){
            \includegraphics[width=.38\linewidth]{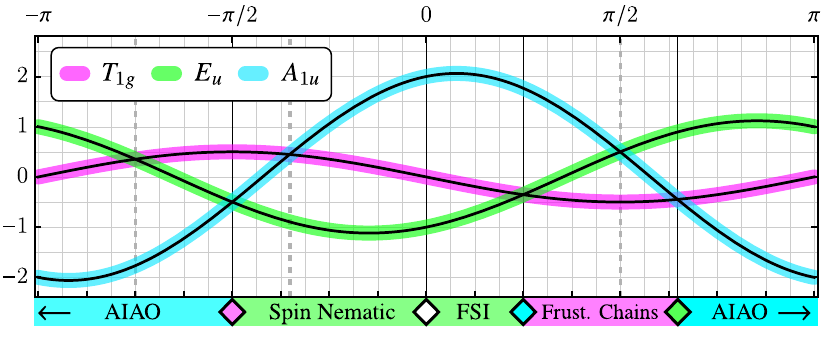}
        }
        \put(-99,99){\large{(a)}}
        \put(-27,99){\large{(b)}}
        \put(-27,63){\large{(c)}}
        \put(-27,29){\large{(d)}}
        \put(-99,31){\large{(e)}}
        \put(+00,99){\large{(f)}}
    \end{overpic}
    \caption{{\bf Overview of the octochlore Ising system.}
    (a)~Octochlore lattice with first ($J_1$) and second ($J_{2a}$ and $J_{2b}$) neighbor interactions. 
    Octahedron centers form a bipartite simple cubic lattice; one ``A'' and one ``B'' octahedron are marked.
    Orthogonal 1D chains are illustrated with red, blue, and green colors. 
    (b,c,d)~The Ising moment configurations can be decomposed into their (b) monopole, (c) dipole, and (d) quadrupole cluster multipole moments, transforming respectively as the $A_{1u}$, $T_{1g}$, and $E_u$ irreducible representations (irreps) of the octahedral point group symmetry (\cref{sec:hamiltonian_irreps}). 
    We highlight the multipole character of each irrep by a 3D spherical harmonic inside each octahedron and a function with the corresponding symmetry below.
    The quadrupolar $E_u$ irrep is inconsistent with the fixed moment constraint $\vert S_i^z\vert = 1$, as indicated by doubled moment size ($\times\, 2$) in panel (d), resulting in fragmentation (\cref{sec:fragmentation_of_Eu}).
    (e)~Variation of irrep energies from \cref{eq:irrep_energies} as a function of $\theta$, with the parameterization of $J_1(\theta)$ and $J_{2a}(\theta)$ given in \cref{eq:theta_J1_J2a}.  
    (f)~Finite-temperature phase diagram of \cref{eq:H} as the function of $\theta$ and temperature, with zero temperature at the outer edge and infinite temperature at the center.
    Black dots mark thermal phase transition critical temperatures obtained from Monte Carlo simulations. 
    An overview of the phase diagram is given in \cref{sec:phase_diagram}.
    The radial temperature scale is given by $\smash{\cramped{r(T/J) = 1-\tanh[(T/10J)^{1/2}]\in[0,1]}}$, with $\cramped{T=0}$ at the outer edge, increasing towards the center.
    \Cref{apx:monte_carlo} gives details of the Monte Carlo simulations and presents a linear temperature scale version of the phase diagram.
    }
    \label{fig:phase_diagram}
\end{figure*}

In an effort to capture the physics of materials with anisotropic spin-spin interactions~\cite{hallasExperimentalInsightsGroundState2018,rauFrustratedQuantumRareEarth2019,smithExperimentalInsightsQuantum2025}, substantial work has been devoted to three-component continuous-spin models on a variety of frustrated lattices.
Such models are readily amenable to ``soft spin'' coarse-grained descriptions in the classical large-spin limit, and for which CSLs arise from flat bands in the momentum-space spectrum of the spin-spin interaction matrix~\cite{yanTheoryMultiplephaseCompetition2017,yanClassificationClassicalSpin2024,yanClassificationClassicalSpin2024a,fangClassificationClassicalSpin2024,bergmanBandTouchingRealspace2008}.
A key strategy of these studies has been the organization of clusters of spins into irreducible representations of the point group symmetry, corresponding to different local ordering patterns.
This approach has clarified that CSLs naturally arise at classical multicritical points where multiple long-range ordered phases become energetically degenerate~\cite{taillefumierCompetingSpinLiquids2017,franciniExactNematicMixed2025,yogendraSymmetrySuperpositionFragmentation2024,noculakClassicalQuantumPhases2023,bentonSpinliquidPinchlineSingularities2016,yanTheoryMultiplephaseCompetition2017,lozano-gomezCompetingGaugeFields2024,chungMappingPhaseDiagram2024,lozano-gomezAtlasClassicalPyrochlore2024,franciniHigherrankSpinLiquids2025,chungGeometricallyFrustratedQuadrupoles2025}.
This approach has led to the identification of numerous examples of so-called higher-rank CSLs~\cite{bentonSpinliquidPinchlineSingularities2016,yanRank2U1Spin2020,heringFractonExcitationsClassical2021,bentonTopologicalRouteNew2021,flores-calderonIrrationalMomentsSignatures2024,lozano-gomezCompetingGaugeFields2024,lozano-gomezAtlasClassicalPyrochlore2024,franciniHigherrankSpinLiquids2025,chungMappingPhaseDiagram2024,davierPinchlineSpinLiquids2025,chungGeometricallyFrustratedQuadrupoles2025}, whose emergent tensor Gauss laws are closely related to those of fracton gauge theories~\cite{pretkoGeneralizedElectromagnetismSubdimensional2017,pretkoFractonPhasesMatter2020,youQuantumLiquidsEmergent2025}.
Such gauge theories locally conserve multipole moments, restricting the gauge charges to sub-dimensional mobility, either rendering them entirely immobile or confining their motion to lines or planes of the crystalline lattice~\cite{pretkoSubdimensionalParticleStructure2017}.
A key unsolved problem for these continuous-spin models is whether a CSL which arises in the large-$S$ limit of a quantum spin Hamiltonian corresponds to a stable small-$S$ \emph{quantum} spin liquid (QSL).
This question remains extremely challenging; for most models, especially in 3D, no controlled and unbiased methods currently exist to determine whether quantum fluctuations stabilize a QSL or instead drive them toward ordered ground states.

Classical Ising models, by contrast, 
correspond to the classical limit of bona fide quantum spin-1/2 models in which off-diagonal spin-flip operators are absent from the Hamiltonian. 
Unlike continuous-spin CSLs, Ising CSLs have discrete gapped excitations.
They can be promoted to QSLs by turning on the spin-flip terms in the Hamiltonian, which perturbatively generate ``ring-exchange''---virtual processes that create, move, and re-annihilate excitations, inducing quantum tunneling between classically degenerate ground states~\cite{hermelePyrochlorePhotonsU12004,gingrasQuantumSpinIce2014,chung2formU1Spin2025}.
This allows for the controlled construction of an effective quantum gauge theory describing the resulting QSL~\cite{hermelePyrochlorePhotonsU12004,chung2formU1Spin2025}, an approach which has proven especially fruitful in rationalizing the behavior of candidate quantum spin ice materials
~\cite{gingrasQuantumSpinIce2014,hallasExperimentalInsightsGroundState2018,rauFrustratedQuantumRareEarth2019,smithExperimentalInsightsQuantum2025}. 
On the other hand, the low-temperature properties of highly frustrated classical 3D Ising models are generally much more difficult to uncover than those of their continuous-spin counterparts, as they are not well approximated by soft-spin descriptions and require carefully tailored numerical methods. 
For this reason, they remain comparatively underexplored beyond pyrochlore spin ice.

In this work, we significantly advance the understanding of highly frustrated 3D Ising models by focusing on a different lattice---a network of corner-sharing \emph{octahedra}, appropriately dubbed the \emph{octochlore lattice}, depicted in \cref{fig:phase_diagram}(a). 
Among lattices of corner-sharing clusters~\cite{moessnerLowtemperaturePropertiesClassical1998,henleyCoulombPhaseFrustrated2010}, the octochlore lattice is the logical next step ``up'' from the pyrochlore lattice---going from a four-coordinated cluster (tetrahedron) to a six-coordinated one (octahedron).
From a materials perspective, the octochlore geometry is realized by the magnetic-moment-bearing M sites of anti-perovskite (or inverse-perovskite) compounds with chemical formula~\ce{M3AX}
~\cite{PhysRevB.97.060401,szaboFragmentedSpinIce2022,KIRCHNER20031247,niewaMetalRichTernaryPerovskite2019,ALHARBI2024416544,gablerMagneticStructureInverse2008,
hirschmannCreatingControllingDirac2022,zhouStabilityTopologicalBehaviors2025,luoAbsenceFragmentedSpin2026}, and also in a class of magnetic trivalent rare-earth (R) fluoride insulators of chemical formula~\ce{AR3F10}~\cite{chamberlainthesis,chamberlainMagneticOrderingKDy3F102003,chamberlainMagneticOrderingRbGd3F102003,chamberlainMagneticOrderingRareearth2005,chamberlainParamagneticPropertiesCubic2006,youHydrothermalSynthesisMixed2010,McMillenCrystalChemistryFluorides2015,demortierUnusualPlanarAnisotropy2025}.
Our work positions these materials as a promising and largely unexplored platform for frustrated magnetism. 
We provide a more detailed overview of these materials in \Cref{sec:materials}.

In retrospect, despite its prominence, the pyrochlore Ising model offers only limited possibilities for uncovering the full breadth of highly frustrated Ising physics in 3D and empowering the experimental discovery of novel quantum magnetic materials.
As a tetrahedron supports only nearest-neighbor Ising spin-spin couplings, the pyrochlore's network of corner-sharing tetrahedra can only exhibit two phases: either 2-in--2-out spin ice or all-in/all-out (AIAO) order~\cite{bramwellFrustrationIsingtypeSpin1998,moessnerReliefGenerationFrustration1998}.
We will demonstrate that the extension to octahedra naturally provides new quadrupolar (nematic) degrees of freedom, in addition to the monopolar (AIAO) and dipolar (spin ice) ones.
Correspondingly, the octochlore lattice affords an extra tuning parameter in the form of a second-neighbor coupling \emph{within} the elementary frustrated octahedral unit, in contrast to the single energy scale for an individual tetrahedron in the pyrochlore lattice.

Previous works studying octochlore spin models focused on either isotropic Heisenberg models~\cite{sklanNonplanarGroundStates2013,bentonTopologicalRouteNew2021} or spin ice physics~\cite{hermelePyrochlorePhotonsU12004,szaboFragmentedSpinIce2022}, leaving the broader Ising phase diagram largely unexplored. 
In this work we study the complete phase diagram of the intra-octahedral couplings in detail. 
We do this by leveraging intuition gained from years of study of spin ice physics on the pyrochlore lattice, along with the more recent insights afforded by the tools of irreducible representation (irrep) analysis~\cite{yanTheoryMultiplephaseCompetition2017,chungMappingPhaseDiagram2024,yanRank2U1Spin2020,lozano-gomezAtlasClassicalPyrochlore2024} and highly efficient cluster Monte Carlo methods~\cite{otsukaClusterAlgorithmMonte2014}. 
The irrep analysis (\cref{sec:hamiltonian_irreps}) decomposes spin configurations on each octahedron into irreducible multipole components~\cite{chungMappingPhaseDiagram2024}; the additional degrees of freedom of the octahedral unit give rise to quadrupolar components not achievable on the tetrahedron, underpinning the comparative richness of this model.
The full phase diagram is shown in \Cref{fig:phase_diagram}(f), with an overview of its features presented in \cref{sec:phase_diagram}.
In addition to a spin ice point and an AIAO phase, we find three parts of the phase diagram exhibiting previously unexplored instances of frustrated physics, including a fracton CSL and two highly frustrated phases with subextensive ground state degeneracies, intermediate between spin liquids and conventional ordered phases.

First, we discover a classical Ising fracton CSL (\cref{sec:fracton_CSL}) which, to the best of our knowledge, is the first thus far reported in 3D (see Refs.~\cite{plackeIsingFractonSpin2024,niggemannClassicalFractonSpin2025} for 2D models).
We identify this as a U(1) CSL analog of the celebrated X-cube model, a paradigmatic realization of fracton topological order~\cite{vijayFractonTopologicalOrder2016}.
Whereas the excitations of spin ice are, famously, magnetic monopoles~\cite{castelnovoMagneticMonopolesSpin2008,castelnovoSpinIceFractionalization2012}, the excitations of the fracton CSL are identified as magnetic quadrupoles, and are restricted to move along straight lines.
We identify the generalization of spin ice's zero-energy loop moves~\cite{melkoLongRangeOrderLow2001,melkoMonteCarloStudies2004} as ``cage moves'', thus characterizing this CSL as a cage-net liquid~\cite{premCageNetFractonModels2019} rather than a string-net liquid~\cite{levinStringnetCondensationPhysical2005}.
Building on this understanding, we develop a highly efficient cluster Monte Carlo algorithm, generalizing one originally designed for spin ice~\cite{otsukaClusterAlgorithmMonte2014,lantagne-hurtubiseSpinIceThinFilms2018}, to explore this CSL down to zero temperature while maintaining thermal equilibrium.
Rather than pinch points~\cite{henleyCoulombPhaseFrustrated2010}, we identify the reciprocal space signatures of this CSL as pinch lines~\cite{bentonSpinliquidPinchlineSingularities2016,davierPinchlineSpinLiquids2025}.

Second, we find a phase whose ground states are independently polarized ferromagnetic chains hosting deconfined 1D spinons, with inter-chain interactions mediated by collisions of spinons moving on intersecting chains.
We dub this the ``frustrated chains phase'' (\cref{sec:frustrated_chains}).
We show that this phase may be viewed as a parent phase for both the spin ice and fracton CSLs, which can each be viewed as a condensate of different bound states of 1D spinons. 
We argue that this phase does not undergo a finite-temperature phase transition, but instead exhibits a \emph{dimensional crossover} from a high-temperature 3D paramagnet to a low-temperature regime of disordered one-dimensional chains.
Near the spin-ice CSL, we further show that the system displays a quasi-critical \emph{avoided} Kasteleyn-like transition~\cite{jaubertThreedimensionalKasteleynTransition2008,jaubertKasteleynTransitionThree2009,jaubertTopologicalConstraintsDefects2009,szaboFragmentedSpinIce2022,pottsSpinIceGeneral2022,powellQuantumKasteleynTransition2022}: upon cooling, it first enters a finite-temperature spin-liquid regime before rapidly crossing over into the disordered-chain regime at lower temperature.

Lastly, we identify a parameter region exhibiting both uniaxial and biaxial spin nematic orderings (\cref{sec:spin_nematic})---upon cooling, the system spontaneously breaks cubic to tetragonal symmetry, stabilizing a uniaxial nematic liquid, while at a lower temperature the system further breaks symmetry to orthorhombic, developing biaxial order.
Analogous to how the AIAO phase is a maximal packing of the monopole quasiparticles of spin ice~\cite{brooks-bartlettMagneticMomentFragmentationMonopole2014}, this nematic phase can be viewed as a maximal packing of the fractonic excitations of the X-cube fracton CSL, i.e., a \emph{fracton crystal}.
We show that the transition to the biaxial nematic phase is fluctuation-driven by an order-by-disorder mechanism~\cite{villainInsulatingSpinGlasses1979,villainOrderEffectDisorder1980,henleyOrderingDisorderGroundstate1987,henleyOrderingDueDisorder1989,wengelSpinglassAntiferromagnetCritical1996,hiziAnharmonicGroundState2009}---the biaxial configurations support deconfined 1D antiferro-spinons which disorder 1D chains and maximize entropy, which arise as a consequence of accidental subsystem symmetries. 
The low-temperature physics is that of effectively decoupled and disordered 1D chains, providing a striking realization of spontaneous dimensional reduction~\cite{mishraDirectionalOrderingFluctuations2004,batistaGeneralizedElitzursTheorem2005,XuReductionEffectiveDimensionality2005,nussinovIntermediateSymmetriesElectronic2006,taharaAntiferromagneticIsingModel2007,nussinovCompassModelsTheory2015,makutaDimensionalReductionQuantum2021} driven by thermal fluctuations in a classical Ising model.

\section{Hamiltonian and Irreducible Multipoles}
\label{sec:hamiltonian_irreps}

The octochlore lattice, depicted in \cref{fig:phase_diagram}(a), is a network of corner-sharing octahedra. 
It is the line graph (or medial lattice) of the bipartite simple cubic lattice~\cite{henleyCoulombPhaseFrustrated2010}, i.e., the vertices of the octochlore lattice correspond to edge-centers of the simple cubic lattice, and the vertices of the cubic lattice correspond to centers of octahedra.
Thus, there are two types of octahedra, which we denote $\A$ and $\B$, similar to the pyrochlore, which is the line graph of the bipartite diamond lattice~\cite{henleyCoulombPhaseFrustrated2010}.
It will prove useful to utilize multiple unit cell choices. 
The primitive simple cubic unit cell contains only three sites, and is thus too small to be capture all single-octahedron spin configurations.
It is therefore convenient to use a conventional 6-site face-centered cubic (fcc) unit cell which captures all single-octahedron configurations.
For Monte Carlo simulations, we shall use an even larger conventional 24-site cubic unit cell containing eight octahedra, i.e., the one shown in \cref{fig:phase_diagram}(a).
Details of these unit cells are given in \cref{apx:conventions}.

We consider Ising-like magnetic moments $\bm{\moment}_i$ located on the sites~$i$ of the octochlore lattice.
The local quantization axes $\hat{\bm{z}}_i$ are aligned along the local $C_{4v}$ easy-axis at the site~$i$, lying along one of the three cubic axes, [001], [010], or [100]. 
We define scalar Ising variables $\smash{\cramped{S_i^z \equiv \hat{\bm{d}}_i \cdot\hat{\bm{z}}_i}}$ taking values $\cramped{S_i^z =\pm 1}$.
Each magnetic moment can be written as $\cramped{\bm{\moment}_i = (1/2)g_{\parallel}\, \mu_{\text{B}} S_i^z \, \hat{\bm{z}}_i}$, with $\mu_{\text{B}}$ the Bohr magneton and $g_{\parallel}$ an ionic Ising $g$-factor.
Note that the quantization axes $\hat{\bm{z}}_i$, and therefore the relative signs of the Ising variables $S_i^z$, are dependent on convention, 
whereas the moment $\bm{d}_i$ is the physical, invariant degree of freedom.
In the 6-site unit cell, we follow the convention of orienting the $\hat{\bm{z}}_i$ in a bipartite all-in/all-out fashion, pointing from the $\A$ to $\B$ cubic sublattices, as in the configuration of arrows shown in \cref{fig:phase_diagram}(b).

\Cref{fig:phase_diagram}(a) illustrates the lattice geometry of corner-sharing octahedra, and indicates the first ($J_1$) and second neighbor ($J_{2a}$ and $J_{2b}$) interactions. 
In this work, we take the inter-octahedron coupling $J_{2b}$ to be zero, retaining only the two intra-octahedron couplings $J_1$ and $J_{2a}$.
Using the 6-site unit cell convention of the $\hat{\bm{z}}_i$ defined above, the Hamiltonian takes the form
\begin{equation}
    H 
    = 
    J_1 
    \sum_{\langle ij \rangle} S_i^z S_j^z 
    + 
    J_{2a} 
    \sum_{\mathclap{\llangle ij \rrangle_a}}
    S_i^z S_j^z .
    \label{eq:H}
\end{equation}
With this unit cell convention, $J_{2a}>0$ in \cref{eq:H} favors ferromagnetic alignment of the moments at opposite corners of an octahedron. 
We parameterize the Hamiltonian with a single angle $\theta$ and a fixed overall energy scale $\cramped{J>0}$, as 
\begin{equation}
    \cramped{J_1 = J \cos\theta}
    \quad \text{and} \quad 
    \cramped{J_{2a} = J \sin\theta}.
    \label{eq:theta_J1_J2a}
\end{equation}
In what follows, temperatures are expressed in units of $J/k_{\mathrm{B}}$, and specific heat and entropy in units of $k_{\mathrm{B}}$.

\subsection{Irreducible Cluster Multipole Moments}

Since the $J_1$ and $J_{2a}$ interactions are internal to a single octahedron, the Hamiltonian can be expressed as the sum of contributions from each octahedron, as we now detail.
Specifically, under the action of the octahedral point symmetry group $O_h$, a configuration of the six Ising dipoles on a single octahedron can be decomposed into a sum of multipole moments.
The six dipoles collectively transform as a 6-dimensional representation of $O_h$, which decomposes into irreducible components $A_{1u} \mixed T_{1g} \mixed E_u$, corresponding to the net monopole, dipole, and diagonal quadrupole moments, respectively, of the spin configuration.
In terms of the local quantization axes $\hat{\bm{z}}_i$ defined above, these take the form (see \cref{apx:irreps_multipoles} for details)
\begin{align}
    \begin{aligned}
    \text{Monopole ($A_{1u}$):}& \,\,\,
    \phantom{Q_{\o}^{\alpha\beta}} 
    \mathllap{\phi_{{\o}}} 
    \!=\! 
    \frac{(-1)^{\o}}{\sqrt{6}}
    \! \sum_{i \in {\o}} S_i^z,
    %
    \\
    \text{Dipole ($T_{1g}$):}& \,\,\,
    \phantom{Q_{\o}^{\alpha\beta}} 
    \mathllap{D_{{\o}}^\alpha}  
    \!=\! 
    \frac{(-1)^{\o}}{\sqrt{2}}
    \! \sum_{i \in {\o}} S_i^z \hat{z}_i^\alpha ,
    \\
    \text{Quadrupole ($E_u$):}& \,\,\,
    Q_{{\o}}^{\alpha\beta} 
    \!=\! 
    \frac{(-1)^{\o}}{\sqrt{2}}
    \! \sum_{i \in {\o}} 
    S_i^z 
    \left(\hat{z}_i^\alpha \hat{z}_i^\beta - \frac{1}{3} \delta^{\alpha\beta}
    \right)\mathclap{.}
    \end{aligned}
    \label{eq:irrep_multipoles}
\end{align}
Here, $\o$ labels octahedra and $\delta$ is the Kronecker delta. 
The sign $(-1)^{\o}$ is positive for $\A$ octahedra and negative for $\B$ octahedra. 
These are included because a spin that points ``out'' of an $\A$ octahedron points ``in'' to a $\B$ octahedron.
Note that the quadrupole tensor $Q_{\o}$ is diagonal and trace-free, with zero off-diagonal components, and thus possesses only two degrees of freedom corresponding to the two-dimensional $E_u$ irrep. 
Also note that these are \emph{cluster} multipole moments~\cite{suzukiMultipoleExpansionMagnetic2019,bouazizSpinModelsCluster2025}---linear combinations of spins which are all time-reversal odd---and should not be confused with on-site multipoles, which are defined as products of spin operators~\cite{pencSpinNematicPhases2011,chungGeometricallyFrustratedQuadrupoles2025}.

The chosen normalizations in \cref{eq:irrep_multipoles} ensure that
\begin{equation}
    \vert\phi_{\o}\vert^2 + \vert \bm{D}_{\o}\vert^2 + \Tr[Q_{\o}^2] = \sum_{i \in \o} (S_i^z)^2.
\end{equation}
Since the Hamiltonian is bilinear in $S_i^z$ and transforms trivially under the symmetry, it can be decomposed on each octahedron into a sum of quadratic invariants (\cref{apx:irreps_multipoles}),
\begin{equation}
    H
    = 
    \frac{1}{2}
    \sum_{\text{oct. }{\o}} \, 
    \left(
    J_{A_{1u}} \vert \phi_{\o}\vert^2 + J_{T_{1g}} \vert \bm{D}_{\o}\vert^2 + J_{E_u} \Tr[Q_{\o}^2]
    \right),
    \label{eq:H_irreps}
\end{equation}
where the energy of each multipole is given by
\begin{equation}
    J_{A_{1u}} =  4J_1 + J_{2a} \,, \quad 
    J_{T_{1g}} = - J_{2a}       \,, \quad
    J_{E_u}    = -2J_1 + J_{2a} .
    \label{eq:irrep_energies}
\end{equation}
The dependence of these energy scales upon the parameter $\theta$ in \cref{eq:theta_J1_J2a} is plotted in \cref{fig:phase_diagram}(e).
The multipole with the lowest energy determines the type of ground state order on each octahedron, dividing the phase diagram broadly into three regions. 
However, the Ising spin length constraint $\cramped{\vert S_i^z \vert = 1} \, \forall \, i$ \emph{forbids}
a configuration with purely quadrupolar spin configuration, i.e. one  with both $\cramped{\phi_{\o}=0}$ and $\cramped{\bm{D}_{\o}= \bm{0}}$, leading to additional frustration in the $E_u$ sector, which we discuss in \cref{sec:fragmentation_of_Eu} below.

\begin{figure}
    \centering
    \vspace{1ex}
    \large{(a) \underline{$E_u \fragmented T_{1g}$}} $\quad (E = -4J_1+J_{2a})$
    \\[1ex]
    \includegraphics[width=\linewidth]{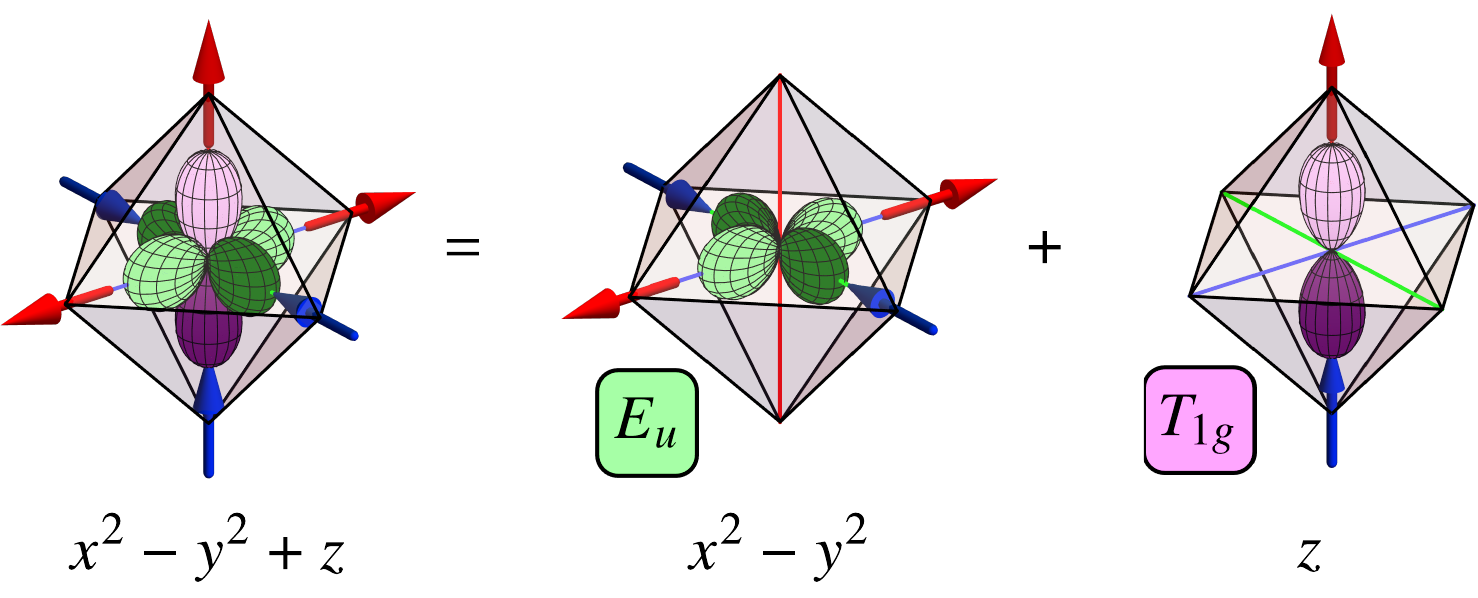}
    \\[3ex]
    \large{(b) \underline{$E_u\fragmented A_{1u}$}} $\quad (E = -4J_1+3J_{2a})$
    \\[1ex]
    \includegraphics[width=\linewidth]{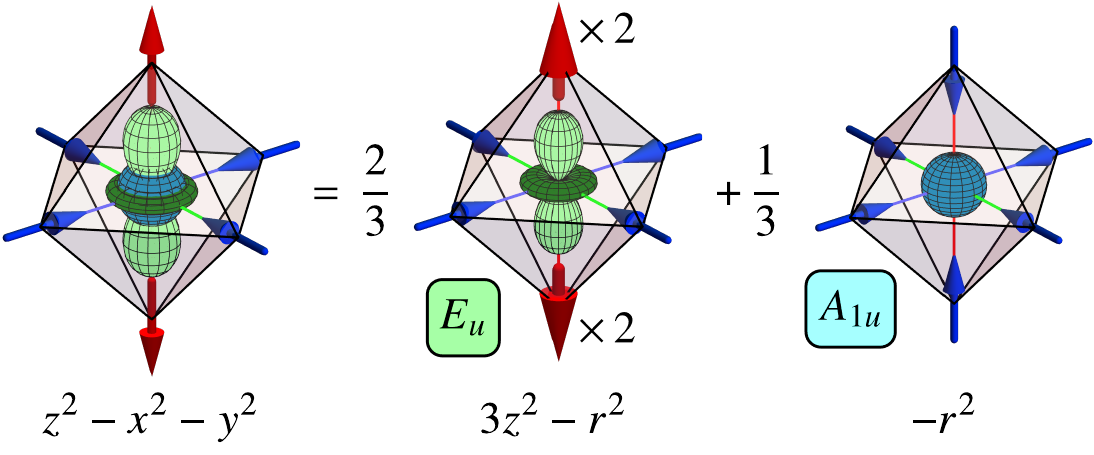}
    \\
    \caption{\textbf{Fragmentation of the $\bm{E_u}$ irrep}. 
    There are no spin configurations with fixed-length Ising moments, $S_i^z \in \pm 1$ on all sites, which are purely quadrupolar. 
    Instead, there are two symmetry classes of Ising configurations which mix $E_u$ with one of the other two irreps. 
    An example of each is given on the left side of (a) and (b).
    On the right, we decompose each as a linear combination of irreducible monopole ($A_{1u}$), dipole ($T_{1g}$), and quadrupole ($E_u$) configurations with \emph{variable} spin lengths $S_i^z \in \{0,\pm 1, \pm 2\}$, where $S_i^z = 0$ corresponds to the absence of a spin. 
    Beneath each configuration, we have written a polynomial with the corresponding symmetry, where $r^2 = x^2 + y^2 + z^2$. 
    Analogous decompositions of all spin configurations are given in \cref{apx:irreps_multipoles}.
    }
    \label{fig:E_irrep_fragmentation}
\end{figure}

\subsection{Fragmentation of the \texorpdfstring{$\bm{E_u}$}{Eu} Quadrupole Moment}
\label{sec:fragmentation_of_Eu}

Because the Hamiltonian \cref{eq:H_irreps} is quadratic in each multipole moment, the single-octahedron ground state will maximize the lowest-energy multipole and minimize the other two, preferably to zero. 
This is readily achieved for the monopole moment $A_{1u}$ irrep---the 6-out configuration shown in \cref{fig:phase_diagram}(b) has zero dipole and quadrupole moments, and is purely monopolar.
The same holds for the dipole moment $T_{1g}$ irrep---the 3-in-3-out configuration shown in \cref{fig:phase_diagram}(c) has zero monopole and quadrupole moment, and is purely dipolar. 
However, this is not the case for the $E_u$ quadrupole moment irrep: of the $2^6$ configurations of six Ising spins with $\vert S_i^z \vert = 1$ on a single octahedron, none have \emph{both} zero monopole and dipole moment (see~\cref{apx:irreps_multipoles} for a complete decomposition of the $2^6$ states). 
Only by violating the spin length constraint can one construct pure-$E_u$ quadrupolar configurations, such as the one shown in \cref{fig:phase_diagram}(d) with the two apical spins taking the unphysical value $\cramped{S_i^z = 2}$, indicated by ``$\times 2$''.
This configuration is ``four-in--four-out'' and so has zero monopole moment, and also has zero net dipole moment along each of the three axes, making it purely quadrupolar at the cost of violating the spin length constraint.
We refer to this obstruction to a pure-quadrupolar ground state as ``frustration of the $E_u$ irrep'' by the spin length constraint~\footnote{
    This frustration is of the same type as occurs in the 2D nearest-neighbor kagome Ising model, where the frustrated cluster is a triangle of three spins, and the irreps are a 1D monopole and 2D dipole moment. The putative spin ice phase of this model should maximize the dipole moment and minimize the monopole moment, but this is not possible due to the spin length constraint. In that case the system is forced to remain trivially paramagnetic at all temperatures~\cite{huseClassicalAntiferromagnetsKagome1992}.
}.

At the single-octahedron level, the consequences of this irrep frustration is that the configurations that maximize the quadrupole moment are ``fragmented''~\cite{brooks-bartlettMagneticMomentFragmentationMonopole2014,lhotelFragmentationFrustratedMagnets2020,szaboFragmentedSpinIce2022}, meaning that they mix $E_u$ with one of the other two irreps.
This is illustrated in \cref{fig:E_irrep_fragmentation}.
There are two symmetry-inequivalent classes of maximal-$E_u$ configurations, mixing $E_u$ with either $T_{1g}$ or $A_{1u}$.
One member of each class is shown on the left hand side of the graphical equations in \cref{fig:E_irrep_fragmentation}(a,b).
We refer to these configurations as $\cramped{E_u \fragmented T_{1g}}$ and $\cramped{E_u \fragmented A_{1u}}$. 
On the right hand side, we have decomposed each spin configuration as a linear combination of an $E_u$ component and either a $T_{1g}$ or an $A_{1u}$ component.
Beneath each we write a polynomial which reflects the symmetry of this irrep---a monopole transforms like $r^2$, dipoles transform like $x$, $y$, and $z$, and the two-component $E_u$ quadrupole transforms like $(3z^2 - r^2)$ and $(x^2 - y^2)$. 
In the $E_u$ sector of the phase diagram in \cref{fig:phase_diagram}(e), i.e. the parameter region where $J_{E_u}< J_{A_{1u}},J_{T_{1g}}$, one of these two fragmented configurations will have lower energy, with \cref{fig:E_irrep_fragmentation}(a) having lower energy when $J_{2a}>0$ and \cref{fig:E_irrep_fragmentation}(b) having lower energy when $J_{2a} < 0$.
This therefore subdivides the $E_u$ sector into two distinct ground state phases about the $\theta=0^\circ$ point.

\section{Phase Diagram}
\label{sec:phase_diagram}

In this section, we summarize the variety of ground states of the Hamiltonian \cref{eq:H} that appear across the phase diagram; the remainder of the paper focuses on those without clear precedent or analogs in the existing literature.
\Cref{fig:phase_diagram}(f) shows the phase diagram parameterized by the angle $\theta$ defined in \cref{eq:theta_J1_J2a}.
The radial axis is temperature $T/J$, with $T=0$ at the outer edge increasing towards infinite temperature at the center of the disk (\cf~caption of \cref{fig:phase_diagram}).
The outer edge of the disk is colored according to the single-octahedron ground state irrep(s): blue for $A_{1u}$, \cref{fig:phase_diagram}(b); magenta for $T_{1g}$, \cref{fig:phase_diagram}(c); hatched green-blue for $\cramped{E_u\fragmented A_{1u}}$, \cref{fig:E_irrep_fragmentation}(a); and hatched green-magenta for $\cramped{E_u\fragmented T_{1g}}$, \cref{fig:E_irrep_fragmentation}(b).
The phase diagram is correspondingly divided into four wedges by solid lines, each corresponding to a different set of ground state spin configurations, with colored diamonds on the outer edge of the disk indicating the phase boundaries.
We have performed extensive classical Monte Carlo simulations to determine finite-temperature phase boundaries, indicated by black dots. Details of the Monte Carlo simulations are provided in \cref{apx:monte_carlo}.

\begin{table*}[t]
    \centering
    \begin{booktabs}{
            width = \linewidth,
            colspec = {X[3.4,l] X[4.9,l] X[2,c] X[4.6,l] X[3.2,l]}
        }
        \toprule
        Phase & Description & $\log_2 (\mathrm{GSD})$ & Order Type & Zero-Energy Locus
        \\
        \toprule
        \footnotesize{$A_{1u}$} & All-In/All-Out & 1  & SSB {[$\mathbbm{Z}_2$]} & 0D: Zone Corner
        \\
        \footnotesize{$T_{1g}$} & Frustrated Chains & $\propto L^2$ & Dimensional Crossover & 2D: $\{hk0\}$ Planes
        \\
        \footnotesize{$E_u \fragmented T_{1g}$} & Fragmented Spin Ice & $\propto L^3$ & Spin Liquid \footnotesize{[U(1) Coulomb]} & 1D: Zone Edges$^{*}$
        \\
        \footnotesize{$E_u \fragmented A_{1u}$} & Spin Nematic \footnotesize{[Uniaxial, Biaxial]} & $\propto L^2$ & SSB~[$O_h$], Dim. Reduction & 1D: Zone Edges$^{*}$
        \\
        \bottomrule
        Phase Boundary & Description & $\log_2(\mathrm{GSD})$ & Order Type & Zero-Energy Locus
        \\
        \toprule
        \footnotesize{$E_u\mixed T_{1g}$} & Spin Ice & $\propto L^3$ & Spin Liquid \footnotesize{[U(1) Coulomb]} & 3D: FB$\,\times\, 2$ $+$ $(000)$
        \\
        \footnotesize{$A_{1u}\oplus T_{1g}$} & X-cube & $\propto L^3$ & Spin Liquid \footnotesize{[U(1) Fracton]} & 3D: FB$\,\times\, 1$ $+$ $\{00\ell\}$
        \\
        \footnotesize{$\cramped{[E_u\fragmented T_{1g}]\mixed [E_u \fragmented A_{1u}]}$} & Paramagnet & $\propto L^3$ & Paramagnet \footnotesize{[$J_{2a} = 0$]} & 1D: Zone Edges$^{*}$ 
        \\
        \footnotesize{$E_u\mixed A_{1u}$} & Decoupled AFM Chains & $\propto L^2$ & 1D Critical \footnotesize{[$J_1 = 0$]} & 2D: $\{hk0\}$ Planes
        \\
        \bottomrule
    \end{booktabs}
    \caption{{\textbf Summary of the phases and phase boundaries that appear in the phase diagram of \cref{fig:phase_diagram}(f).}
    Quadrants are labeled by the relevant ground state irreps, with the two fragmented $E_u$ ground states indicated using the notation in \cref{fig:E_irrep_fragmentation}.
    Boundaries are labeled by the degenerate combination of irreps. We provide a description of each, the system-size scaling of the ground state degeneracy (GSD), and order type of the low-temperature phase: a phase with spontaneous symmetry breaking (SSB), a classical spin liquid, or otherwise.
    The last column lists the zero-energy locus of the interaction matrix band structure.
    Cases with completely flat bands (FB) indicate both the number of flat bands and the momentum-space locus where the dispersive bands touch them.
    Cases marked $*$ are fragmented, so the flat locus is not directly reflective of the ground state degeneracies (\cf~\cref{sec:degen_and_flat_bands}). 
    }
    \label{tab:summary_degeneracies}
\end{table*}

\subsection{Zero-Energy Flat Bands and Ground State Degeneracies}
\label{sec:degen_and_flat_bands}

A quadratic spin Hamiltonian can be written as $H = E_0 + (1/2)\sum_{ij} \mathcal{J}_{ij} S_i^z S_j^z$, where $E_0$ is the ground state energy and $\mathcal{J}$ is the interaction matrix encoding two-spin interactions, shifted to have zero minimum eigenvalue.
The interaction matrix can be diagonalized by Fourier transform, its eigenvalues forming a set of bands $\mathcal{J}_n(\bq)$ where $\bq$ are reciprocal space wavevectors and $n$ indexes the bands, equal to the number of degrees of freedom in the primitive unit cell.
The zero-energy (null) eigenvectors of the interaction matrix play two complementary roles in the low-temperature physics:
first, if a null eigenvector exists satisfying the non-linear spin length constraint then it defines a ground state spin configuration;
second, any two spin configurations differing by a null eigenvector have the same energy; in other words, the null eigenvectors generate zero-energy deformations or zero modes.

If a zero-energy spin configuration exists, then the null eigenvectors play a decisive role in the low-temperature ground state physics.
If they occur only at isolated momenta, the ground state manifold contains only a few states and the zero-energy deformations are spatially extended over the entire system, implying that the system will undergo a thermodynamic phase transition to long-range order at low temperature. 
If, in the opposite extreme, an entire zero-energy band is flat, then the zero modes are localized in direct space~\cite{bergmanBandTouchingRealspace2008}, implying an extensive ground state degeneracy, giving rise to a classical spin liquid~\cite{henleyPowerlawSpinCorrelations2005,yanClassificationClassicalSpin2024,yanClassificationClassicalSpin2024a}.
Intermediate cases between isolated band minima and fully flat bands---a flat locus whose dimension is between zero and the dimension of space---are comparatively less explored, but indicate strong frustration~\footnote{
    Here we give a few examples. The nearest-neighbor Ising AFM model on the 3D FCC lattice exhibits flat 1D $\{10\ell\}$ lines on the zone boundary, indicating plane-based zero modes giving rise to subextensive ground state degeneracy~\cite{danielianGroundStateIsing1961,danielianLowTemperatureBehaviorFaceCentered1964,wengelSpinglassAntiferromagnetCritical1996}. It undergoes a first order transition at finite temperature which selects ground states consisting of AFM-ordered planes~\cite{polgreenMonteCarloSimulation1984,stubelFinitesizeScalingMonte2018} driven by an order-by-disorder mechanism~\cite{villainInsulatingSpinGlasses1979,villainOrderEffectDisorder1980,henleyOrderingDisorderGroundstate1987,henleyOrderingDueDisorder1989,wengelSpinglassAntiferromagnetCritical1996,hiziAnharmonicGroundState2009}.
    Examples of a codimension-1 zero-energy locus (flat lines in 2D or flat planes in 3D) occur in compass models~\cite{nussinovCompassModelsTheory2015} which exhibit spin nematic orders~\cite{mishraDirectionalOrderingFluctuations2004,batistaGeneralizedElitzursTheorem2005,nussinovIntermediateSymmetriesElectronic2006,nussinovDiscreteSlidingSymmetries2005,wenzelMonteCarloSimulations2008,wenzelReexaminingDirectionalorderingTransition2010,wenzelUnveilingNatureThreeDimensional2011,nussinovCompassModelsTheory2015,franciniExactNematicMixed2025}.
}.
A $k$-dimensional zero-energy locus implies direct-space zero modes associated to $(d-k)$-dimensional subsystems in $d$-dimensions, the number of which scales with the linear system size $L$ as $L^k$, suggesting (but not necessarily guaranteeing) a ground state degeneracy scaling as $L^k$.

On the other hand, if a zero-energy spin configuration does \emph{not} exist, then ground state configurations must be linear combinations of higher-energy eigenvectors of $\mathcal{J}$.
In this case the band structure analysis does not give direct insight into the low-temperature physics, and a more detailed analysis is required. 
In examples appearing in the present octochlore model and others we are aware of, the lack of a zero-energy spin configuration generally enhances the degeneracies beyond what is predicted by the band structure~\footnote{
    The most well-known examples where the band structure fails to capture the low-energy physics are the antiferromagnetic nearest-neighbor kagome and triangular lattice Ising model.
    In both cases it is not possible to satisfy $S_1^z + S_2^z + S_3^z = 0$ on any triangle when $\vert S_i^z \vert = 1$. 
    In the kagome case the system remains trivially paramagnetic at all temperatures~\cite{huseClassicalAntiferromagnetsKagome1992}, while the triangular lattice model realizes a $\mathbbm{Z}_2$ classical spin liquid~\cite{wannierAntiferromagnetismTriangularIsing1950}.
}.

The octochlore model with Hamiltonian given by \cref{eq:H} provides a clean platform where the full breadth of these behaviors are exhibited in a single model.
\Cref{tab:summary_degeneracies} summarizes the four phases and phase boundaries appearing in the phase diagram~\cref{fig:phase_diagram}(f), including both the reciprocal space locus of the zero-energy band structure as well as the zero-temperature ground state degeneracies~\footnote{
    For simplicity, we refer to the low-temperature regimes as ``phases'', even though adding additional further-neighbor interactions like $J_{2b}$ will lead to some selection of an order-unity number of ground states out of these massive ground state manifolds.
    From the perspective of an extended phase diagram including such interactions, the parameter regions with (sub)extensive ground state degeneracies will be phase boundaries or multicritical loci.
}.
The $A_{1u}$ AIAO phase exhibits a single band minimum at the zone center, correspondingly having only two symmetry-breaking long range ordered ground states.
Two of the phase boundaries---$T_{1g}\oplus E_u$  and $T_{1g}\oplus A_{1u}$---have completely flat bands, indicating localized zero modes and spin liquidity.
We also find examples between these two extremes, giving rise to rich frustrated physics intermediate between long-range order and spin liquidity.
The $T_{1g}$ phase and the $E_u \oplus A_{1u}$ phase boundary exhibit flat plane degeneracies, which imply one-dimensional (chain-based) zero modes.
Each portion of the phase diagram just described has zero-energy ground states, because the local energetic constraints enforced by \cref{eq:H_irreps}---that higher-energy irreps are completely suppressed---can be satisfied exactly, so the band structure is reflective of the low-energy physics.
However, as explained in \cref{sec:fragmentation_of_Eu}, the $E_u$ irrep is fragmented, meaning that ground states necessarily mix in some higher-energy irreps and so have positive energy relative to the minimum of $\mathcal{J}$.
Therefore, when the $E_u$ irrep has the lowest energy the zero-energy locus of the band structure is not a reliable guide to the low-temperature physics.
Indeed, the two $E_u$ phases and the phase boundary between them all exhibit a one-dimensional zero-energy locus, implying only planar zero modes and ground state degeneracies scaling as $L^1$, whereas in these three cases the ground state degeneracies scale as either $L^2$ or $L^3$.

In the rest of this Section, we give a brief overview of the variety of low-temperature behaviors observed in the phase diagram and listed in \cref{tab:summary_degeneracies}, before diving into detailed investigations in the remainder of the paper.
For ease of presentation, we start with the four fine-tuned phase boundaries (marked by colored diamonds in \cref{fig:phase_diagram}(e,f)), before discussing the four phases (indicated by colored arcs on the outer edge of the disk in \cref{fig:phase_diagram}(f)).
In this Section, we denote the Hamiltonian relative to the ground state energy by $\widetilde{H}$, defined as
\begin{equation}
    \widetilde{H}(J_1,J_{2a}) \equiv  H(J_1,J_{2a}) - E_0(J_1,J_{2a}),
    \label{eq:H_zero_energy}
\end{equation}
where $H$ is the Hamiltonian in \cref{eq:H} and $E_0$ is the ground state energy, each a function of the interaction parameters $J_1$ and~$J_{2a}$.

\subsection{Overview of Phase Boundaries}

$\bm{\diamond}$ $\bm{E_u \mixed T_{1g}}$ \textbf{(Spin Ice CSL)} --- Denoted in the phase diagram by a blue diamond ($\theta=45^\circ$), at this point, the Hamiltonian relative to the ground state energy reduces to
\begin{equation}
    \widetilde{H}_{\text{spin ice}} 
    =6J_{2a}
    \sum_{\o} \vert \phi_{\o}\vert^2
    \quad
    (J_1 = J_{2a}>0).
    \label{eq:H_Spin_Ice}
\end{equation}
The single-octahedron ground states satisfy a zero-divergence 3-in--3-out ``ice rule'' constraint, and the system realizes a U(1) Coulomb phase~\cite{henleyCoulombPhaseFrustrated2010}~\footnote{
    The fact that this is a U(1) spin liquid refers to the fact that in  the ground state manifold, satisfying a zero-divergence constraint, the conserved flux through closed surfaces is integer-valued, as in a U(1) gauge theory~\cite{hermelePyrochlorePhotonsU12004}.
}, 
or equivalently a loop-gas~\cite{jaubertAnalysisFullyPacked2011} or string net condensate~\cite{levinStringnetCondensationPhysical2005}.
Minimal excitations are point particles carrying $A_{1u}$ ``monopole moments''~\cite{castelnovoMagneticMonopolesSpin2008}, as indicated by the energy cost of creating $\phi_{\o}$ monopoles in $\widetilde{H}_{\text{spin ice}}$, \cref{eq:H_Spin_Ice}.
The interaction matrix exhibits two zero-energy flat bands, with one quadratically dispersing band that touches the flat bands at the zone center. 
This band structure corresponds to a long-wavelength free-energy of the form $\int (\nabla\cdot\bm{D})^2$, where~$\bm{D}$ is the coarse-grained dipole moment~\cite{henleyCoulombPhaseFrustrated2010,moessnerSpinIceCoulomb2021}. 
Microscopically, the flat bands give rise to zero-energy collective loop flips of head-to-tail spins~\cite{melkoMonteCarloStudies2004,melkoLongRangeOrderLow2001,jaubertAnalysisFullyPacked2011,szaboFragmentedSpinIce2022}.

$\bm{\diamond}$ $\bm{ E_u \mixed A_{1u}}$ \textbf{(Decoupled Antiferromagnetic Chains)} --- Denoted in the phase diagram by a magenta diamond, at this point the Hamiltonian relative to the ground state energy reduces to
\begin{equation}
    \widetilde{H}_{\text{AFM chains}} 
    =
    2\vert J_{2a}\vert
    \sum_{\o} \vert \bm{D}_{\o}\vert^2 
    \quad
    (J_1 = 0, J_{2a}<0).
    \label{eq:H_AFM_Chains}
\end{equation}
This point has $J_{1}=0$, and the Hamiltonian splits into decoupled 1D antiferromagnetic (AFM) chains, illustrated in \cref{fig:phase_diagram}(a) by colored lines along each of the Cartesian direction.
Each chain has two N\'{e}el ground states, but they do not order at any finite temperature. 
Excitations are deconfined antiferro-spinons, which  can move freely along a chain by flipping spins at zero energy cost.
It is clear from \cref{eq:H_AFM_Chains} that the excitations are charged in the $T_{1g}$ irrep, meaning that these quasiparticles carry a dipole moment.
The direction of this local dipole moment switches as the quasiparticle hops on the  N\'{e}el background, but the staggered dipole moment $\bm{D}_{\o}$ (staggered by the alternating sign $(-1)^{\o}$ in \cref{eq:irrep_multipoles}) is conserved.

$\bm{\diamond}$ $\bm{A_{1u} \mixed T_{1g}}$ \textbf{(X-cube CSL)} --- Denoted in the phase diagram by a green diamond, 
this point is the focus of \cref{sec:fracton_CSL}. 
At this point the Hamiltonian relative to the ground state energy reduces to
\begin{equation}
    \widetilde{H}_{\text{X-cube}} 
    =
    3 J_{2a}
    \sum_{\o} \Tr[Q_{\o}^2] 
    \quad 
    (J_{2a} = -2J_{1}>0).
    \label{eq:H_X-cube}
\end{equation}
We will demonstrate that $\widetilde{H}_{\text{X-cube}}$ realizes a previously unreported \emph{fracton CSL}. 
We identify this phase as a classical U(1) version of the X-cube topological order~\cite{vijayFractonTopologicalOrder2016,slagleXcubeModelGeneric2018}.
Using a tailored cluster Monte Carlo algorithm, we demonstrate that this CSL is stable down to zero temperature. 
Unlike the spin ice CSL, whose ground state manifold is a loop-gas or string-net condensate, the X-cube CSL can be described as a cage-net condensate~\cite{premCageNetFractonModels2019}, where the minimal zero-energy moves involve closed cage-like structures rather than loops.
The excitations in this phase carry a conserved $E_u$ quadrupole moment 
In particular, excitations carrying a $(3z^2 - r^2)$ quadrupole charge are restricted to move along the $z$ direction, and similarly for the $x$ and $y$ directions, making them sub-dimensional ``lineon'' quasiparticles.

$\bm{\diamond}$ $\bm{[E_u\fragmented T_{1g}]\mixed [E_u \fragmented A_{1u}}]$ \textbf{(Paramagnet)} --- Denoted in the phase diagram by a white diamond ($J_1>0$, $J_{2a}=0$), this point is the nearest-neighbor Ising AFM on the octochlore lattice~\cite{chuiSimpleThreedimensionalIsing1977,taharaAntiferromagneticIsingModel2007}. 
At this point, the two fragmented $E_u$ configurations in \cref{fig:E_irrep_fragmentation} become degenerate. There is no finite-temperature phase transition, so the ground state is paramagnetic~\cite{taharaAntiferromagneticIsingModel2007}.
In particular, notice that an $E_u \fragmented T_{1g}$ configuration
(\cref{fig:E_irrep_fragmentation}(a))  can be converted into an $E_u \fragmented A_{1u}$ configuration (\cref{fig:E_irrep_fragmentation}(b)), and vice versa, by flipping a single spin.
It follows that in a ground state consisting of a mixture of these two octahedral configurations, there will exist a finite density of single spins that can freely flip without changing the energy at zero temperature. 
Furthermore, from the perspective of the neighboring fragmented spin ice CSL discussed below, this means that single-charge monopoles can be created at zero energy cost.
These features imply that the system remains a trivial paramagnet with only short-range correlations down to zero temperature, and therefore does not realize a CSL (analogous to the nearest-neighbor kagome Ising antiferromagnet, \cf~\cite{Note2}).

\subsection{Overview of Phases}

$\bm{\circ}$ $\bm{A_{1u}}$ \textbf{(All-in/All-out)} --- In this phase, denoted in the phase diagram by a cyan arc ($J_{2a}<2\vert J_1 \vert, J_1<0$), the single-octahedron ground states are either 6-in or 6-out, such as shown in \cref{fig:phase_diagram}(b), with ``all-in/all-out'' (AIAO) global ground states.
This is the only portion of the phase diagram that exhibit an order-unity ground state degeneracy and ``trivial'' magnetic long-range order.
The low temperature phase breaks the global $\mathbbm{Z}_2$ time-reversal and inversion symmetries (but preserves their product).
The symmetry-breaking order parameter is the scalar
\begin{equation}
    \phi = \frac{1}{\sqrt{6}} 
    \left(\frac{1}{N_{\o}}\sum_{\o} (-1)^{\o}\,\phi_{\o}\right),
    \label{eq:monopole_OP}
\end{equation}
where $N_{\o}$ is the number of octahedra in the lattice, normalized so that $\vert\phi\vert^2\leq 1$, with $\phi_{\o}$ defined in \cref{eq:irrep_multipoles}.
This phase sits opposite the spin ice CSL in the phase diagram. 
In particular, it contains the point where the Hamiltonian is given by $-\widetilde{H}_{\text{spin ice}}$ [\cf~\cref{eq:H_Spin_Ice}], marked in the phase diagram with a cyan disk.
Therefore, its ground state can be said to maximize the divergence of the dipole configuration at every octahedron.
One may view each ground state as a maximally packed crystal of $\pm 3$-charged monopole quasiparticles~\cite{brooks-bartlettMagneticMomentFragmentationMonopole2014} on the two cubic sublattices in an \ce{NaCl}-type arrangement, as shown in \cref{fig:fracton-crystal}(c).
Indeed, the scalar order parameter in \cref{eq:monopole_OP} may be viewed as a measure of the ``monopole contrast'' between the two sublattices, $\cramped{\phi \propto (\sum_{\o\in A} \phi_{\o} - \sum_{\o\in B} \phi_{\o})}$, i.e. the staggered monopole charge density.

$\bm{\circ}$ $\bm{\cramped{E_u\fragmented T_{1g}}}$ \textbf{(Fragmented Spin Ice)} --- Denoted in the phase diagram by a hatched green-magenta arc, in this phase the single-octahedron ground states are the fragmented configurations shown in \cref{fig:E_irrep_fragmentation}(a).
These are a subset of the 3-in--3-out spin ice configurations. 
Reference~\cite{szaboFragmentedSpinIce2022} showed that these configurations are sufficient to preserve an extensive ground state manifold, giving a CSL that is a restricted version of spin ice dubbed ``fragmented spin ice'' (FSI).
The FSI is a U(1) Coulomb phase like spin ice, with algebraic correlations and ordinary two-fold pinch points. 
This spin liquid retains the spin ice monopole charge quasiparticle excitations, but the zero-charge $T_{1g}$ states (\cref{fig:phase_diagram}(c)) are also energetic excitations.
These zero-charge excitations can be viewed as an energetic penalty for particular types of loop configurations in the loop-gas picture of the spin ice ground state manifold~\cite{jaubertAnalysisFullyPacked2011} 
(\cf~\cref{sec:kasteleyn}).
The stability of the spin liquid means that these energetic restrictions on loop configurations are an irrelevant perturbation to the long-wavelength Coulomb description.

$\bm{\circ}$ $\bm{T_{1g}}$ \textbf{(Frustrated Chains)} --- Denoted in the phase diagram by a magenta arc, this phase is the focus of \cref{sec:frustrated_chains}. 
In this phase, the single-octahedron ground states are configurations illustrated in \cref{fig:phase_diagram}(c).
This phase sits opposite the decoupled AFM chains point, so its ground states can be formally described as maximal packings of antiferro-spinons.
Naively, one might expect this phase to exhibit long-range magnetic order characterized by the net dipole moment order parameter,
\begin{equation}
    \bm{D} =  \frac{1}{\sqrt{6}}\left(\frac{1}{N_{\o}} \sum_{\o} (-1)^{\o}\,\bm{D}_{\o}\right),
    \label{eq:dipole_OP}
\end{equation}
where $N_{\o}$ is the number of octahedra in the lattice, normalized so that $\vert\bm{D}\vert^2 \leq 1$, with $\bm{D}_{\o}$ defined in \cref{eq:irrep_multipoles}.
The situation is not so simple, however.
At the special point with $J_1 = 0$ the system decouples into 1D ferromagnetic chains, every chain having two ground states; the minimal excitations are chain-based domain walls, which behave as deconfined quasiparticles---1D ferro-spinons---and the system does not order at any finite temperature.
We show that turning on $\cramped{J_1\neq 0}$ does not lift the subextensive ground state degeneracy, and we therefore dub this the \emph{frustrated chains} phase. 
In fact, the chains interact only indirectly via ``collisions'' between spinons moving on orthogonal chains.
We argue that there is no finite-temperature phase transition in this phase, instead the system exhibits a continuous dimensional crossover from the high-temperature paramagnetic phase to a low-temperature phase of weakly interacting 1D chains. 
Noting that the spin ice and X-cube CSL's appear at the boundaries of this phase, we show that their existence and physical properties can be rationalized by viewing them as condensates of 1D ferro-spinon bound states.
Near the ends of the phase we demonstrate that dimensional crossover becomes quasi-critical, and explain this as a consequence of an avoided Kasteleyn-like transition~\cite{jaubertThreedimensionalKasteleynTransition2008,jaubertKasteleynTransitionThree2009,jaubertTopologicalConstraintsDefects2009,szaboFragmentedSpinIce2022,pottsSpinIceGeneral2022,powellQuantumKasteleynTransition2022}.

$\bm{\circ}$ $\bm{\cramped{E_u \fragmented A_{1u}}}$ \textbf{(Spin Nematic)} --- Denoted in the phase diagram by a hatched green-cyan arc, this phase is discussed in \cref{sec:spin_nematic}. 
Here, the single-octahedron ground states are the fragmented configurations shown in~\cref{fig:E_irrep_fragmentation}(b).
This phase sits opposite the fracton CSL, in particular, it contains the point where the Hamiltonian is given by $-\widetilde{H}_{\text{X-cube}}$ [\cref{eq:H_X-cube}], marked in the phase diagram with a green disk.
Therefore, its ground states can be viewed as a maximal packing of the $E_u$-charged lineon quasiparticle excitations of the X-cube CSL---a \emph{fracton crystal}.
We show that this system harbors a subextensive ground state degeneracy and exhibits low-energy emergent subsystem symmetries, where certain subsets of chains can be flipped independently in the ground state manifold. 
Using Monte Carlo simulations, we identify two finite-temperature transitions upon cooling: the first signals the development of  uniaxial nematic order, with symmetry reducing from cubic to tetragonal;  the second brings about a biaxial nematic order, further reducing the symmetry to orthorhombic. 
These are characterized by the global quadrupole tensor order parameter which measures the staggered quadrupole density,
\begin{equation}
    Q^{\alpha\beta} = \frac{\sqrt{3}}{4}\left(\frac{1}{N_{\o}} \sum_{\o} (-1)^{\o} Q_{\o}^{\alpha\beta}\right),
    \label{eq:quadrupole_OP}
\end{equation}
where $N_{\o}$ is the number of octahedra in the lattice, normalized so that $\Tr[Q^2]\leq 1$,
with $Q^{\alpha\beta}_{\o}$ defined in \cref{eq:irrep_multipoles}.
The transition to uniaxial order can be readily described in the Landau symmetry-breaking paradigm, where the system lowers its free energy by developing quadrupolar order.
The transition to biaxial order, however, is fluctuation-driven: by ordering along a second axis, the system is able to support deconfined antiferro-spinons which disorder chains along the third axis, thereby maximizing entropy and minimizing the free energy.
At low temperatures, the system behaves as effectively decoupled and disordered one-dimensional chains, providing a clean realization of spontaneous dimensional reduction~\cite{mishraDirectionalOrderingFluctuations2004,batistaGeneralizedElitzursTheorem2005,XuReductionEffectiveDimensionality2005,nussinovIntermediateSymmetriesElectronic2006,taharaAntiferromagneticIsingModel2007,nussinovCompassModelsTheory2015,makutaDimensionalReductionQuantum2021}.

This completes our review of the phase diagram, shown in \cref{fig:phase_diagram}(f) and summarized in \cref{tab:summary_degeneracies}.
Some of the phases and boundaries are relatively simply understood: the spin ice CSL and the AIAO phase opposite to it are straightforward generalizations of their pyrochlore analogs; the fragmented spin ice CSL is a restricted version of spin ice~\cite{szaboFragmentedSpinIce2022}; the AFM chains point correspond to decoupled 1D Ising chains; and the boundary of the two fragmented $E_u$ phases is a trivial paramagnet~\cite{taharaAntiferromagneticIsingModel2007}.
In the remainder of this paper, we investigate the remaining three parts of the phase diagram, each of which harbors particularly rich magnetic frustration physics which has not been studied previously. 
\Cref{sec:frustrated_chains} focuses on the $T_{1g}$ frustrated chains phase, from which we develop an intuitive understanding of the origin of the spin ice and fracton CSLs occurring at its endpoints, the nature of their fractionalized excitations, and the structure of their ground state manifolds. 
\Cref{sec:fracton_CSL} then explores in detail the $T_{1g} \oplus A_{1u}$ fracton CSL and establishes its essential properties.
Finally, aided by a solid understanding of the fracton CSL, \cref{sec:spin_nematic} studies the opposite side of the phase diagram, the spin nematic $E_u\fragmented A_{1u}$ phase.

\section{Frustrated Chains Phase (\texorpdfstring{$\bm{T_{1g}}$}{T1g})}
\label{sec:frustrated_chains}

In this section we investigate the $T_{1g}$ phase, indicated by the magenta arc at the top of the phase diagram \cref{fig:phase_diagram}(f).
We dub this the frustrated chains phase, as it has a subextensively degenerate ground state manifold of polarized chains.
We expose the origin of inter-chain interactions via spinon collisions, and the fate of the equilibrium state at low temperatures.
In the process, we uncover the mechanism by which the spin ice and X-cube spin liquids arise at the end points of this phase through the condensation of two distinct bound states of 1D spinons.

\subsection{Interchain Coupling via 1D Spinon Interactions and Exact Subsystem Zero Modes}
\label{sec:frustrated_chains_interactions}

We begin by studying the zero-temperature ground state manifold and its elementary excitations. 
At the center of this phase is a special point where $J_1 = 0$, marked with a magenta disk in \cref{fig:phase_diagram}(f).
At this point, the Hamiltonian decouples into 1D ferromagnetic Ising chains, running along each of the cubic $\langle100\rangle$ directions. 
Each chain has two fully-polarized ground states, so the zero-temperature entropy is proportional to the number of chains, $\propto L^2$, where $L$ is the linear size of the system.
The system has an exact $\mathbbm{Z}_2$ subsystem symmetry~\cite{XuReductionEffectiveDimensionality2005,batistaGeneralizedElitzursTheorem2005,nussinovIntermediateSymmetriesElectronic2006,youSubsystemSymmetryProtected2018,shirleyFoliatedFractonOrder2019,batistaGeneralizedElitzursTheorem2005,nussinovCompassModelsTheory2015}---flipping spins on any chain is an exact symmetry of the Hamiltonian.
While these exact symmetries do not survive when $J_1 \neq 0$, a remnant of them still plays a key role in the low-temperature physics in the form of exact zero modes, as explained below.

\begin{figure}[t]
    \centering
    \begin{overpic}[width=\linewidth]{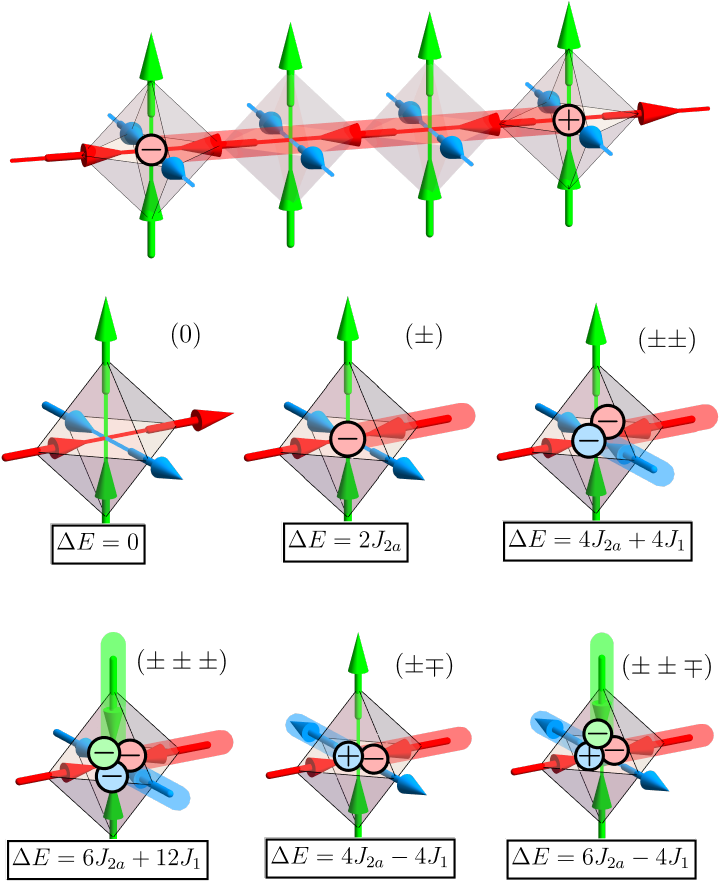}
        \put(01,93){\large{(a)}}
        \put(01,65){\large{(b)}}
        \put(32,65){\large{(c)}}
        \put(60,65){\large{(d)}}
        \put(01,28){\large{(e)}}
        \put(32,28){\large{(f)}}
        \put(60,28){\large{(g)}}
    \end{overpic}
    \\[3ex]
    \phantom{i}\hfill
    \begin{overpic}[width=.95\linewidth]{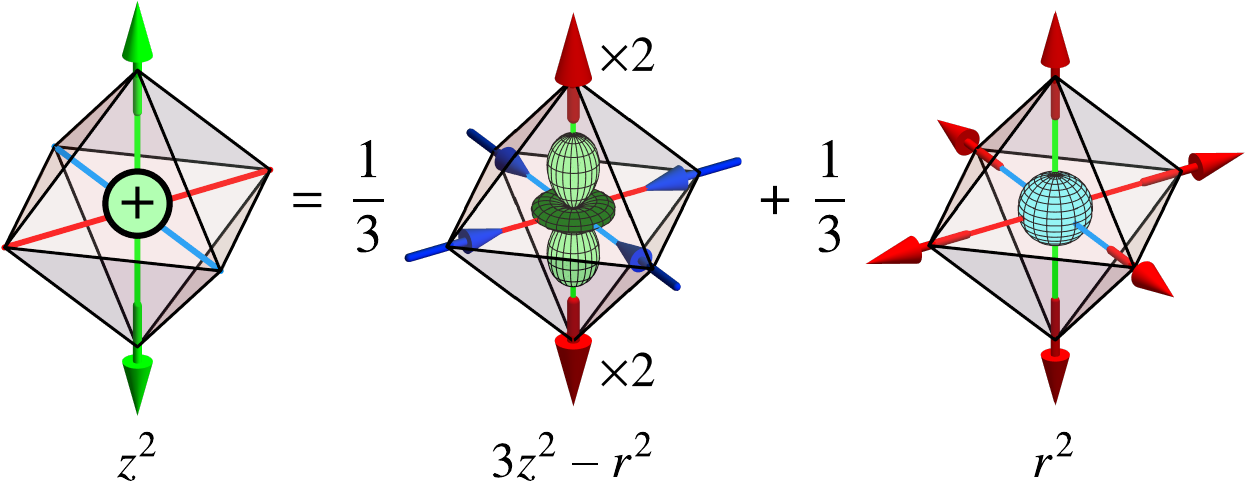}
        \put(-4,35){\large{(h)}}
    \end{overpic}
    \\[-1ex]
    \caption{
        \textbf{1D spinon interactions in the $\bm{T_{1g}}$ phase.}
        (a) When $J_1=0$, the system decouples into 1D Ising chains.  
        For $J_{2a}> 0$, each chain is ferromagnetic, and the fundamental excitations are positive and negative charged spinons created from a ground state by flipping a straight open string (chain segment) of head-to-tail spins, denoted by the red tube. 
        (b) For small $\vert J_1 \vert$, the energy of the polarized chain states does not change, and the system retains a subextensive ground state degeneracy. 
        (c) The single-spinon energy gap is $2J_{2a}$, independent of $J_1$.
        (d-g) A finite positive (negative) $J_1$ induces attraction (repulsion) between spinons of opposite charge. 
        (h) A single spinon carries both a conserved $A_{1u}$ monopole charge and a conserved uniaxial $E_u$ quadrupole charge, with the local uniaxial director aligned along its direction of motion. 
    }
    \label{fig:spinon_interactions}
\end{figure}

One-dimensional Ising chains are the prototypical example of fractionalized Ising systems and spin liquid behavior: single spin flips create pairs of domain walls which can separate at zero energy cost, i.e. they are deconfined quasiparticles; rather than exhibiting long-range order, 1D chains are instead critical, with a diverging correlation length as $T\to 0$.
By the standard Peierls argument, their configurational entropy massively outweighs their energetic cost, preventing the spontaneous breaking of the chain's $\mathbbm{Z}_2$ symmetry.
We will refer to these fractionalized, deconfined 1D domain walls as \textit{spinons}.
\Cref{fig:spinon_interactions}(a) illustrates a pair of spinons at the ends of a domain of reversed spins (red tube).
The positive (negative) spinon corresponds to a domain wall with two red spins pointing out of (in to) an octahedron, i.e. locally appearing as a ``source'' (``sink'').

Now consider turning on a non-zero $J_1$. 
This couples different chains and explicitly breaks the exact $\mathbbm{Z}_2$ subsystem symmetry present at $J_1=0$.
However, it does not lift the ground-state degeneracy, because the single octahedron ground states are unchanged when $J_1$ is introduced: 
for both $J_1=0$ and $\cramped{J_1\neq 0}$, the ground states are precisely the six pure-$T_{1g}$ configurations, such as the one shown in \cref{fig:phase_diagram}(c), in which antipodal moments are ferromagnetically aligned. 
These states are related by cubic symmetry and therefore remain exactly degenerate, as the symmetric $J_1$ interaction cannot energetically distinguish between them.
Since the global ground-state manifold is obtained by gluing together these single-octahedron ground states, every configuration of polarized chains that was a ground state at $\cramped{J_1=0}$ remains a ground state for $\cramped{J_1 \neq 0}$. 
The system therefore retains its subextensive ground-state degeneracy---two states per chain---throughout the entire $T_{1g}$ phase. 
We therefore refer to this as the \emph{frustrated chains phase}.

Rather than lifting the ground state degeneracy, the $J_1$ interaction couples chains by inducing a contact interaction between spinons at the intersections between chains.
In \cref{fig:spinon_interactions}(b-g), we illustrate for each single-octahedron spin configuration its associated description in terms of spinons, with spins and spinons from  each of the three intersecting chains colored red, blue, and green.
Below each configuration is its energy $\Delta E$ relative to the ground state.
The $J_1$ interaction does not shift the zero-spinon (\cref{fig:spinon_interactions}(b)) or single-spinon (\cref{fig:spinon_interactions}(c)) energies.
Rather, it only changes the multi-spinon energies (\cref{fig:spinon_interactions}(d-g)).
Let $\smash{\cramped{\s_{\o}^{\alpha}= +1}}$ ($-1$) denote a spinon (anti-spinon) at octahedron $\o$ on the chain in direction $\alpha\in\{x,y,z\}$, and $\cramped{\smash{\s_{\o}^{\alpha}=0}}$ denote no spinon~\footnote{
    In terms of the spin variables $S_i^z$, the spinon charge is given by $\smash{\cramped{\s_{\o,\alpha} = \frac{1}{2} (S^z_{\o+\alpha}+S^z_{\o-\alpha})\in \{-1,0,1\}}}$, where $\cramped{\o\pm\alpha}$ denotes the spin in the direction $\pm \alpha$ relative to the center of octahedron $\o$, and we have used the 6-site basis used in the Hamiltonian \cref{eq:H} (\cf \cref{apx:conventions}).
}.
Then the Hamiltonian can be rewritten exactly as
\begin{equation}
    H = E_0 + 2J_{2a} \sum_{\o}\sum_{\alpha} \vert \s_{\o}^{\alpha}\vert^2 + 4J_1 \sum_\o \sum_{\alpha<\beta} \s_{\o}^{\alpha} \s_{\o}^{\beta},
    \label{eq:H_fchains_spinons}
\end{equation}
where $E_0$ is the ground state energy (\cf~\cref{eq:H_zero_energy}).
Positive (negative) $J_1$ induces attraction (repulsion) between oppositely charged spinons from different chains.
This realizes an unusual form of frustration in which inter-chain coupling neither lifts the subextensive degeneracy nor alters the single-particle energies, but instead appears exclusively through interactions between otherwise free fractionalized excitations~\footnote{
    An analogous situation also occurs in the  $J_1$-$J_{2a}$ checkerboard lattice Ising model, where $\cramped{J_{2a}>J_1}$ selects a subextensive set of polarized-chain ground states~\cite{henrySpinwaveAnalysisTransversefield2012}, though that problem has not been studied from the perspective of spinon interactions. 
}.

\begin{figure*}
    \centering
    \begin{overpic}[width=.99\linewidth]{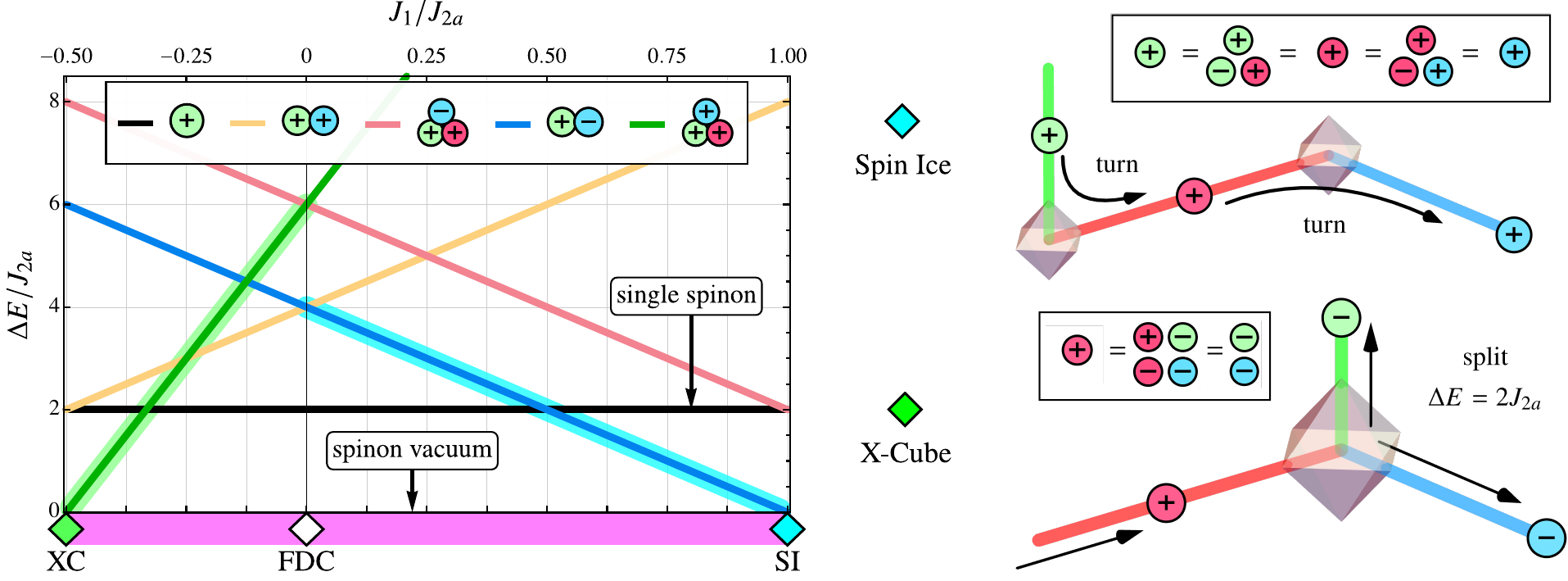}
        \put(-1,34){\large{(a)}}
        \put(55,34){\large{(b)}}
        \put(55,14){\large{(c)}}
    \end{overpic}
    \caption{
        \textbf{ Spin liquids as spinon bound state condensates.}
        (a) Variation of multi-spinon energies (\cf~\cref{fig:spinon_interactions}) as a function of $J_1/J_{2a}$. 
        When $J_1 = 0$ the Hamiltonian reduces to ferromagnetic decoupled chains (FDC).
        Pairs of oppositely charged spinons [\cref{fig:spinon_interactions}(f)] condense at the spin ice point $\cramped{J_1/J_{2a}=1}$, while triple-charges bound states [\cref{fig:spinon_interactions}(e)] condense at the X-cube point $\cramped{J_1/J_{2a} = -1/2}$, giving rise to two different CSLs.
        (b) At the spin ice point, a single spinon can turn a corner onto an orthogonal chain without costing any energy; there is no distinction between the colors of spinons, and they are free to move in 3D. 
        (c) At the X-cube point, a single spinon is equivalent to a bound state of two anti-spinons from the orthogonal directions. Thus, individual spinons are confined to move along 1D lines---they are lineons.
        The bound state can be split by paying energy $\Delta E = 2J_{2a}$ to turn a single spinon into two anti-spinons.
    }
    \label{fig:spinon_condensation}
\end{figure*}

While the $\cramped{J_1 \neq 0}$ coupling between chains explicitly breaks the subsystem symmetries of the decoupled-chain limit, some remnants of these symmetries are retained in the coupled system.
In particular, these symmetries survive as \emph{accidental} symmetries
---symmetries of the ground state manifold which are not symmetries of the Hamiltonian~\cite{javanparastOrderbydisorderCriticalityXY2015}---since the ground states are unaffected by coupling between spinons.
Moreover, the interaction matrix retains exact chain-based zero modes (\cf~\cref{sec:degen_and_flat_bands})---any fully-polarized chain can be flipped at zero energy cost. 
To see this, first note that when $\cramped{J_1 = 0}$ the interaction matrix spectrum reduces to that of decoupled chains, each contributing a cosine dispersion.
The minima of these bands collectively form zero-energy loci in the $(hk0)$ and equivalent cubic planes, and the corresponding eigenvectors correspond to uniform polarized configurations on each chain.
When perturbed with $\cramped{J_1 \neq 0}$, the fact that the Hamiltonian \cref{eq:H_fchains_spinons} only couples positive-energy excitations implies that the zero-energy flat planes are preserved, and correspondingly the ground state manifold and chain-flipping zero modes are retained.
These chain-based zero modes imply that the spinon quasiparticles are energetically deconfined on each chain: flipping a polarized segment only costs energy at its endpoints, so there is no string tension between spinons (domain walls between segments with different polarization), as is manifest in the spinon-interaction form of the Hamiltonian, \cref{eq:H_fchains_spinons}.
As we will see, the persistence of these zero modes throughout the excitation spectrum has much stronger consequences than ground-state accidental symmetries alone, precluding the order-by-disorder mechanism that ordinarily lifts accidental degeneracies~\cite{villainOrderEffectDisorder1980}.

\subsection{Spin Liquids as Spinon Bound State Condensates}
\label{sec:CSLs_as_Spinon_Condensates}

Before we explore the finite-temperature behavior of this phase, it will prove extremely instructing to examine carefully the behavior of the spinon interaction term in ~\cref{eq:H_fchains_spinons} and what happens at the boundaries of the frustrated chains phase, where the $T_{1g}$ irrep becomes degenerate with either the $A_{1u}$ or $E_u$ irrep and the system passes from this frustrated chains regime to two different CSLs.

\Cref{fig:spinon_condensation}(a) shows the evolution of the multi-spinon energies listed in \cref{fig:spinon_interactions}(b-g) as $J_1$ is varied. 
When $\cramped{J_1 > 0}$, opposite-charge spinons attract and can form bound states with lower energy than separated spinons. 
As shown in \cref{fig:spinon_interactions}(f), with increasing $J_1/ J_{2a}$, the energy of the spinon/anti-spinon bound states decreases as $\cramped{J_1/J_{2a}\to 1}$ until they \emph{condense}---they become degenerate with the spinon vacuum.  
Condensation of the spinon/anti-spinon bound state implies that the three colors of spinons, corresponding to the three chain directions, become \emph{indistinguishable}, as illustrated in \cref{fig:spinon_condensation}(b): a green spinon moving along its chain (by flipping spins) can freely transmute to a red spinon by nucleating an anti-green/red pair from the vacuum, allowing it ``turn a corner'' and move along a red chain; it may then further transmute to a blue spinon to move along a blue chain, and so on.
This implies that the excitation spectrum contains only a single type of spinon---a fully mobile three-dimensional quasiparticle.
Viewing spinon trajectories as strings of flipped spins, as in \cref{fig:spinon_interactions}(a), condensation of the spinon/anti-spinon bound states allows strings from different chains to be joined end-to-end without paying any energy at the ``corner'' where they meet. 
These can form arbitrary closed loops in 3D, corresponding to processes where a pair of spinons are created, propagated around a loop and re-annihilated, giving rise to a classical string-net condensate~\cite{levinStringnetCondensationPhysical2005} or Coulomb phase~\cite{henleyCoulombPhaseFrustrated2010}.
This is precisely the spin ice point in the phase diagram, where the spinons of the frustrated chains phase have become the ``magnetic monopoles'' of the spin ice 
CSL~\cite{castelnovoMagneticMonopolesSpin2008}.

Within the frustrated chains phase, spinons are excitations on top of a pure-$T_{1g}$ ground state.
Each spinon individually carries both an $E_u$ and $A_{1u}$ charge which is conserved as it moves along its chain (by flipping spins).
This is illustrated in \cref{fig:spinon_interactions}(h): a spinon moving along a $z$-directed chain, characterized by two $z$-directed spins locally pointing ``out'' of an octahedron, carries a positive $+r^2$-type  $A_{1u}$ monopole moment and a $(3z^2-r^2)$-type $E_u$ quadrupole moment.
Condensing the spinon/anti-spinon bound states is equivalent to condensing the $E_u$ quadrupole charge, meaning that it is no longer a conserved quantity.
This implies that the resulting fully mobile 3D quasiparticles only conserve the $A_{1u}$ monopole charge.
Correspondingly, the spin ice Hamiltonian \cref{eq:H_Spin_Ice} only penalizes the $A_{1u}$ monopole charge locally, while the $T_{1g}$ dipole and $E_u$ quadrupole moments are free to fluctuate at zero energy cost.
Since the propagation direction of the frustrated chain spinon is encoded in the orientation of its local quadrupole moment, condensation of the $E_u$ sector effectively removes this directional label, rendering the spinon fully mobile in three dimensions.

These arguments illustrate how spin ice arises from coupling intersecting ferromagnetic chains by condensation of spinon bound states.
The natural next question is what type of CSL is produced by a different pattern of spinon condensation at the opposite end of the frustrated chains phase, with $\cramped{J_1 < 0}$. 
As $\cramped{J_1/J_{2a} \to -1/2}$, the \emph{triple spinon} $\cramped{(\pm\pm\pm)}$ bound states, shown in \cref{fig:spinon_interactions}(e), condense into the vacuum. 
Generalizing the above arguments, it is clear that this will result in a CSL which is not a string-net, but rather constructed from ``cages'' of straight edges intersecting in trios at corners and whose quasiparticles carry quadrupolar charges.
This yields the fractonic X-cube CSL, the topic of \cref{sec:fracton_CSL}, with completely distinct physics to spin ice but whose properties can also be readily deduced from the spinon bound state condensation picture.

\subsection{Dimensional Crossover and Avoided Kasteleyn-Like Transitions}
\label{sec:kasteleyn}

Having explored the ground state properties of the frustrated chains phase, its fractionalized excitations, and the bound state condensation mechanism giving rise to two distinct spin liquids at the phase boundaries, we now turn to explore the finite-temperature behavior using classical Monte Carlo simulations.

Given a large ground state manifold, one generally expects a finite-temperature transition driven by order-by-disorder~\cite{villainOrderEffectDisorder1980}, where some ground states admit greater fluctuations than others and so have lower free energy at finite temperature.
Generally speaking, a low-temperature series expansion about different ground states yields differences in finite-temperature free energy, and the system selects the ground state with the lowest free energy~\footnote{
    For continuous spin systems this is generally straightforward---one performs a spinwave expansion by examining the effects of smooth continuous deformations of the ground state spin texture. 
    Ising systems are not amenable to such a linearized treatment, but in principle one can organize the low-temperature expansion in terms of single spin flips on top of a given ground state~\cite{danielianLowTemperatureBehaviorFaceCentered1964,slawnyLowtemperatureExpansionLattice1979}.
}.
However, this procedure is not applicable to the frustrated chains phase described by the spinon-interaction Hamiltonian \cref{eq:H_fchains_spinons}---because spinons are domain walls \emph{between} ground states, it is not meaningful to ask which ground state they propagate on top of. 
They are not created by local deformations (spin flips) of a polarized chain state, rather they are created by flipping a non-local string of spins.
The problem of expanding around a particular ground state is therefore ill-posed: it requires summing over configurations ``near'' a polarized chain state, but any such configuration containing even a single spinon pair is simultaneously ``near'' the ground state reachable by winding that spinon pair around the periodic boundary at zero energy cost.
The lack of such an expansion suggests that energetic interactions between fractionalized excitations should not be able to mediate selection among the ground state vacuum manifold.

\begin{figure*}[t]
    \centering
    \hspace{.01\textwidth}
    \begin{overpic}[width=0.56\textwidth]{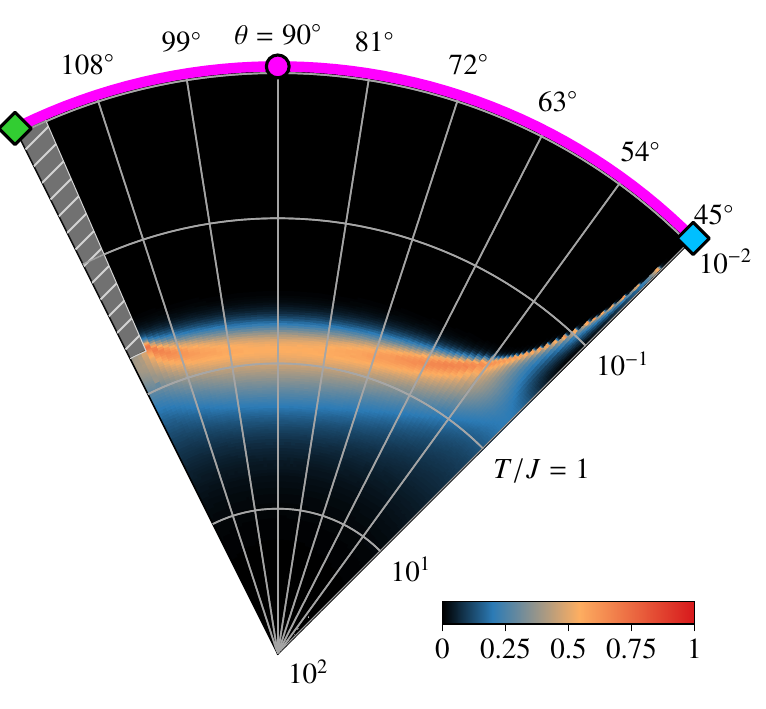}
        \put(0,86){\large (a)}
        \put(63,16){\large Specific heat}
        \put(50.0,71.0){\large \textcolor{white}{avoided}}
        \put(45.0,66.0){\large \textcolor{white}{Kasteleyn-like}}
        \put(49.2,61.0){\large \textcolor{white}{transition}}
        \linethickness{1.5pt}
        \put(67,63){\color{white}\vector(1,-1.1){10}}
        \put(09.9,46){\rotatebox{-63.4349488229}{numerically inaccessible}}
        \put(4,51){
            \rotatebox{-63.4349488229}{\includegraphics[width=0.04\linewidth]{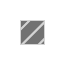}}
        }
    \end{overpic}
    \hfill
    \begin{overpic}[width=0.415\textwidth]{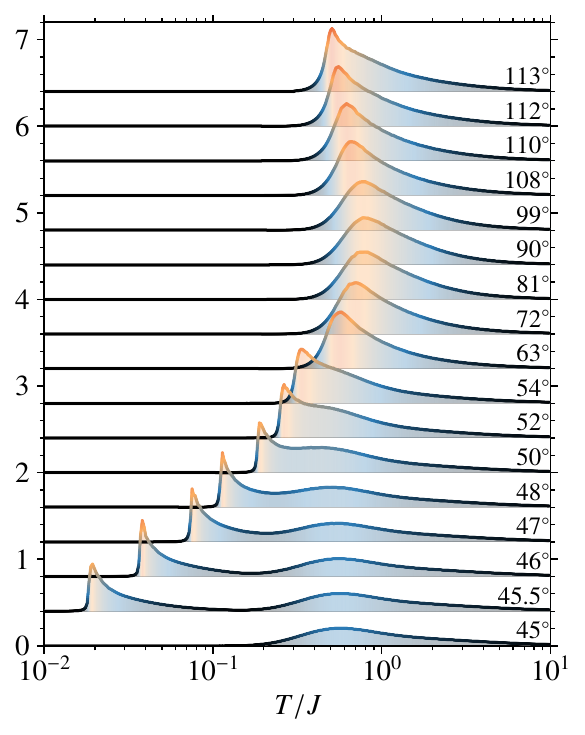}
        \put(-6,95){\large (b)}
        \put(8,93){Specific heat cuts}
    \end{overpic}
    \caption{
    \textbf{Specific heat in the frustrated chains phase.} 
    (a) Angular map of the specific heat throughout the frustrated chains phase (\cf~\cref{fig:phase_diagram}) taken from Monte Carlo simulations. 
    In the vicinity of the decoupled chains point ($J_1=0$ at $\theta = 90^\circ$) the specific heat exhibits a smooth crossover. 
    When approaching the spin ice point ($\theta = 45^\circ$), this smooth crossover splits into two features, a high-temperature crossover upon entering into a finite-temperature spin ice CSL regime and a sharp lower-temperature anomaly dropping towards zero temperature. 
    We label this anomaly as an avoided Kasteleyn---a finite maximum in the specific heat corresponding to a rapid crossover arising from being close to a Kasteleyn transition within an extended phase diagram.
    At the opposite end near the X-cube point we are unable to thermalize to low temperatures. 
    (b) Cuts of the specific heat, demonstrating the evolution from a smooth crossover near the decoupled chains point to the emergence of a sharp low-temperature peak near the spin ice point. 
    }
    \label{fig:frustrated_chains_arc}
\end{figure*}

There are a number of reasons to expect that this system does not order at any temperature.
For one, the existence chain-based zero modes discussed in \cref{sec:frustrated_chains_interactions}---flipping polarized chains costs zero energy---imply that the spinon excitations remain deconfined.
The spinon-interaction form of the Hamiltonian, \cref{eq:H_fchains_spinons}, is conceptually very similar to ``extended spin ice''~\cite{rauSpinSlushExtended2016} and related models~\cite{mizoguchiClusteringTopologicalCharges2017,mizoguchiMagneticClusteringHalfmoons2018,tokushukuTrimerClassicalSpin2019}.
Those models are constructed from a classical spin liquid by adding short-range energetic interactions between the deconfined quasiparticle excitations. 
Since these interactions are only between positive-energy excitations they do not drive order-by-disorder selection within the ground state manifold; if the system does not order in the absence of such interactions, then it will not order when made weakly interacting. 
By the same logic, we do not expect a finite-temperature phase transition in the frustrated chains phase, because the decoupled chains remain disordered at finite temperature.
Instead, we expect a ``dimensional crossover''; a smooth crossover from the high-temperature 3D paramagnetic phase to a low-temperature phase of weakly interacting 1D chains---intra-chain correlations diverge at low temperature while inter-chain correlations tend to zero, with the entropy per spin vanishing at $\cramped{T=0}$ without any finite-temperature phase transition.

This lack of order is consistent with what we find in classical Monte Carlo simulations.
\Cref{fig:frustrated_chains_arc}(a) shows the specific heat as a function of the parameter angle $\theta$ (\cref{eq:theta_J1_J2a}) and temperature~$T/J$, where the wedge shape shown can be directly compared to the phase diagram in~\cref{fig:phase_diagram}(f). 
\Cref{fig:frustrated_chains_arc}(b) shows cuts of the specific heat at various angles.
At the decoupled chains point ($\theta = 90^{\circ}$), the specific heat necessarily matches the exact analytic expression for finite-length 1D Ising chains, which in the thermodynamic limit converges to a smooth Schottky peak corresponding to the thermal depopulation of gapped non-interacting spinons of energy $2J_{2a}$ (\cref{apx:1D_Chains}).  
Coupling the chains with a moderate $\cramped{J_1 \neq 0}$, the specific heat displays a qualitatively similar sooth crossover form, with no sharp anomaly to indicate a phase transition.

In the neighborhood of the classical spin ice and X-cube spin liquids at the ends of the frustrated chains phase, the behavior of the specific heat differs qualitatively from that of decoupled 1D chains. 
It is precisely near these endpoints that the spinon bound states have significantly lower energy than their constituent spinons, see \cref{fig:spinon_condensation}(a).
We focus on the end of the phase near the spin ice point ($\theta = 45^{\circ}$).
There, the short-loop worm update algorithm for spin ice models~\cite{melkoMonteCarloStudies2004,jaubertAnalysisFullyPacked2011}, supplemented with a Metropolis acceptance criterion accounting for the finite energy cost of loop updates, allows our Monte Carlo simulations to access the very low-temperature regime. 
As seen in \cref{fig:frustrated_chains_arc}(b), the specific heat exhibits two distinct features---a higher-temperature Schottky-like bump and a lower-temperature sharp asymmetric peak.
This sharp feature emerges at $T=0$ from the spin ice point ($J_1/J_{2a} = 1$, or $\theta = 45^\circ$) and rises to higher temperature when moving towards the decoupled chains point ($J_1 = 0$, $\theta = 90^\circ$).
As it rises, it broadens and merges with the Schottky-like crossover bump. 
This two-peak specific heat structure was also reported for $\cramped{J_{2a}/J_1 = 1.05}$ ($\theta\approx 46.4^{\circ}$) in Ref.~\cite{szaboFragmentedSpinIce2022}.
There, it was suggested, based on the lopsided shape of the specific heat curve, that the low-temperature peak could signal a Kasteleyn-like transition~\cite{jaubertThreedimensionalKasteleynTransition2008,jaubertKasteleynTransitionThree2009,jaubertTopologicalConstraintsDefects2009,pottsSpinIceGeneral2022}, a scenario which we develop more fully here.

The sharp low-temperature specific-heat anomaly is most naturally understood in relation to the nearby spin ice CSL and its description as a spinon bound state condensate (\cref{sec:CSLs_as_Spinon_Condensates}).
Within the loop-gas picture of spin ice~\cite{jaubertAnalysisFullyPacked2011}, a spinon/anti-spinon bound state corresponds to a ``kink'' where a string makes a 90-degree turn from one chain direction to another~\footnote{
    These strings are only well-defined relative to a fixed spin ice ground state; here it is most sensible to use one of the polarized-chain states which are ground states of the frustrated chains phase, such as the state with all spins polarized along the positive Cartesian directions.
}.
Microscopically, a string costs energy $\cramped{J_{\text{spinon}} = 2J_{2a}}$ for each endpoint and energy $\cramped{J_{\text{kink}}=4( J_{2a}-J_1)>0}$ for each kink (\cf~\cref{fig:spinon_interactions}(f)).
The condensation of these bound states at the spin ice point corresponds to the vanishing of this bending energy, $\cramped{J_{\text{kink}}\to 0}$, thus allowing loops to assume arbitrary shapes at zero temperature.
The energy scale $J_{\text{kink}}$ may therefore be viewed as an energetic cost for \emph{bending} loops.
Sufficiently close to the spin ice point there is a well-defined separation of energy scales, $\cramped{0<J_{\text{kink}}\ll J_{\text{spinon}}}$. 
Upon cooling below $\cramped{T \sim J_{\text{spinon}}}$, the system undergoes a thermal depopulation of spinons (string ends) and falls into the spin ice manifold of closed-string states, signaled in the specific heat by a broad Schottky-like crossover, below which the density of string ends (spinons) is exponentially suppressed.
As the temperature falls further below $\cramped{T \sim J_{\text{kink}}}$ the kinks in the loop-gas become energetically expensive, causing the strings to ``straighten out'' and release their remaining entropy.
The system then enters the subextensively degenerate manifold of polarized chains states, now viewed as a small subset of the spin ice ground state manifold.

This physics is strongly reminiscent of the Kasteleyn transition of pyrochlore spin ice in an external field~\cite{jaubertThreedimensionalKasteleynTransition2008,jaubertKasteleynTransitionThree2009,jaubertTopologicalConstraintsDefects2009,pottsSpinIceGeneral2022}. 
A Kasteleyn transition is a type of confinement transition which occurs between a topological loop-gas or Coulomb phase and a zero-entropy phase of fully polarized strings~\cite{kasteleynDimerStatisticsPhase1963}. 
It exhibits a characteristic lopsided specific heat curve which diverges when approaching from above but is discontinuous when approaching from below~\cite{jaubertThreedimensionalKasteleynTransition2008,jaubertKasteleynTransitionThree2009,jaubertTopologicalConstraintsDefects2009}.
The low-temperature specific heat peak near the spin ice point exhibits a very similar lopsided profile, as seen in \cref{fig:frustrated_chains_arc}
(c)~\cite{szaboFragmentedSpinIce2022}.

The Kasteleyn transition is generally induced by a potential (such as an external field) which energetically prefers strings to polarize in one direction~\cite{jaubertThreedimensionalKasteleynTransition2008,jaubertKasteleynTransitionThree2009,jaubertTopologicalConstraintsDefects2009,pottsSpinIceGeneral2022,powellQuantumKasteleynTransition2022}.
The minimal excitation is an infinite string of flipped spins pointing opposite to the field, and the Kasteleyn transition corresponds to the change in sign of the free energy of a single string~\cite{jaubertThreedimensionalKasteleynTransition2008,jaubertKasteleynTransitionThree2009,jaubertTopologicalConstraintsDefects2009}, 
\begin{equation}
    F_{\text{string}}^{\text{Kast.}} = \ell \sigma - T S(\ell). 
    \label{eq:F_string_Kastleyn}
\end{equation}
Here $\ell$ is the length of the string, $\sigma$ is the string tension (proportional to the field), and $S(\ell)$ is the configurational entropy of a string of length $\ell$.
While the energy cost of a system-spanning string is infinite, the energy density and entropy density are both finite and compete.
Below the transition the system is completely frozen, resulting in the discontinuous behavior of the specific heat when approaching the transition from below.

In the octochlore model, analogous physics arises with the string tension $\sigma$ replaced by the bending potential $J_{\text{kink}}$, which causes strings to straighten out but does not select a particular ground state. 
The energy and configurational entropy of a closed string both scale with the number of kinks, 
\begin{equation}
    F_{\text{string}} = N_{\text{kink}} J_{\text{kink}} - T S(N_\text{kink}),
    \label{eq:F_string_kink}
\end{equation}
which should be compared to \cref{eq:F_string_Kastleyn}.
However, unlike the ordinary Kasteleyn transition, the minimal excitation is not an infinite string, but rather a four-corner rectangular loop of arbitrary size. 
This means that below the transition the system can continue to fluctuate, albeit very weakly because a string with few kinks has very few configurational states compared to a string with many kinks. 
While this is not a Kasteleyn transition in the strict sense, it is clearly a closely related type of confinement transition, which we will describe as Kasteleyn-like~\footnote{
    Some similar models have been explored as lattice models of semi-flexible polymers~\cite{floryStatisticalThermodynamicsSemiflexible1956,petschekSemiflexibleSelfavoidingPolymers1985}.
}.

Crucially, however, a Kasteleyn transition can only occur in the limit where the strings are not allowed to have endpoints, i.e. when $\cramped{J_{\text{spinon}} \to \infty}$ with $J_{\text{kink}}$ held finite~\cite{jaubertThreedimensionalKasteleynTransition2008,jaubertKasteleynTransitionThree2009,jaubertTopologicalConstraintsDefects2009}.  
The existence of finite-energy spinons (string ends) always allows the system to escape a polarized ground state.
The lengthscale introduced by the finite density of single spinons in the system cuts off the growth of correlations and rounds off the transition, turning it into a smooth crossover which becomes arbitrarily sharp as $\cramped{J_{\text{spinon}}/J_{\text{kink}} \to \infty}$~\cite{jaubertThreedimensionalKasteleynTransition2008,jaubertKasteleynTransitionThree2009,jaubertTopologicalConstraintsDefects2009,timoninSpinIceField2011}.
This limit is only reached in our phase diagram when $\cramped{J_{\text{kink}} \to 0}$ i.e. at the spin ice point, so the Kasteleyn transition never actually occurs. 
We thus interpret the sharp peak in the specific heat at low temperatures near the spin ice point as an \emph{avoided} Kasteleyn-like transition---a quasi-critical crossover~\cite{timoninSpinIceField2011}.

Analogous quasi-critical behavior occurs in pyrochlore classical spin ice in a magnetic field.
Historically, Ref.~\cite{harrisLiquidGasCriticalBehavior1998} studied the problem of pyrochlore spin ice in a [001] field and mistakenly identified the sharp specific heat feature with a line of first-order transitions.
It was not until later that it was understood that this was an avoided Kasteleyn transition~\cite{jaubertThreedimensionalKasteleynTransition2008,jaubertKasteleynTransitionThree2009,jaubertTopologicalConstraintsDefects2009,timoninSpinIceField2011}, because the actual transition can only be reached in the limit that the monopole energy is taken to infinity.
Indeed, \cref{fig:frustrated_chains_arc}(c) can be directly compared to Figure 3 in Ref.~\cite{harrisLiquidGasCriticalBehavior1998}.
In the octochlore model, $J_{\text{kink}}$ plays a role akin to the field, but the actual Kasteleyn-like transition only occurs in an \emph{extended} phase diagram in which $J_{\text{spinon}}$ and $J_{\text{kink}}$ are tuned independently, so that one can take $\cramped{J_{\text{spinon}}\to \infty}$ while keeping $J_{\text{kink}}$ finite.
The sharp specific heat anomaly observed in our phase diagram which sharpens upon approaching the spin ice point can therefore be viewed as a quasi-critical line~\cite{timoninSpinIceField2011}---a sharp but finite-width peak in thermodynamic quantities which arises as the ``shadow'' of a nearby critical phase transition~\footnote{
    The quasi-critical crossover due to finite monopole energy is analogous to applying a weak symmetry-breaking field turning a symmetry breaking phase transition into a crossover which becomes arbitrary sharp as the field is weakened.
    The Kasteleyn transition may be viewed as symmetry breaking in the following manner: the spin ice ground state manifold has a generalized symmetry called a 1-form symmetry corresponding to the conservation of flux lines through closed surfaces~\cite{gaiottoGeneralizedGlobalSymmetries2015,mcgreevyGeneralizedSymmetriesCondensed2023,chung2formU1Spin2025}, which is spontaneously broken in the ``flux-condensed'' disordered phase in which strings proliferate.
    However, this symmetry is explicitly broken by the presence of charges, which are sources and sinks of flux lines. 
}
.

To summarize, on the $J_1 > 0$ side of the frustrated chains phase we find no signal of a finite-temperature thermodynamic phase transition.
Near the decoupled chains point the system exhibits a smooth dimensional crossover from the 3D paramagnet to weakly interacting 1D chains, while near the spin ice end of the phase it first enters the spin ice manifold before exiting by a quasi-critical crossover into the 1D chains regime.

\begin{figure*}[t]
    \centering
    \begin{overpic}[width=0.36\textwidth]{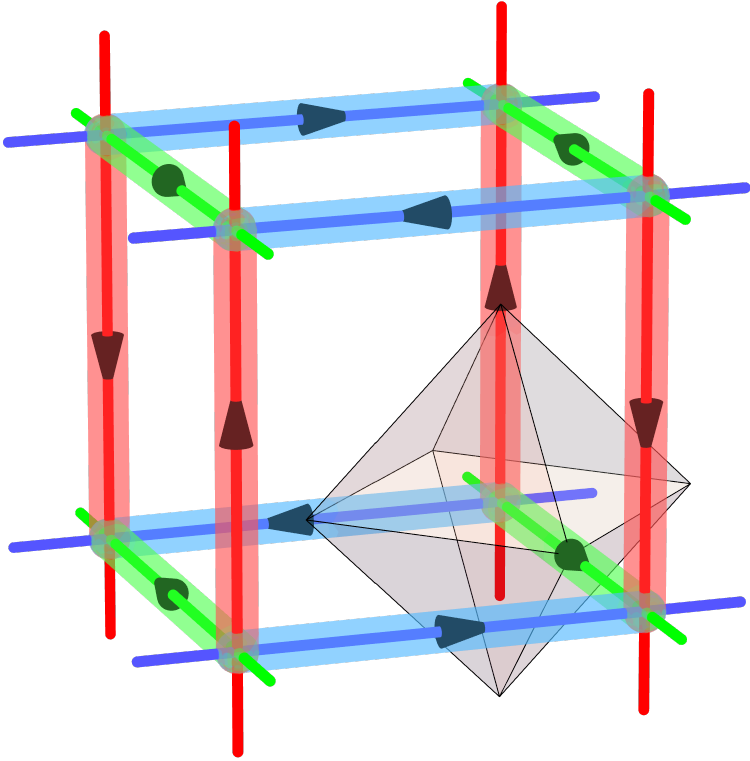}
        \put(0,94){\large{(a)}}
    \end{overpic}
    \hfill 
    \raisebox{.05\height}{
        \begin{overpic}[width=0.24
        \textwidth]{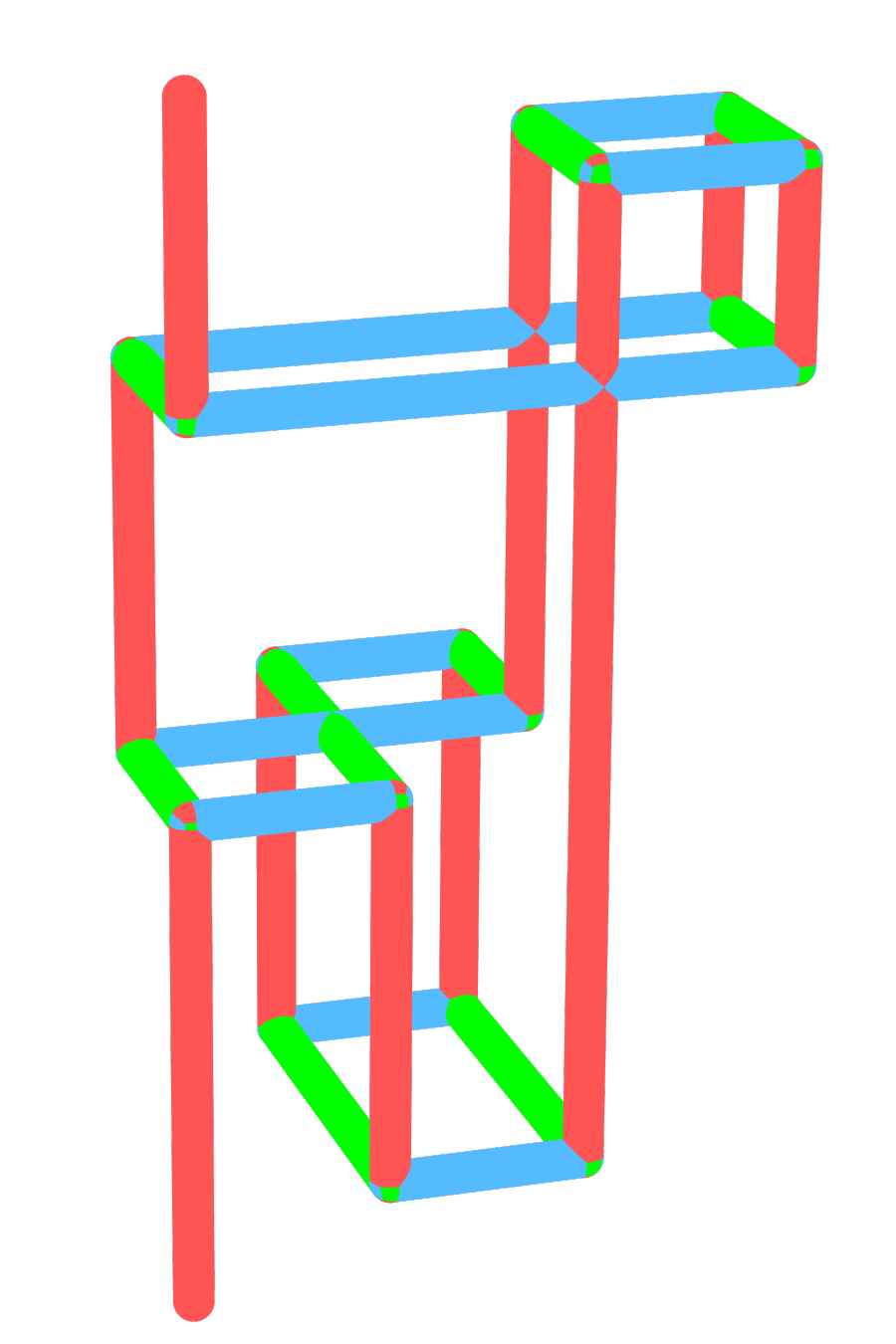}
            \put(-4,91){\large{(b)}}
            \put(16,91){\textcolor{red}{\scriptsize{$\uparrow$ PBC}}}
            \put(16.5,2.5){\textcolor{red}{\scriptsize{$\downarrow$ PBC}}}
        \end{overpic}
    }
    \hfill
    \begin{overpic}[width=0.36\textwidth]{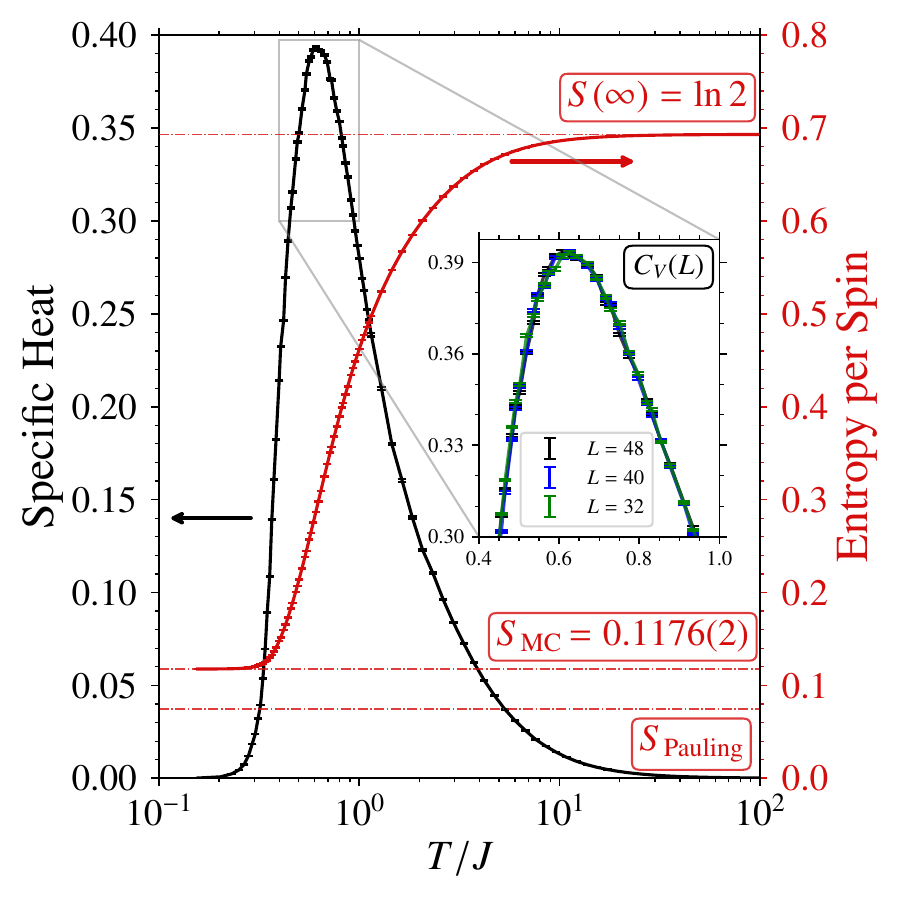}
        \put(-4,95){\large{(c)}}
    \end{overpic}
    \caption{
    \textbf{ Cage-net fracton spin liquid.}
    (a) 
    Ising spins (black arrowheads) may be viewed as lying along edges of the parent cubic lattice (thin colored lines).
    The minimal local zero-energy collective spin flip is a closed cage of twelve spins on the edges of a cube in the arrangement shown. 
    (b) An illustration of a larger ``cage-net'' flippable motif, which may close around the periodic boundaries (PBC). 
    At zero temperature, such collective flips ``tunnel'' the system between degenerate vacua.
    (c) Specific heat (left axis indicated by black left-pointing arrow) and entropy per spin (right vertical axis, indicated by right-pointing red arrow) of the fracton spin liquid obtained from cluster Monte Carlo simulation with linear dimension $L=48$ using a 3-site unit cell (\cref{apx:conventions}). 
    The inset shows a lack of system-size scaling of the specific heat maximum with respect to system size for $\cramped{L=32}$, $40$, $48$, indicating a smooth crossover and no transition. 
    A description of the cluster algorithm which builds cage nets is presented in \cref{apx:cage_net_cluster_algorithm}.
    Simulation details are provided in \cref{apx:monte_carlo}.
    }
    \label{fig:cages_and_thermodynamics}
\end{figure*}

It is worth commenting on what happens when we go beyond the spin ice point into the fragmented spin ice phase ({i.e.} when $\theta<45^\circ$).
This occurs when the kink energy becomes negative, $J_{\text{kink}}<0$, so the fragmented spin ice phase may be loosely thought of as ``maximally kinked spin ice''. 
However this is only heuristic---the stability of the spin ice physics in this phase depends strongly on the microscopic spin length constraint and lattice details through the fragmentation of the $E_u$ irrep discussed in \cref{sec:fragmentation_of_Eu}.
In the continuum description of spin ice, negative bending energy signals an instability whose fate can only be resolved by carefully examining the microscopic Ising model.

Finally, we turn to the $J_1 < 0$ side of the frustrated chains phase, and the limit where $J_1/J_{2a} \to -1/2$ and the triple-spinon bound states condense, as discussed in~\cref{sec:CSLs_as_Spinon_Condensates}.
This point is studied in great detail in the following  \cref{sec:fracton_CSL}, where we show that it realizes a \emph{fractonic} classical spin liquid whose zero-energy condensed objects are not loops but ``cages'', such as illustrated in \cref{fig:cages_and_thermodynamics}(a,b). 
While we are able to efficiently explore this CSL with a cluster Monte Carlo algorithm (\cref{apx:cage_net_cluster_algorithm}), that algorithm fails when the Hamiltonian is even slightly perturbed for $\cramped{J_1/J_{2a} \gtrsim -1/2}$.
We were able to explore the vicinity of the spin ice CSL using a worm algorithm where loops are constructed via a random walk process~\cite{melkoMonteCarloStudies2004,jaubertAnalysisFullyPacked2011}. 
The cage structures shown in \cref{fig:cages_and_thermodynamics}(a,b) cannot, however, be constructed from random walks. 
In the absence of an efficient cluster algorithm analogous to the worm update, we are unable to thermally equilibrate the system at low temperatures in the frustrated-chain phase near the fracton CSL, corresponding to the gray-shaded region in \cref{fig:frustrated_chains_arc}(b).
Nevertheless, the data available shown in \cref{fig:frustrated_chains_arc}(a) shows that at this end of the phase the specific heat is developing a feature similar to that at the spin ice end. 
The same physical principles underlying the avoided Kasteleyn-like transition in the vicinity of the spin ice point suggest that an analogous quasi-critical crossover should emerge in the vicinity of this CSL, signaling an avoided \textit{fractonic Kasteleyn-like transition}.
Such a putative transition must be controlled by the energy cost of cage corners, i.e. the triple-spinon bound states illustrated in \cref{fig:spinon_interactions}(e), which play the analogous role of the loop kinks near the spin ice point.
Each cage corner costs energy $J_{\text{corner}} = 6(J_{2a} + 2J_1)$.
If single spinons were entirely forbidden, we would expect a Kasteleyn-like transition described by a free energy of the form
\begin{equation}
    F_{\text{cage}} = N_{\text{corner}} J_{\text{corner}} - T S(N_\text{corner}), 
    \label{eq:F_cage_corner}
\end{equation}
with $N_{\text{corner}}$ the number of cage corners.
However, since single spinons are allowed with finite energy, we expect that this putative transition is rounded off, resulting in a finite-height but sharp specific heat feature analogous to the one observed in the vicinity of the spin ice point.
The thermally equilibrated specific-heat data that we were able to obtain are consistent with this expectation: the feature appears as a red ridge forming around $\theta \sim 108^\circ$ in \cref{fig:frustrated_chains_arc}(a), progressively developing into a peak (compare the specific heat curve at $\theta=113^\circ$ with that at $\theta = 52^\circ$ in \cref{fig:frustrated_chains_arc}(b)).
We leave the exploration of this physics to future work, and proceed now to explore the fractonic spin liquid in detail.

\section{Cage-Net Fracton Spin Liquid (\texorpdfstring{$\bm{T_{1g}\mixed A_{1u}}$}{T1g+A1u})} 
\label{sec:fracton_CSL}

In this section, we provide a detailed account of the properties of the spin liquid that arises at the point 
$\cramped{ J_1 / J_{2a} = -1/2}$.
As explained in \cref{sec:CSLs_as_Spinon_Condensates} and illustrated in \cref{fig:spinon_condensation}(a), this point corresponds to the condensation of the triple spinon bound state shown in \cref{fig:spinon_interactions}(e).
As we will see, the condensation of the triple spinon bound states gives rise to a \emph{fracton} classical spin liquid (CSL). 
As shown in the inset of \cref{fig:spinon_condensation}(c), condensation of triple spinon bound states implies that a single spinon (red) is equivalent to a bound state of two anti-spinons of the other two colors (green, blue).
Unlike the spin ice spin liquid which arises by condensing spinon/anti-spinon pairs and has fully mobile quasiparticles, condensing triple spinon bound states does not liberate individual spinons from their chains.
In other words, these excitations are ``lineons''~\footnote{
    The term ``lineon'' refers to a quasiparticle confined to a line, ``planeon'' refers to a quasiparticle confined to a plane, and ``fracton'' refers to a quasiparticle that cannot move (confined to a point). 
    However, the term ``fracton'' is often used loosely for any quasiparticle with sub-dimensional mobility.
}, quasiparticles with sub-dimensional mobility along a line~\cite{pretkoSubdimensionalParticleStructure2017}.
Such sub-dimensional mobility is the hallmark of fractonic phases of matter~\cite{nandkishoreFractons2019,pretkoFractonPhasesMatter2020}.
The triple-spinon $(\pm\pm\pm)$ condensation pattern implies that lineons can be created from (or annihilated to) the vacuum in red-blue-green trios.

\subsection{Emergent Cage-Net Gauge Structure}
\label{sec:cage_net_tunneling}

A classical spin liquid is characterized by a local constraint, i.e. its \emph{emergent gauge structure}. 
According to \cref{eq:H_X-cube}, the local constraint is that each octahedron has zero $E_u$ quadrupole moment, $Q_{\o}=0$. 
Equivalently, the CSL is characterized by its ground state (vacuum) manifold and the zero-energy collective spin flips that tunnel the system between different vacua. 
These tunneling moves should be viewed as processes by which charges nucleate from the vacuum, move around, and recombine to create a new vacuum configuration. 
The simplest example of this is the spin ice CSL: a monopole/anti-monopole pair can be created from the vacuum and split apart at energy cost $2\times 2J_{2a}$.
Then, one of these spinons can  propagate around a closed path (by flipping spins) and re-annihilate with its partner to return the system to a vacuum configuration. 
The resulting collective spin flip is often called a loop move, and forms the basis for efficient numerical Monte Carlo simulation of spin ice physics~\cite{melkoLongRangeOrderLow2001,melkoMonteCarloStudies2004,
jaubertAnalysisFullyPacked2011}.

Applying this logic to the triple-spinon-condensed CSL, we begin by nucleating a spinon trio at one octahedron site, paying energy $3 \times 2J_{2a}$ to split this bound state apart.
Each of the individual spinons can move apart by hopping one octahedron along $+\hat{\bm{x}}$, $+\hat{\bm{y}}$, and $+\hat{\bm{z}}$. 
As illustrated in \cref{fig:spinon_condensation}(c), each single spinon is equivalent to a bound state of two anti-spinons of the other two colors.
These pairs can each be split at energy cost $2J_{2a}$.
For the spinon located at $+\hat{\bm{x}}$, we move the resulting anti-spinons along $+\hat{\bm{y}}$ and $+\hat{\bm{z}}$, and similarly for the other two spinons.
This results in an anti-spinon pair---equivalent to a single spinon each---located at the three octahedra at $\hat{\bm{x}}+\hat{\bm{y}}$, $\hat{\bm{x}}+\hat{\bm{z}}$, and $\hat{\bm{y}}+\hat{\bm{z}}$.
These three spinons can then to the octahedron at $\hat{\bm{x}}+\hat{\bm{y}}+\hat{\bm{z}}$ and mutually annihilate to return to the vacuum. 
This process results in the minimal vacuum-to-vacuum spin flip, or ``fluctuator''~\cite{yanClassificationClassicalSpin2024,yanClassificationClassicalSpin2024a}, involving twelve spins arranged on the edges of a cube as shown in \cref{fig:cages_and_thermodynamics}(a); arrows indicate the relative directions the spins must point for this cluster to be flippable at zero energy cost.
Note that as a process of flipping single spins, this 
vacuum-to-vacuum fluctuation requires passing through intermediate states with at least four spinons, with total energy cost $4 \times 2J_{2a}$.

We refer to this structure as a ``cage''.
Larger vacuum-to-vacuum fluctuators can be formed by such spinon proliferation and recombination processes, and we refer to these in general as ``cage moves'', by analogy with the ``loop moves'' of spin ice~\cite{melkoLongRangeOrderLow2001, melkoMonteCarloStudies2004,jaubertAnalysisFullyPacked2011}. 
In particular, one can combine many copies of the minimal cube to form large contractible flippable cages. 
General cages can be built from straight oriented segments representing lineon trajectories, which are glued together at corners where three lineons are created or annihilated.
This includes, as a special case, a single straight line, meaning that fully polarized chains can be flipped at zero energy cost. 
For a system with periodic boundaries, this corresponds to creating a spinon/anti-spinon pair, sending one of them around a non-contractible handle of the torus, then re-annihilating the pair.
\Cref{fig:cages_and_thermodynamics}(b) illustrates a cage which includes a non-contractible winding. 
These structures correspond precisely to the generalized Wilson loops of fracton gauge theories~\cite{devakulCorrelationFunctionDiagnostics2018}.
Thus, they give rise to a fracton CSL---a classical \emph{cage-net condensate}~\cite{premCageNetFractonModels2019}.

\subsection{Thermodynamics from Cluster Monte Carlo Simulations}
\label{sec:montecarlo}

Having established the vacuum-to-vacuum ``tunneling'' processes characterizing this putative CSL, it remains to demonstrate the thermodynamic stability of the underlying Ising model against the development of any ordering instabilities.
This means that the system possesses a non-zero ground state entropy, does not undergo any finite-temperature ground state selection by an order-by-disorder mechanism, and exhibits algebraically decaying correlations.
Demonstrating these properties requires performing a Monte Carlo simulation of the statistical Boltzmann distribution.
As is the general case with Ising models, single spin flip Markov chain updates face exponentially small acceptance rates at low temperature. 
This problem is doubly compounded in the present fracton spin liquid by the highly correlated sequence of spin flips required to achieve a vacuum-to-vacuum cage move and the associated high energy barrier.
In an ice model, one only pays the price of creating a pair of spinons, after which they can undergo diffusive random walks at zero energy cost until they meet again and annihilate. 
In the fracton CSL, a single spin flip creates two lineons, which cannot move off of a chain; instead, they can only be split as in \cref{fig:spinon_condensation}(c), costing additional energy.
Even after a number of lineons are created, they can only be recombined if their associated lines intersect in trios, to form a cage net such as those in \cref{fig:cages_and_thermodynamics}(a,b).
This makes single spin flip Monte Carlo untenable for low-temperature simulations.

To overcome this problem, we have developed a highly efficient cluster algorithm capable of generating cage-nets like the one shown in \cref{fig:cages_and_thermodynamics}(b).
The algorithm generalizes the graph decomposition algorithm developed by Otsuka for constructing string-nets to simulate spin ice~\cite{otsukaClusterAlgorithmMonte2014}, itself a generalization of the Swendsen-Wang algorithm~\cite{swendsenNonuniversalCriticalDynamics1987}. 
In order to construct cage-nets, the algorithm operates by ``breaking up'' each octahedron into straight segments, corners, or endpoints, designed to decompose each spin configuration into networks of straight segments which may either terminate in trios or have open ends. 
At zero temperature, open ends are forbidden and only closed cage-nets can be created, such as the one shown in \cref{fig:cages_and_thermodynamics}(b), while at finite temperatures these cages can have open ends, creating or destroying lineons.
A detailed explanation of the algorithm is provided in \cref{apx:cage_net_cluster_algorithm}.

The thermodynamic results of the simulations using the cluster algorithm are presented in \cref{fig:cages_and_thermodynamics}(c), which shows the temperature dependence of the specific heat and entropy. 
The specific heat exhibits a smooth Schottky-like peak with a maximum near $\cramped{T/J \sim 0.6}$, corresponding to the depopulation of lineon quasiparticles at low temperatures.
We find no evidence of any sharp anomaly that might suggest a phase transition, with the specific heat curve exhibiting no quantitative changes with system size, as shown in the inset  of \cref{fig:cages_and_thermodynamics}(c). 
The entropy, obtained by integrating the specific heat (\cref{apx:monte_carlo}), demonstrates extensive degeneracy at zero temperature, with a rather large entropy per spin of approximately $\cramped{0.1176(2)}$.
For comparison, we also compute a Pauling-type estimate as follows: viewed as a vertex model on the cubic lattice with $2^6=64$ total allowed vertex configurations, the ground states allow 10 vertex configurations (eight $T_{1g}$ ground states plus two $A_{1u}$ ground states); treating these constraints as independent, the Pauling estimate of the number of ground states of the entire system is $\cramped{\Omega \approx 2^{N_e} (10/64)^{N_v}}$, where $N_e$ is the number of edges (equal to the number of spins) and $N_v$ is the number of vertices (equal to a third of the number of spins), giving an estimate of the zero-temperature entropy per spin of $\ln \Omega/N_e\approx 0.074$~\footnote{
    The Pauling entropy is expected to give a lower bound on the ground state entropy because it overcounts the single-vertex constraints by treating them as independent, while some constraints are redundant when considering larger clusters. Corrections could be obtained systematically through the so-called numerical linked cluster expansion (NLC)~\cite{singhCorrectionsPaulingResidual2012}.
    }.

\begin{figure*}
        \centering
    \phantom{aa}
    \begin{overpic}[width=0.31\linewidth]{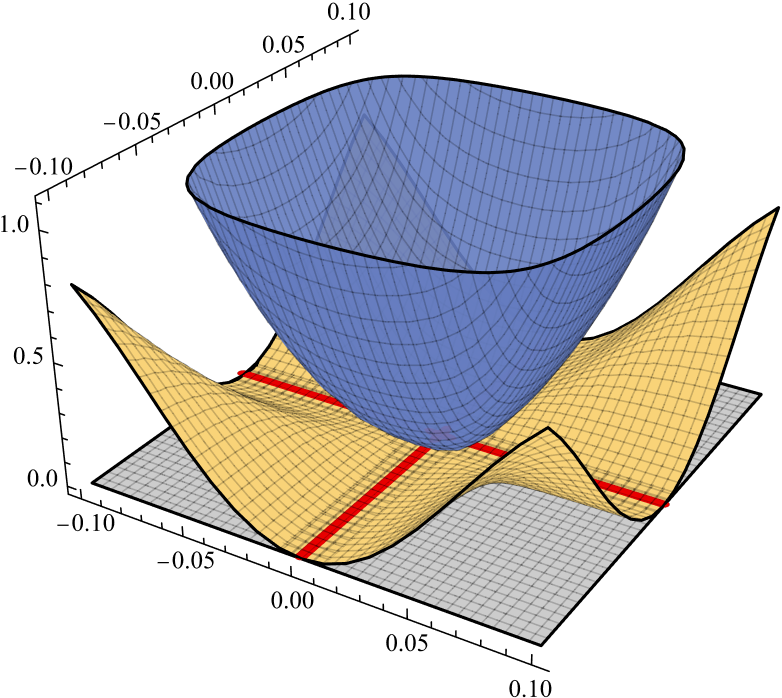} 
        \put(28,3){$(h00)$}
        \put(10,85){$(0k0)$}
        \put(-7,42){\rotatebox{90}{$E(\bq)$}}
    \end{overpic}   
    \hfill
    \begin{overpic}[width=.28\linewidth]{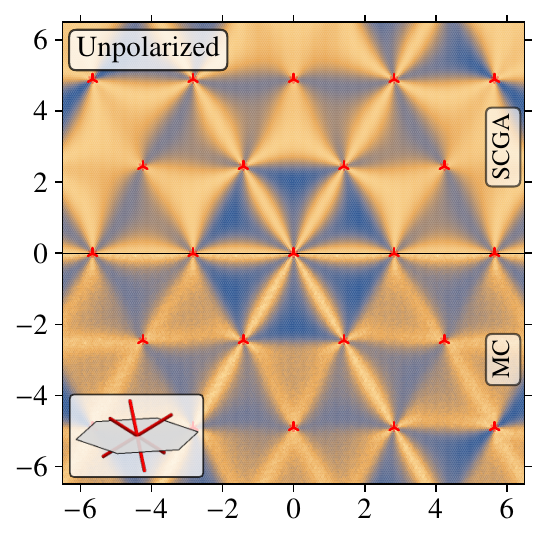}
        \put(63,70){\scriptsize $(0\bar{2}2)$}
        \put(32,70){\scriptsize $(\bar{2}02)$}
        \put(72,55){\scriptsize $(2\bar{2}0)$}
        \put(23,55){\scriptsize $(\bar{2}20)$}
        \put(63,32){\scriptsize $(20\bar{2})$}
        \put(32,32){\scriptsize $(02\bar{2})$}
        \put(40,-6){$u \, (1\bar{1}0)/\sqrt{2}$}
        \put(-5,35){\rotatebox{90}{$v\,(\bar{1}\bar{1}2)/\sqrt{6}$}}
    \end{overpic}
    \hfill
    \begin{minipage}[t]{0.34\linewidth}
    \centering
    \begin{minipage}[t]{0.82353\linewidth}
        \begin{overpic}[width=\linewidth]{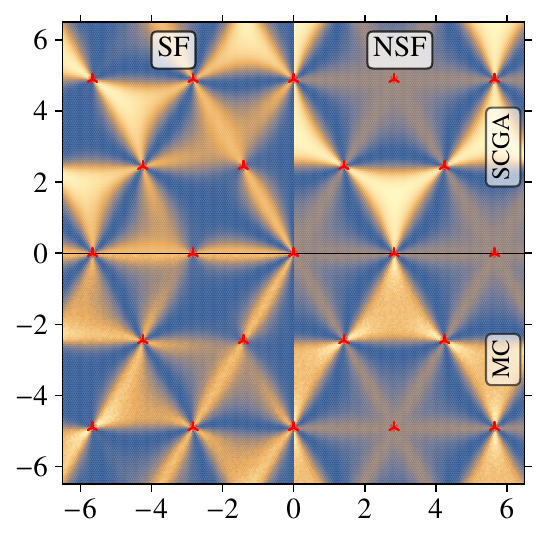}
            \put(40,-6){$u \, (1\bar{1}0)/\sqrt{2}$}
            \put(-5,38){\rotatebox{90}{$v\,(\bar{1}\bar{1}2)/\sqrt{6}$}}
            \put(14,32){\scriptsize $\textcolor{white}{(\bar{2}4\bar{2})}$}
            \put(23,16){\scriptsize $\textcolor{white}{(04\bar{4})}$}
            \put(-240,93){\large (a)}
            \put(-120,93){\large (b)}
            \put(-5,93){\large (c)}
        \end{overpic}
    \end{minipage}%
    \hspace{-2mm}%
    \begin{minipage}[t]{0.15\linewidth}
        \vspace{-4.85cm}
        \includegraphics[width=\linewidth]{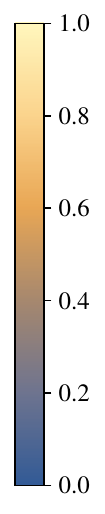}
    \end{minipage}
\end{minipage}
    \\[2ex]
    \caption{\textbf{X-cube as nodal line spin liquid.} (a) The long-wavelength band structure of the interaction matrix for the fracton X-cube spin liquid, with one flat band (gray) and two dispersing bands (orange, blue). All three touch at the zone center, while one dispersive band (orange) touches the flat band along the $\{00h\}$ lines, highlighted in red, making this a nodal line spin liquid~\cite{davierPinchlineSpinLiquids2025,bentonSpinliquidPinchlineSingularities2016}. (b) Unpolarized neutron scattering structure factor in the reciprocal space plane orthogonal to the $(111)$ direction, $\bq=u(1\bar{1}0)/\sqrt{2}+v(\bar{1}\bar{1}2)/\sqrt{6}$. 
    This plane cuts through the intersection points of the pinch lines as shown in the inset, revealing a multifold pinch point scattering pattern with pinch points at $\bq$ with $u=n\sqrt{2}$, $v=m\sqrt{6}$, $n,m\in\mathbb{Z}$. Reciprocal space indices are given in units of $2\pi/a_0$ with $a_0$ the edge length of the 24-site cubic conventional cell (\cref{apx:conventions}).
    (c) Further refinement of the cross section into polarized scattering channels---spin flip (SF) and non-spin-flip (NSF) (\cref{apx:interaction_matrix_SCGA_structure_factors}).
    Both (b) and (c) show structure factors obtained with the self-consistent Gaussian approximation (SCGA, top half, \cref{apx:interaction_matrix_SCGA_structure_factors}) taken at $T_{\textrm{SCGA}}=0.02J$ and Monte Carlo simulations (MC, bottom half) at $T_{\textrm{MC}}=0.16J$.
    }
    \label{fig:X-cube_bands_structure_factors}
\end{figure*}

\subsection{Rank-2 Tensor CSL: U(1) X-cube Model}
\label{sec:rank-2tensor}

By writing the Hamiltonian as in \cref{eq:H_X-cube}, it is clear that the ground states satisfy a set of microscopic local constraints on each octahedron $\o$, given by
\begin{equation}
    Q_{\o}^{xx} = 
    Q_{\o}^{yy} = 
    Q_{\o}^{zz} = 0,
    \label{eq:X-cube-constraint_Q=0}
\end{equation}
i.e. zero $E_u$ quadrupole moment of the local spin configuration.
We can alternatively express the three diagonal components of $Q$ as the $(x^2-y^2)$ and equivalent quadrupole moment components in the three planes,
\begin{align}
    q^z_{\o} &= Q^{xx}_{\o}-Q^{yy}_{\o}, \nonumber \\
    q^y_{\o} &= Q^{zz}_{\o}-Q^{xx}_{\o}, \nonumber \\
    q^x_{\o} &= Q^{yy}_{\o}-Q^{zz}_{\o}.
    \label{eq:X-cube-qs}
\end{align}
These are not linearly independent---$\cramped{\sum_\alpha q^\alpha_{\o} = 0}$, reflecting $\cramped{\Tr[Q] = 0}$---and so form an over-complete basis for the two $E_u$ degrees of freedom, but with the advantage that the three directions are treated equivalently. 
Explicitly, in terms of the local moment directions $\hat{\bm{\moment}}_i = S_i^z \hat{\bm{z}}_i$ these can be written as
\begin{equation}
    \begin{aligned}
    &
    q^z_{\o}
    \propto 
    [
    d^x(\bm{r}_{\o}+\delta \hat{\bm{x}}) 
    - 
    d^x(\bm{r}_{\o}-\delta \hat{\bm{x}}) 
    ]
    \\
    &\qquad 
    -
    [
    d^y(\bm{r}_{\o}+\delta \hat{\bm{y}}) 
    - 
    d^y(\bm{r}_{\o}-\delta \hat{\bm{y}}) 
    ]
    ,
    \end{aligned}
    \label{eq:X-cube-constraints-spins}
\end{equation}
and similarly for $q_{\o}^x$ and $q_{\o}^y$, where $\bm{r}_{\o}$ is the location of the center of the octahedron $\o$ and $\delta$ is the distance from the center of an octahedron to a corner. 
Coarse-graining by taking the limit $\delta \to 0$ in \cref{eq:X-cube-constraints-spins}, we obtain the following three Gauss laws relating the coarse-grained quadrupolar charge density $\bm{\q}$ to the dipole moment density field $\bm{\D}$,
\begin{align}
    \q_z &= \partial_x \D_x - \partial_y \D_y,\nonumber \\
    \q_y &= \partial_z \D_z - \partial_x \D_x,\nonumber \\
    \q_x &= \partial_y \D_y - \partial_z \D_z,
    \label{eq:q_continuum}
\end{align}
which is the continuum equivalent of \cref{eq:X-cube-qs}.
These three constraints can be put in the form of a tensor gauge theory by rearranging the components of the dipole moment vector field into a hollow symmetric rank-2 tensor ``electric'' field $\mathscr{E}$, and writing the Gauss constraint as
\begin{equation}
   \q_\alpha = \epsilon_{\alpha\beta\gamma}\partial_\beta \mathscr{E}_{\gamma \alpha} = 0
   \quad 
   \text{with }
   \quad
   \mathscr{E} = \begin{pmatrix}
       0 & \D_z & \D_y \\
       \D_z & 0 & \D_x \\
       \D_y & \D_x & 0
   \end{pmatrix}
   ,
   \label{eq:gauss_law}
\end{equation}
where $\epsilon_{\alpha\beta\gamma}$ is the Levi-Civita symbol and only the $\beta$ and $\gamma$ indices are contracted. 
Summing over $\alpha$, one finds that $\cramped{\sum_\alpha \q_\alpha = 0}$,  reflecting the analogous redundancy of the microscopic equivalents, \cref{eq:X-cube-qs}, showing that there are only two linearly independent charges---the two components of the $E_u$ quadrupole moment.

The Gauss law in~\cref{eq:gauss_law} is known to encode the gauge structure of the U(1) X-cube field theory~\cite{seibergExoticU1Symmetries2020,slagleQuantumFieldTheory2017,slagleXcubeModelGeneric2018,vijayFractonTopologicalOrder2016} (called the ``$\hat{A}$ theory'' in Ref.~\cite{seibergExoticU1Symmetries2020}).
The quantum X-cube model, originally introduced in its $\mathbbm{Z}_2$ version by Vijay, Haah, and Fu in Ref.~\cite{vijayFractonTopologicalOrder2016}, is a minimal lattice model realizing fracton topological order, and can be viewed as a generalization the $\mathbbm{Z}_2$ toric code model which realizes Abelian topological order~\cite{kitaevFaulttolerantQuantumComputation2003}.
As a type of lattice gauge theory, the X-cube model can straightforwardly extended to $\mathbbm{Z}_N$ and U(1) gauge groups~\cite{maFractonTopologicalOrder2017,slagleQuantumFieldTheory2017,seibergExoticU1Symmetries2020,leeZ_NGeneralizationsThreedimensional2025}.
Its $\mathbbm{Z}_N$ incarnations are often presented in terms of a stabilizer Hamiltonian containing two sets of commuting operators---``star operators'' at each vertex and ``cube operators'' for each cube~\cite{vijayFractonTopologicalOrder2016,slagleXcubeModelGeneric2018,paiFractonFusionStatistics2019,lakeSubdimensionalCriticalityCondensation2021}.
The octochlore fracton CSL we have been discussing in this section realizes a classical version of the U(1) X-cube model.
In the classical model (viewed as the classical limit of a quantum spin-1/2 model), the local constraints given in  \cref{eq:X-cube-constraint_Q=0,eq:X-cube-qs} enforce that the electric field (whose role is played by the spin-1/2 $S^z$ operators) has zero $(\cramped{a^2-b^2})$ moment in each plane at each vertex ($\cramped{a\neq b \in \{x,y,z\}}$), which are equivalent to the constraints enforced by the star operators in the X-cube models.
The cube operators, on the other hand, correspond to spin-flipping operators implementing the cube flip shown in \cref{fig:cages_and_thermodynamics}(a), which are absent in the classical Hamiltonian.

The correspondence can be made precise as follows: in a U(1) lattice gauge theory the electric field $\hat{E}$ takes integer values on each link of the lattice.
Let $\cramped{\hat{q}^z_v = (\Delta_x \hat{E} - \Delta_y \hat{E})_v}$ denote the $\smash{\cramped{(x^2-y^2)}}$ moment of the electric field configuration at vertex~$v$, where $\Delta_x$ denotes the discrete finite-difference along the $x$~direction, etc. (compare to \cref{eq:X-cube-constraints-spins}).
In the matter-free theory, the operators implementing U(1) gauge transformations are $\cramped{\hat{A}_v^\alpha(\theta) = \exp(i \theta \hat{q}^\alpha_v)}$. 
In the $\mathbbm{Z}_N$ equivalent gauge theories, where $\theta$ is restricted to be a multiple of $2\pi/N$, these are generated by powers of $\cramped{\hat{A}_v^\alpha = \exp(i 2\pi \hat{q}^\alpha_v/N)}$~\cite{ohRank2ToricCode2022}, called star operators.
In the theory with matter, these operators measure the amount of electric charge at vertex $v$.
The octochlore X-cube CSL realizes the U(1) version, but with a half-odd-integer-valued electric field, which is further constrained to the values $\cramped{\hat{E} = \pm 1/2}$ on each link because the spin-1/2 operators $\hat{S}^z$ plays the role of~$\hat{E}$~\cite{hermelePyrochlorePhotonsU12004}.

In the quantum X-cube models, spin flips violate two of the three star stabilizers at a single vertex, creating lineon quasiparticles restricted to move in one direction~\cite{vijayFractonTopologicalOrder2016}.
This also happens in the CSL: flipping a spin along the $z$-axis generates $q^x_{\o} =- q^y_{\o} \neq 0$ on the two neighboring octahedra; each of these quasiparticles therefore carries $\cramped{(z^2-x^2)}$ and $\cramped{(z^2-y^2)}$ quadrupolar charge, summing to an overall uniaxial $(3z^2 - r^2)$ quadrupolar charge; as we demonstrated previously, these are lineon quasiparticles which are constrained to move along the $z$ axis.
Furthermore, in the X-cube models the three star operators at each vertex multiply to the identity, enforcing that the lineon quasiparticles annihilate in trios~\cite{lakeSubdimensionalCriticalityCondensation2021}, which in the classical model corresponds to the statement that $\cramped{\sum_\alpha q_\o^\alpha = 0}$, or equivalently that $\Tr[Q_\o] = 0$.
To our knowledge, it has not been previously noted that the lineon quasiparticles of the X-cube model carry quadrupolar charge but this fact is clear in our treatment of the Hamiltonian using multipoles~\footnote{
    This fact is indirectly implied by the field theory construction of Ref.~\cite{seibergExoticU1Symmetries2020}, which assumes that the Noether charges and currents of a global U(1) symmetry transform in a given irrep of the tetrahedral group $S_4\simeq T_d\subseteq \mathrm{SO}(3)$, then gauges that symmetry to obtain a tensor gauge theory. The Gauss law \cref{eq:gauss_law} corresponds to the ``$\hat{A}$'' theory, whose charges transform in the two-dimensional irrep of $S_4$, which descends from the 5-dimensional symmetric tensor (quadrupole) irrep of SO(3).
    }.
In the classical model, we have another route to understand this quadrupolar charge via the spinon condensation mechanism explained in \cref{sec:CSLs_as_Spinon_Condensates}.
As illustrated in \cref{fig:spinon_interactions}(h), a single spinon carries a uniaxial quadrupolar charge and an $A_{1u}$ monopolar charge.
At the X-cube point, the triple-spinon bound state condensation corresponds to condensing the monopolar charge into the vacuum, so it is no longer conserved; the X-cube quasiparticles retain a conserved quadrupole component, and are restricted to move along the uniaxial director axis.

\subsection{Nodal Lines and Spectroscopic Signatures}

As discussed in \cref{sec:degen_and_flat_bands}, classical spin liquids are usually associated with a set of zero-energy flat bands in the spectrum of the interaction matrix sharing a topological band touching with the dispersing bands~\cite{yanClassificationClassicalSpin2024,yanClassificationClassicalSpin2024a,lozano-gomezAtlasClassicalPyrochlore2024,fangClassificationClassicalSpin2024,davierCombinedApproachAnalyze2023}.
The flat bands give rise to characteristic diffuse patterns in reciprocal space structure factors, while the band touchings give rise to ``pinched'' singular features reflecting power-law decay correlations in direct space~\cite{henleyPowerlawSpinCorrelations2005,henleyCoulombPhaseFrustrated2010}.
These pinched features are experimentally detectable in neutron scattering experiments, and can be resolved most clearly using polarization analysis~\cite{chungProbingFlatBand2022}, as was famously done in \ce{Ho2Ti2O7}~\cite{fennellMagneticCoulombPhase2009}.

The bands of the interaction matrix of \cref{eq:H_X-cube} are shown in \cref{fig:X-cube_bands_structure_factors}(a) for the $\cramped{(hk0)}$ plane in the vicinity of the zone center (see \cref{apx:interaction_matrix_SCGA_structure_factors} for details). 
Since there are three spins per primitive unit cell, there are three bands.
One of three bands (gray) is flat, with both dispersing bands (orange, blue) touching it at the zone center. 
One of the dispersing bands (orange) is degenerate with the flat band along the $\{00h\}$ lines of reciprocal space, a situation referred to as a ``nodal line spin liquid''~\cite{bentonSpinliquidPinchlineSingularities2016,davierPinchlineSpinLiquids2025}.
The origin of these nodal lines follows from real-space topological redundancies in the constraints~\cite{bergmanBandTouchingRealspace2008}.
In terms of the charges $q_{\o}^\alpha$ defined in \cref{eq:X-cube-qs}, the Hamiltonian \cref{eq:H_X-cube} has the form $\cramped{\widetilde{H}\propto \sum_{\o}\vert \bq_{\o}\vert^2}$, so ground states satisfy the constraint $\cramped{q_{\o}^\alpha =0}$ on every octahedron, with one octahedron per primitive cell.
For a system of $L^3$ unit cells, this appears to impose two independent constraints per cell (since $\cramped{\sum_\alpha q_{\o}^\alpha = 0}$)---zero $(3z^2-r^2)$ moment and zero $(x^2-y^2)$ moment. 
Naively, this yields $2L^3$ quadratic constraints, corresponding to two dispersing bands, leaving $L^3$ unconstrained degrees of freedom forming a single flat band.
However, this local counting overlooks global redundancies in a periodic system which reduce the number of linearly independent constraints and enforce zero-energy touchings between the flat and dispersive bands~\cite{bergmanBandTouchingRealspace2008}. 
In particular, for every constant-$z$ plane $p_z$, one has $\sum_{\o \in p_z} q_{\o}^z = 0$, with analogous relations for constant-$x$ and constant-$y$ planes. 
This gives $3L$ apparent redundancies, one per lattice plane, which give rise to extra zero modes along the nodal line touchings---$L$ along each of the $(h00)$, $(0h0)$, and $(00h)$ lines.
This actually over-counts the redundancies by one, corresponding to the intersection of the three nodal lines at the zone center.
Summing the planar redundancies over all planes is equivalent to summing the local identities $\cramped{\sum_\alpha q_{\o}^\alpha = 0}$ over all octahedra, so the number of independent redundancies is $3L - 1$.
The total number of zero modes is therefore $\cramped{L^3 + (3L -1)}$, which we can break down as (\cf~\cref{fig:X-cube_bands_structure_factors}(a)): $L^3$ modes in the flat band, $\smash{\cramped{3\times(L-1)}}$ modes along the nodal lines \emph{away} from the zone center and, finally, the two remaining modes that correspond to the zone center where both dispersive bands touching the flat band.

The nodal line singularities should give rise to characteristic ``pinch lines'' in reciprocal space correlation functions~\cite{bentonSpinliquidPinchlineSingularities2016,yanClassificationClassicalSpin2024a}.
However, in general the full structure of these singular features need not necessarily reflected in the spin structure factor (the Fourier transform of $\langle S_i^z S_j^z\rangle$ summed over all sites, see~\cref{apx:interaction_matrix_SCGA_structure_factors}).
Indeed, like some other known examples~\cite{yanRank2U1Spin2020,lozano-gomezAtlasClassicalPyrochlore2024}, that is the case here---the spin structure factor (not shown) does not exhibit pinch lines.
However, their signatures are visible in neutron scattering structure factors, which refine the spin structure factor by weighting the contribution from different sublattices inequivalently depending on the wavevector (\cref{apx:interaction_matrix_SCGA_structure_factors}).
They appear as singular features along a grid of lines in reciprocal space parameterized as $\{2n,2m,l\}$, where $n,m$ are integers and $l$ varies continuously.
The singular nature of the nodal lines is most readily revealed by taking a reciprocal-space plane that intersects them transversely, rather than one in which they lie. 
Planes orthogonal to the $(111)$ directions are therefore particularly useful because they can cut through all three sets of lines at once, especially the one that cuts through the zone center where all three lines intersect, as illustrated in the inset of \cref{fig:X-cube_bands_structure_factors}(b).
We compute the unpolarized neutron structure factor in this plane using both the self-consistent Gaussian approximation (SCGA, \cref{apx:interaction_matrix_SCGA_structure_factors}) and Monte Carlo, shown in \cref{fig:X-cube_bands_structure_factors}(b). 
This reveals a characteristic pinch point pattern, including multi-fold pinches characteristic of higher-rank spin liquids~\cite{premPinchPointSingularities2018}.
Further refinement is in principle possible using polarization techniques~\cite{fennellMagneticCoulombPhase2009}.
For example, \cref{fig:X-cube_bands_structure_factors}(c) shows the calculated spin-flip (SF) and non-spin-flip (NSF) channel cross sections~\cite{fennellMagneticCoulombPhase2009,chungProbingFlatBand2022,loveseyPolarizationEffectsMagnetic1987}.

This concludes our study of the X-cube CSL. 
It is worth comparing the relative simplicity of classical X-cube CSL identified here to other models considered in the literature.
These include models with up to twelve-spin interactions~\cite{vijayFractonTopologicalOrder2016}, highly fine-tuned interactions up to fifth-nearest-neighbor~\cite{niggemannClassicalFractonSpin2025,plackeIsingFractonSpin2024}, special anisotropic spin interactions~\cite{slagleFractonTopologicalOrder2017}, or multiple spins per site~\cite{halaszFractonTopologicalPhases2017}.
While the octochlore fracton CSL may not be realistic from a materials perspective, in the sense that the second-neighbor interaction is twice the first-neighbor interaction strength, it is perhaps the simplest model for fracton spin liquidity so far explored.
Indeed it is likely the minimal possible extension of spin ice (which only requires a single nearest-neighbor interaction on the pyrochlore lattice) to a fracton spin liquid.
We proceed next to explore the last part of the phase diagram, directly opposite to the X-cube CSL point, where the ground state may be viewed as a maximal packing of the quadrupolar-charged X-cube lineon quasiparticles.

\begin{figure*}[t]
    \centering
    \begin{overpic}[width=\linewidth]{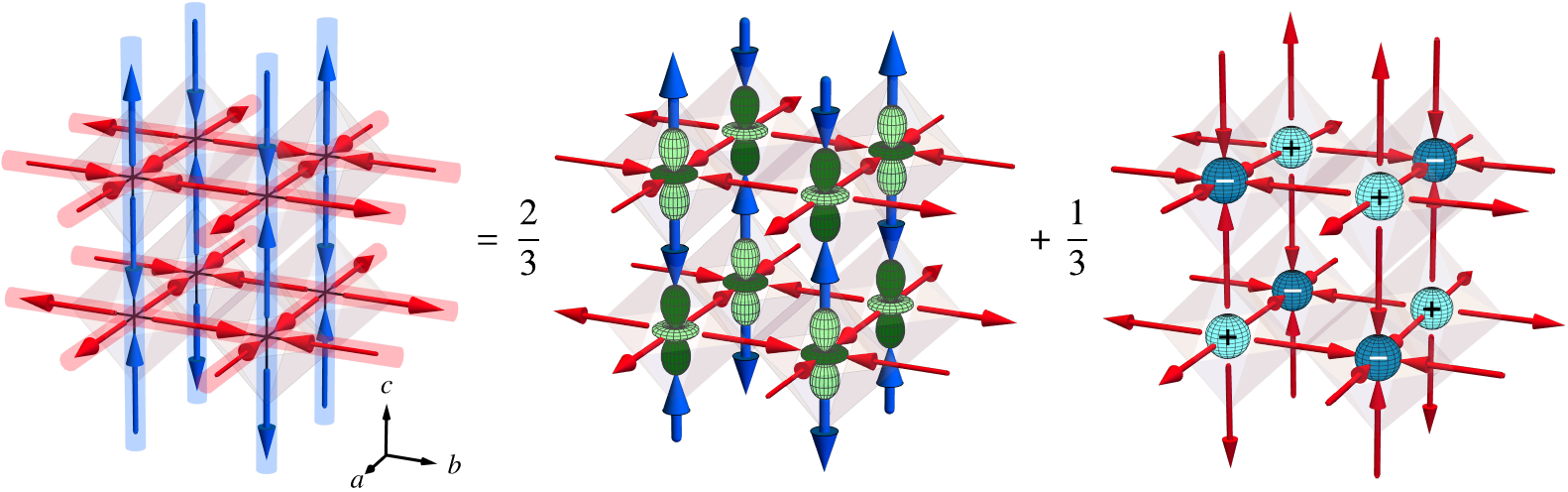}
        \put(01.0,27.0){(a)}
        \put(36.0,27.0){(b)}
        \put(72.0,27.0){(c)}
        \put(06.0,07.0){$\A$}
        \put(14.5,05.5){$\B$}
        \put(40.3,05.5){$\A$}
        \put(49.5,04.0){$\B$}
        \put(75.5,05.9){$\A$}
        \put(85.0,03.9){$\B$}
        \put(58.5,27.0){\small $\times 2$}
    \end{overpic}
    \\[1ex]
    \begin{overpic}[width=.95\linewidth]{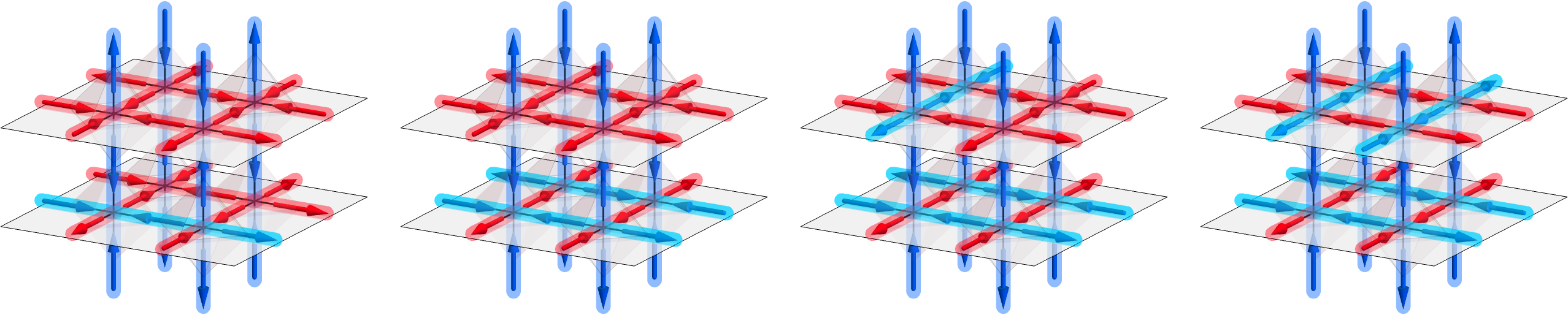}
        \put(00,17){(d)}
        \put(25,17){(e)}
        \put(50,17){(f)}
        \put(75,17){(g)}
    \end{overpic}
    \caption{\textbf{``Fracton crystal'' ground states and emergent subsystem symmetries in the nematic phase}. (a) One of the six uniaxial fracton crystal ground states, which are obtained by uniformly tiling the system with single-octahedron ($c^2-a^2-b^2)$ states (\cf~\cref{fig:E_irrep_fragmentation}(b)). 
    In this state all chains are N\'eel polarized, with chain order parameter \cref{eq:chain_neel_op} given by  $\N_{\c} =-1$ (blue) for all $c$-chains and $\N_{\c} =+1$ (red) for all $a$- and $b$-chains.
    (b,c) This state can be decomposed into a linear combination of (b) fully packed quadrupoles and (c) fully packed monopoles. 
    Just as the AIAO states are maximal packings of the spin ice monopole quasiparticles, i.e. ``monopole crystals'', this state may be viewed as a maximal packing of X-cube fracton quasiparticles, i.e. a ``fracton crystal''.
    It carries a monopole crystal component due to the single-octahedron ground state configurations being fragmented (\cref{sec:fragmentation_of_Eu}).
    In (b), large blue arrows represent spins with $S_i^z = \pm 2$ as in \cref{fig:E_irrep_fragmentation}(b).
    (d-g) Within each plane orthogonal to the uniaxial axis each chain can be flipped to produce another ground state. Once one chain has been flipped all parallel chains in the same plane can be flipped, but perpendicular ones cannot, giving rise to an unusual constrained emergent subsystem symmetry. 
    }
    \label{fig:fracton-crystal}
\end{figure*}

\section{\texorpdfstring{Spin Nematic Phase ($\bm{E_u\fragmented A_{1u}}$}{Eu-A1u})}
\label{sec:spin_nematic}

We turn now to the remaining portion of the phase diagram, the $\cramped{E_u \fragmented A_{1u}}$ phase, indicated by the blue-green hatched arc on the bottom right of the phase diagram in \cref{fig:phase_diagram}(f).
As we will show, in this region of the phase diagram the ground state manifold is subextensively degenerate. 
This gives rise to interesting magnetic frustration physics intermediate between spin liquids and ordered phases---a spin nematic phase exhibiting two-stage dimensional reduction upon decreasing temperature.

\subsection{Fracton Crystal Ground States and Accidental Subsystem Symmetries}
\label{sec:fracton_crystal_spinon_excitations_subsystem_symmetry}

This phase contains the point directly opposite to the X-cube fracton CSL point (\cref{sec:fracton_CSL}), whose Hamiltonian is $-\widetilde{H}_{\text{X-cube}}$ (\cref{eq:H_X-cube}).
At the single-octahedron level, ground states in this phase must be built from the fragmented quadrupolar-plus-monopolar configurations in \cref{fig:E_irrep_fragmentation}(b).
These configurations maximize the $E_u$ quadrupole moment locally, meaning that they maximize the number of X-cube lineon quasiparticles on every site. 
This is analogous to the AIAO phase, which sits across from the spin ice CSL and whose ground states maximize the number of monopoles on every octahedron.
Correspondingly, we can view ground states of this phase as maximal packings of X-cube fractons.
To see this explicitly, we construct a representative ground state from a uniform tiling of the fragmented configuration from \cref{fig:E_irrep_fragmentation}(b).
There are six such states related by symmetry, an example of which is shown in \cref{fig:fracton-crystal}(a).
In the same way that the single octahedron can be decomposed into fragmented $E_u$ and $A_{1u}$ components (cf \cref{fig:E_irrep_fragmentation}(b)), these states can be decomposed into a linear combination of a maximal packing of staggered quadrupole moments, as shown in \cref{fig:fracton-crystal}(b), and staggered monopole moments, as shown in \cref{fig:fracton-crystal}(c).
The $A_{1u}$ component in \cref{fig:fracton-crystal}(c) corresponds to a close-packed staggered arrangement of single-charge monopoles, and can therefore be viewed as a monopole crystal~\cite{brooks-bartlettMagneticMomentFragmentationMonopole2014,jaubertMonopoleHolesPartially2015}.
Correspondingly, we may view the $E_u$ component of this decomposition in \cref{fig:fracton-crystal}(b) as forming a \emph{fracton crystal}.

These states saturate the $E_u$ order parameter, \cref{eq:quadrupole_OP}, with $\cramped{\Tr[Q^2] = 1}$.
They also have a sizeable $A_{1u}$ order parameter, \cref{eq:monopole_OP}, with $\cramped{\vert \phi \vert = 1/3}$, owing to the fragmentation of the single-octahedron states (\cref{sec:fragmentation_of_Eu}).
We can further extract the two linearly independent components of the tensor $E_u$ order parameter as
\begin{equation}
    \begin{aligned}
    Q_{3c^2-r^2} 
    &=
    \frac{1}{\sqrt{6}}\left(2Q^{cc} - Q^{aa}-Q^{bb}\right)
    ,
    \\
    Q_{a^2-b^2} 
    &=
     \frac{1}{\sqrt{2}}\left(Q^{aa} - Q^{bb}\right)
    ,
    \end{aligned}
    \label{eq:quadrupole_OP_2_components}
\end{equation}
with $a,b,c, \in \{x,y,z\}$, where $\smash{Q_{3c^2-r^2}}$ measures uniaxiality along the $c$ direction and $\smash{Q_{a^2-b^2}}$ measures biaxiality in the perpendicular $ab$-plane. 
The ground states illustrated in \cref{fig:fracton-crystal}(a) maximize $\smash{Q_{3c^2-r^2}}$ while having $\smash{\cramped{Q_{a^2-b^2}=0}}$, so we characterize these as \emph{uniaxial nematic states}---they break global rotational symmetry from cubic to tetragonal, singling out the $c$-axis as the uniaxial director axis (a state is uniaxial if two eigenvalues of $Q$ are degenerate; since $Q$ is a diagonal matrix its eigenvalues are just its diagonal elements)~\footnote{
    Our use of the term ``nematic'' should be distinguished from the more common examples of spin nematicity in quantum spin models with spin quantum number $S>1/2$ (including classical $\cramped{S\to \infty}$ models) which are characterized by lack of time reversal symmetry breaking, so $\langle S_i^\alpha \rangle = 0$, but breaking of spin or lattice rotational symmetry as measured by on-site quadrupole operators $\smash{\cramped{Q_i^{\alpha\beta}=(S_i^\alpha S_i^\beta - (1/3) \vert \bm{S}_i\vert^2 \delta_{\alpha\beta})}}$~\cite{pencSpinNematicPhases2011}. The classical Ising spin nematic phase we identify breaks both time-reversal symmetry and lattice rotation, but preserves the combination of time-reversal and inversion. 
}.
We will refer to these as \emph{uniaxial fracton crystal} ground states.

\begin{figure*}[t]
    \centering
    \includegraphics[width=.95\linewidth]{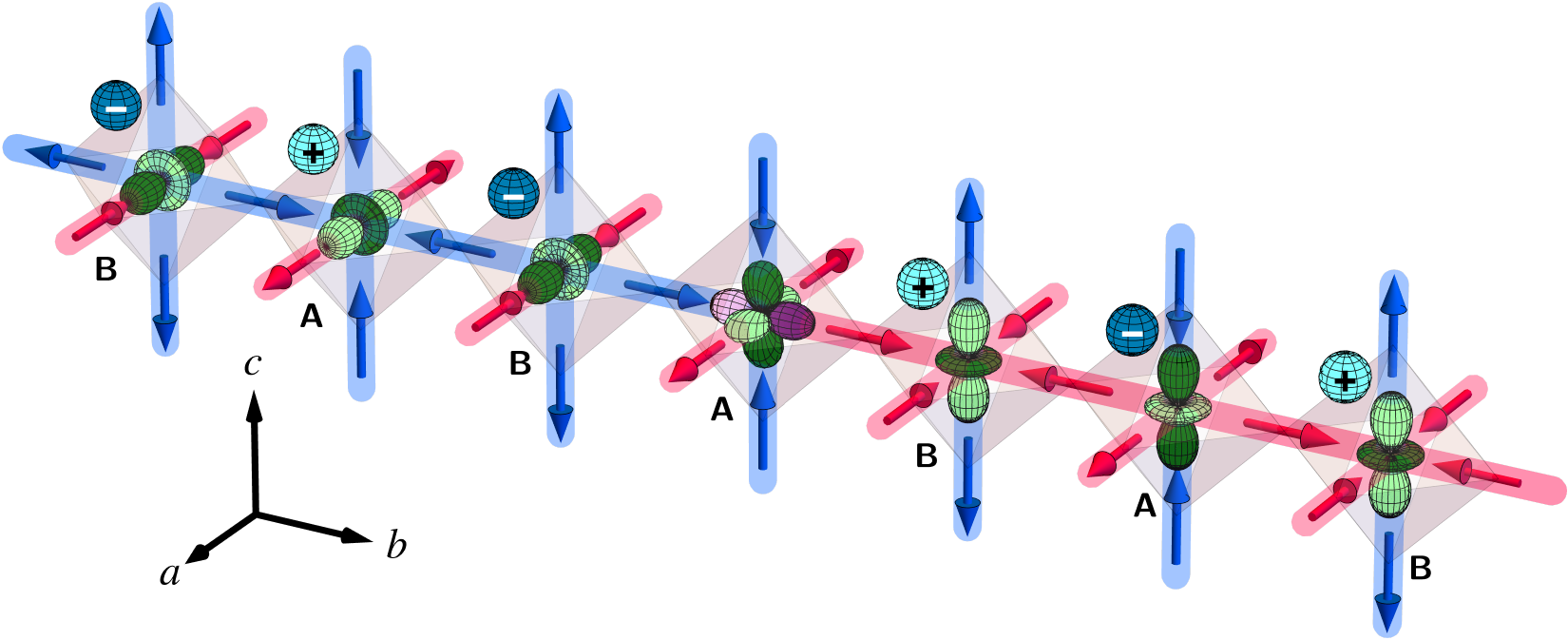}
    \caption{
        \textbf{Deconfined antiferro-spinons in the biaxial nematic phase.} 
        The low-temperature biaxial nematic phase of the $E_u \fragmented A_{1u}$ quadrant of the phase diagram is entropically stabilized by deconfined 1D AFM spinons. 
        Here we show a single $b$-chain, where the $c$-chains are polarized with $\N_{\c} = -1$ (blue) and $a$-chains are polarized with $\N_{\c} = +1$ (red), as in \cref{fig:fracton-crystal}(a).
        Each $b$-chain has two N\'eel polarized ground states; domain walls between two N\'eel domains are deconfined 1D antiferro-spinons, free to hop along the chain at zero energy cost.
        Within each N\'eel domain the local $E_u$ quadrupole moments are aligned along either the $a$-axis or the $c$-axis, but the $A_{1u}$ monopole moments have opposite staggered pattern; thus a mixture of the two domains created by deconfinement of antiferro-spinons promotes biaxial quadrupolar order while suppressing the monopole order one, as is seen to occur below $\TCbi$ in \cref{fig:nematic-thermodynamics}(b,c). 
    }
    \label{fig:nematic_AFspinon_chain_deconfinement}
\end{figure*}

While the six uniaxial fracton crystal states are the simplest ground states constructed from a uniform tiling of the single-octahedron ground states, the complete ground state manifold is much larger. 
To demonstrate this, first note that for each single-octahedron ground state [\cref{fig:E_irrep_fragmentation}(b)] antipodal pairs of spins are anti-aligned. 
It follows that any state constructed from these single-octahedron configurations must consist of antiferromagnetically polarized chains, with each chain in one of its two N\'eel-polarized states.
Let us therefore define a N\'{e}el order parameter  $\N_{\mathsf{c}}$ for each chain~$\mathsf{c}$ in terms of the local moment directions $\hat{\bm{\moment}}_i = S_i^z \hat{\bm{z}}_i$ as
\begin{equation}
    \N_{\c} = (-1)^{\c}\frac{1}{L}\sum_{i \in \c} (-1)^i \hat{\bm{\moment}}_i\cdot\hat{\bm{e}}_{\c}.
    \label{eq:chain_neel_op}
\end{equation}
Here, $\hat{\bm{e}}_{\c} \in \{\hat{\bm{x}},\hat{\bm{y}},\hat{\bm{z}}\}$ is the direction of the chain and $L$ is the number of spins in the chain.
The sign $(-1)^i$ alternates along the chain so that $\N_{\c}$ measures the staggered magnetization, while the sign $(-1)^{\c}$ alternates along the two sublattices of the square lattice formed by the chains when viewed along their length, chosen so that the ``all-out'' state correspond to $\N_{\c}=+1$ on every chain.
The uniaxial fracton crystal ground state shown in \cref{fig:fracton-crystal}(a) differs from the ``all-out'' state by flipping every chain along the $z$ axis---it has $\cramped{\N_{\c} = -1}$ for $c$-oriented chains (colored blue) and $\cramped{\N_{\c} = +1}$ for $a$- and $b$-oriented chains (colored red).

We can deduce a simple rule that is satisfied if and only if a state is a ground state: all chains must be N\'eel polarized, and at the intersection of three chains the polarizations are not all the same (i.e., two blue and one red or vice versa), guaranteeing that the local octahedral configuration is of the form $\cramped{\pm(c^2 - a^2 - b^2)}$ (\cf~\cref{fig:E_irrep_fragmentation}(b)).
In particular, starting from the uniaxial fracton crystal state with uniaxial director along the $c$-axis, as in \cref{fig:fracton-crystal}(a), we can flip a single chain in any $ab$-plane.
We show in \cref{fig:fracton-crystal}(d)  the result of flipping a single $b$-oriented chain.
Once we have flipped such a chain, we can also flip any parallel chain in the same plane, as shown in \cref{fig:fracton-crystal}(e).
However, we cannot flip an orthogonal chain in the same plane, which would violate the ground state rule by creating an intersection of three chains with the same $\N_{\c}$, i.e. an all-in or all-out octahedron.  
In each plane orthogonal to the $c$-axis, therefore, we are free to flip all chains along either the $a$-axis or the $b$-axis, but not both.
Once a single chain along the $a$-axis has been flipped, no $b$-oriented chains can be flipped within the same plane, and vice-versa.
The choice of which direction to flip chains can be chosen independently in every $ab$-plane, as illustrated in \cref{fig:fracton-crystal}(f,g).
Overall, this gives a subextensive ground state degeneracy scaling as $\smash{2^{L^2}}$.

\begin{figure*}[t]
    \centering
    \begin{overpic}[width=\linewidth]{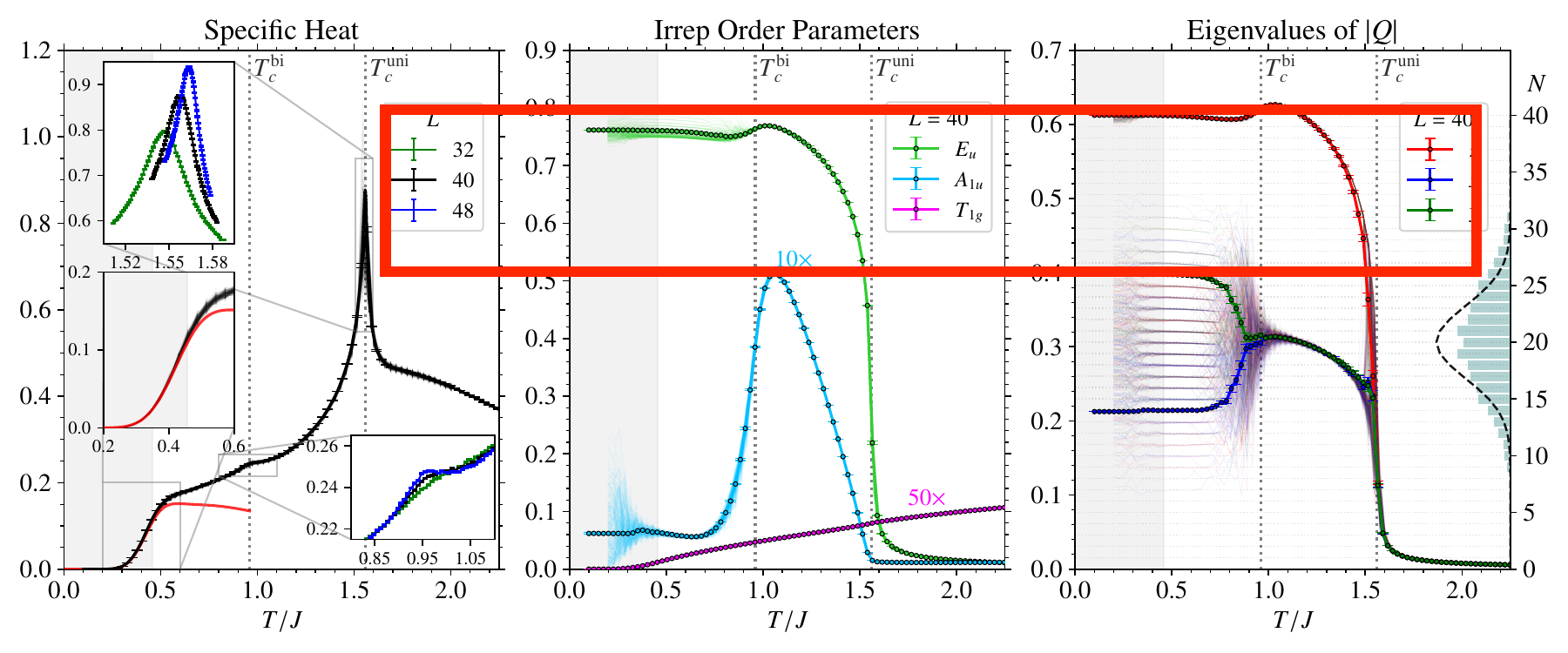}
         \put( 1,40){\large (a)}
         \put(34,40){\large (b)}
         \put(67,40){\large (c)}
         \put(8.3, 22.3){\textcolor{red}{\footnotesize 1D}}
         \put(7.2, 21.0){\textcolor{red}{\footnotesize chains}} 
         \put(72, 6.4){\small Biaxial}
         \put(81.1, 6.4){\small Uniaxial}
         \put(90.5, 6.4){\small PM}
         \put(38.5,15){\rotatebox{90}{\textcolor{gray!80}{\small finite-size 1D chains}}}
    \end{overpic}
    \caption{
    \textbf{Specific heat and order parameters in the $\bm{E_u\fragmented A_{1u}}$ spin nematic phase.} 
    Here we show thermodynamic data obtained from 250 annealed Monte Carlo simulations at the point directly opposite to the X-cube CSL denoted ``fully packed fractons'' in \cref{fig:phase_diagram}(f). 
    This point is representative of the behavior we find in the entire phase.
    (a) Specific heat, showing a sharp anomaly at the higher-temperature critical temperature $\TCuni$, and a much weaker anomaly at $\TCbi$ (insets show system size scaling).
    At the lowest temperatures we can fit to the exact solution for decoupled 1D chains (\cref{apx:1D_Chains}).
    (b) Thermal expectation value for each multipole order parameter's magnitude, \crefrange{eq:monopole_OP}{eq:quadrupole_OP}: $\langle \vert \phi\vert\rangle$, $\langle \vert \bm{D}\vert\rangle$, and $\langle \Tr[Q^2]^{1/2}\rangle$. 
    Note that $A_{1u}$ is scaled by $10\times$ and $T_{1g}$ by $50\times$. 
    The behavior of the $E_u$ and $A_{1u}$ order parameters at $\TCuni$ is consistent with a Landau expansion, \cref{eq:landau_free_energy}.
    The erratic behavior at the lowest temperatures is a finite-size effect arising from the polarization of 1D chains, which would remain disordered in the thermodynamic limit.
    (c) Absolute values of eigenvalues of the quadrupole tensor order parameter.
    The intermediate phase is uniaxial, with two degenerate eigenvalues, while the low-temperature phase is biaxial, with three distinct eigenvalues.
    In the biaxial phase the two subleading eigenvalues split and plateau at different levels, corresponding to different numbers of 1D chains deconfined in each direction; gray horizontal lines indicate the plateau values predicted from \cref{eq:plateau_Q}.
    The histogram on the right-hand side compares the distribution of eigenvalues to a binomial distribution (dashed line).
    Simulation details are provided in \cref{apx:monte_carlo}. 
    }
    \label{fig:nematic-thermodynamics}
\end{figure*}

This situation is reminiscent of the frustrated chains phase discussed in \cref{sec:frustrated_chains}, which also exhibits zero-energy chain flips and corresponding subextensive ground state degeneracy.
However, here the ability to flip chains is significantly constrained, because flipping chains along the $a$- or $b$-axes in a single plane are mutually incompatible.
To cleanly define the accidental symmetries, we must divide the ground state manifold into sectors labeled by a choice of either the $a$- or $b$-axis for each plane orthogonal to the uniaxial director $c$-axis~\footnote{
    There are also ground states which do not live in these sectors---for example, starting from the state in \cref{fig:fracton-crystal}(a) flip all chains in one $bc$-plane, then $c$-axis chains in that plane can be flipped, and all $b$-axis chains can be flipped. However, these ground states are not naturally reached from the uniaxial states and do not appear to influence the low-temperature behavior, \cf~\cref{sec:biaxial}.
}. 
Within each such ground state sector, flipping chains along the chosen axes is an accidental subsystem symmetry.
As we will see, at low temperatures the system spontaneously will select one such sector.

It is important to draw a key distinction between the zero-energy chain-flips in this portion of the phase diagram compared to those which occur in the $T_{1g}$ frustrated chains phase. 
In the frustrated chains phase, the chain-flips are zero modes arising from the flat planar degeneracy of the interaction matrix, in the sense explained in \cref{sec:degen_and_flat_bands}.
In the $E_u\boxplus A_{1u}$ phase, by contrast, the ground states are not zero-energy states because the fragmented ground states necessarily mix the positive-energy $A_{1u}$ irrep with the zero-energy $E_u$ irrep.
Consequently, the zero-energy locus of the interaction matrix is not a good guide to the low-energy physics of this phase.
As listed in \cref{tab:summary_degeneracies}, the interaction matrix has zero-energy flat lines in reciprocal space, which only imply zero modes which flip entire planes of spins. 
Indeed, such planar zero modes do exist---if an entire plane has $\cramped{(a^2-b^2)}$-type order, such as the $ac$ and $bc$ planes in \cref{fig:fracton-crystal}(a), then such a plane can always be flipped at zero energy cost, which is the analog of flipping a polarized chain in the $T_{1g}$ frustrated chains phase.
However, the chain-based flips in the ground state manifold discussed above and illustrated in \cref{fig:fracton-crystal}(d-g) arise as a consequence of fragmentation, rather than as zero modes of the interaction matrix. 
In particular, they arise because the fragmented ground states are locally of the form $c^2-a^2-b^2$ on each octahedron; flipping an $a$-chain changes the configuration to $c^2+a^2-b^2$, which is still a fragmented ground state. 
Therefore, the chain-based flips in the ground state manifold are \emph{accidental} symmetries---symmetries of the ground state manifold which are not symmetries of the Hamiltonian---rather than exact zero modes.

Unlike exact zero modes, accidental symmetries only act within the ground-state manifold and are generically expected to be broken by order-by-disorder~\cite{villainOrderEffectDisorder1980}.
Nevertheless, accidental symmetries acting on one-dimensional chains have particularly strong consequences because they imply that the minimal excitations above the ordered ground states are deconfined quasiparticles (1D domain walls) moving along the chains, which carry enormous configurational entropy.
This is much more impactful on the low-energy physics than zero modes acting on two-dimensional manifolds, which implies the existence of string-like 2D domain walls, which are confined by their string tension.
In the case at hand, since the chains are all N\'eel ordered in the ground state, the accidental chain-based symmetries flipping chains imply the existence of fractionalized \emph{antiferro}-spinon excitations---domain walls between the N\'eel domains on a single chain.
This is illustrated in \cref{fig:nematic_AFspinon_chain_deconfinement}, which shows two N\'eel domains on a single chain, meeting at a domain wall. 
Just like ferro-spinons, these low-energy quasiparticles are deconfined, they can be moved along the chain at zero energy cost by flipping a series of spins. 
Notably, however, this deconfinement requires that the spins surrounding the chain are frozen in their ground state configuration, and is therefore only operative at very low temperatures when defects are dilute.
Flipping spins along the uniaxial director axis is not an accidental symmetry, implying that the antiferro-spinons cannot move along this axis without paying energy proportional to their separation---they are confined along the director ($c$-axis).
Furthermore, because the accidental subsystem symmetries can only perform unconstrained flips of \emph{parallel} chains in each plane, the deconfinement of antiferro-spinons can only occur in one of two directions in each plane.
This suggests that the low-temperature physics will involve the spontaneous selection of one of the above-defined ground state sectors, driven by a free energy gain from deconfinement of antiferro-spinons.

\subsection{Transition to Uniaxial Nematic Order}
\label{sec:uniaxial}

We now investigate the finite-temperature behavior of the system in the $E_u \fragmented A_{1u}$ region of the phase diagram using Monte Carlo simulations.
Despite the subextensive degeneracy of the ground-state manifold discussed above, upon cooling, the system first undergoes a conventional finite-temperature transition out of the paramagnet and into a uniaxial nematic phase, signaled by a prominent sharp peak in the specific heat.
The corresponding transition temperatures are marked in the phase diagram \cref{fig:phase_diagram}(f).
The maximum critical temperature occurs near the point in the phase diagram directly opposite the X-cube CSL, which is labeled ``fully packed fractons'' in \cref{fig:phase_diagram}(f).
We therefore focus on this representative point, whose thermodynamic behavior is qualitatively characteristic of the entire phase.
\Cref{fig:nematic-thermodynamics} shows the temperature evolution of (a) the specific heat, (b) the multipole order parameters defined in \cref{eq:monopole_OP,eq:dipole_OP,eq:quadrupole_OP}, and (c) the magnitudes of the eigenvalues of the quadrupole tensor order parameter.
The specific heat displays a prominent peak at $\TCuni/J\approx 1.56$, indicating a thermodynamic phase transition. 
We first focus on the physics of this transition and the phase just below it.
A second transition occurs at lower temperature, which will be analyzed separately in \cref{sec:biaxial}.

\begin{figure*}[t]
    \centering
    \begin{overpic}[width=\linewidth]{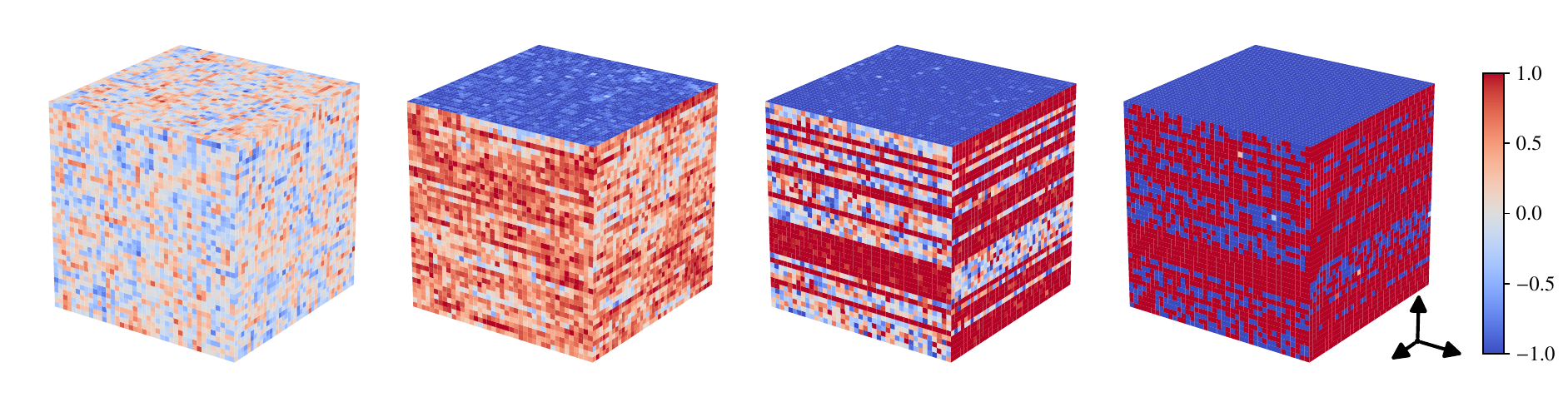}
        \put(03.0,23){(a)}
        \put(25.5,23){(b)}
        \put(47.5,23){(c)}
        \put(70.0,23){(d)}
        \put(03.0,-1){
            \begin{minipage}[t]{.2\textwidth}
                \centering
                $T/J = 2.00$ \\[1pt]
                Paramagnetic Phase
            \end{minipage}
        }
        \put(25.5,-1){
            \begin{minipage}[t]{.2\textwidth}
                \centering
                $T/J = 1.30$ \\[1pt]
                Uniaxial Phase
            \end{minipage}
        }
        \put(48.5,-1){
            \begin{minipage}[t]{.2\textwidth}
                \centering
                $T/J = 0.75$ \\[1pt]
                Biaxial Phase \\
                {\small (disordered chains)}
            \end{minipage}
        }
        \put(71.5,-1){
            \begin{minipage}[t]{.2\textwidth}
                \centering
                $T/J = 0.30$ \\[1pt]
                Polarized Chains \\
                {\small (finite size effect)}
            \end{minipage}
        }
        \put(87.6, 2.2){$a$}
        \put(93.2, 2.2){$b$}
        \put(91.1, 6){$c$}
        \put(94.5, 22.5){$\N_{\c}$}
    \end{overpic}
    \\[6ex]
    \caption{
    \textbf{Spin nematic phase and dimensional reduction.} 
    Here we plot snapshots of the chain N\'{e}el order parameters $\N_{\c}$, \cref{eq:chain_neel_op}, where each chain is associated with a point on the surface of a cube, taken from a single $L=40$ annealed simulation.
    Positive values of $\N_{\c}$ are denoted in red and negative values in blue, while $\N_{\c}\approx 0$ is denoted in white. 
    For reference, the all-out (all-in) state would have all faces red (blue). 
    (a) At high temperatures in the paramagnetic phase, all chains are disordered. 
    (b) Just below the transition into the uniaxial phase with tetragonal symmetry, chains along the $c$-axis are strongly polarized ($\N_{\c} \approx -1$), while chains along the $a$- and $b$-axes exhibit weaker and opposite  polarization ($\N_{\c} \approx +0.5$).
    This is consistent with a thermally disordered version of the uniaxial fracton crystal state depicted in \cref{fig:fracton-crystal}(a), which would produce a completely blue face and two completely red faces.
    (c) Just below the transition into the biaxial phase with orthorhombic symmetry, in each plane orthogonal to the $c$-axis a secondary axis is chosen along which the chains are polarized opposite to the $c$-axis polarization, while the chains along the third axis become disordered due to deconfinement of 1D AFM spinons (\cf~\cref{fig:nematic_AFspinon_chain_deconfinement}), i.e. the system has undergone dimensional reduction to disordered 1D chains. 
    (d) At the lowest temperatures the disordered chains randomly polarize in our finite-size simulation, resulting in the erratic behavior of order parameters seen in \cref{fig:nematic-thermodynamics}(b,c).
    }
    \label{fig:nematic_cubes}
\end{figure*}

To identify the symmetry of the ordered phase, we examine the multipolar order parameters and the eigenvalues of the quadrupole tensor. 
Just below $\TCuni$, \cref{fig:nematic-thermodynamics}(b) shows that the $E_u$ order parameter develops rapidly, while the $A_{1u}$ order parameter exhibits a weaker activation. 
\Cref{fig:nematic-thermodynamics}(c) shows that two eigenvalues of the quadrupole tensor are degenerate, demonstrating that this transition spontaneously breaks the cubic rotational symmetry down to a tetragonal subgroup.
This characterizes this intermediate-temperature regime as a uniaxial nematic phase.

The formation of this intermediate-temperature uniaxial phase can be understood from a Landau free energy expansion in terms of the scalar $A_{1u}$ and $E_u$ order parameters,
\begin{align}
    F[\phi,Q] &= 
    r_E \Tr[Q^2] + 
    r_A \phi^2 + 
    w\, \phi \Tr[Q^3]
    \nonumber
    \\
    &+ 
    g \phi^2\Tr[Q^2] + 
    u_A \phi^4 + 
    u_E \Tr[Q^2]^2 +
    \cdots 
    \label{eq:landau_free_energy}
\end{align}
where the ellipses denote terms involving the $T_{1g}$ dipole order parameter as well as sixth-order and higher terms~\footnote{
    Up to fourth order the only allowed $T_{1g}$ couplings are  $r_T\vert \bm{D}\vert^2$, $u_T \vert \bm{D}\vert^4$, $g' \vert \bm{D}\vert^2 \phi^2$, $g'' \vert \bm{D}\vert^2 \Tr[Q^2]$, with no linear-cubic couplings.
    We expect both $r_A,r_T > 0$ at the transition since these irreps are gapped in the microscopic Hamiltonian, but with $\cramped{r_A \ll r_T}$ since the $\cramped{c^2-a^2-b^2}$ single-octahedron fragmented ground state configurations, \cref{fig:E_irrep_fragmentation}(b), carry an $A_{1u}$ component.
}.
While we will not embark into a detailed analysis of the critical behavior, the form of the free energy explains the principal features observed in Monte Carlo simulations. 
Microscopically, the energetics of an isolated octahedron favor states with dominant $E_u$ character, making quadrupolar order the primary instability upon cooling. 
The linear-cubic coupling $w\phi \Tr[Q^3]$ then induces a secondary $A_{1u}$ component when $E_u$ order develops, naturally accounting for the simultaneous activation of both order parameters in \cref{fig:nematic-thermodynamics}(b) and for the comparatively weaker magnitude of the $A_{1u}$ signal~\footnote{
    Indeed the sign of $w$ must be fixed microscopically by the fact that the signs of these two contributions are opposite---the $(c^2-a^2-b^2)$ state in \cref{fig:E_irrep_fragmentation}(b) has $\cramped{\Tr[Q_{\o}^3]>0}$ and $\cramped{\phi_{\o}<0}$, implying that~$\cramped{w>0}$). 
}.
Furthermore, the minima of this quartic free energy functional always correspond to uniaxial orders (of $\cramped{3c^2-r^2}$ type), which have $\cramped{\Tr[Q^3]\neq 0}$, whereas purely biaxial orders (of $\cramped{a^2-b^2}$ type) have $\cramped{\Tr[Q^3]=0}$ and are only selected by tuning a sixth-order $\Tr[Q^3]^2$ term which is irrelevant under scaling~\footnote{
    We could alternatively define a two-component vector order parameter for the $E_u$ irrep using \cref{eq:quadrupole_OP_2_components}, or equivalently a single complex order parameter,
    \begin{equation*}
        \tilde{q} = Q_{3z^2-r^2}+iQ_{x^2-y^2} = \vert \tilde{q} \vert e^{i\vartheta}.
    \end{equation*}
    In terms of $\tilde{q}$, we can rewrite the quadrupole tensor invariants as $\cramped{\Tr[Q^2]=\vert \tilde{q} \vert^2}$ and $\cramped{\Tr[Q^3] = \vert \tilde{q}\vert^3 \cos(3\vartheta)}/\sqrt{6}$. 
    The six pure-uniaxial $\cramped{\pm(3c^2-r^2)}$ order parameters correspond to $\cramped{\vartheta \in (2\pi/6)\mathbbm{Z}}$, while the six pure-biaxial $\cramped{\pm(a^2 - b^2)}$ order parameters with $\cramped{\Tr[Q^3]=0}$ correspond to $\cramped{\vartheta \in (2\pi/6)(\mathbbm{Z}+1/2)}$.
    This free energy is then formally equivalent to that describing the development of so-called ``$\psi_2$'' (uniaxial) and ``$\psi_3$'' (biaxial) orders in anisotropic pyrochlore spin models, in which biaxial states are selected by a (dangerously) irrelevant sixth-order $\cos(6\vartheta)$ term~\cite{javanparastOrderbydisorderCriticalityXY2015,franciniExactNematicMixed2025,rauFrustratedQuantumRareEarth2019,hallasExperimentalInsightsGroundState2018}.
    The activation of the scalar order parameter has also been observed to occur in the pyrochlore by the same linear-cubic coupling~\cite{franciniExactNematicMixed2025}.
}.

These observations are consistent with a simple microscopic picture: a typical microstate in this phase may be viewed as a thermally disordered version of one of the six uniaxial fracton crystal ground states.
In fact, this picture can be directly observed in real-space snapshots taken from Monte Carlo simulations. 
The key observables that expose the behavior are the chain-based N\'eel polarizations $\N_{\c}$ (\cref{eq:chain_neel_op}).
In \cref{fig:nematic_cubes} we visualize the values of $\N_{\c}$ for all chains by associating each chain to a point on the surface of a cube.
Each panel shows a snapshot of a typical configuration at different temperatures from a single simulation which has been slowly annealed. 
Blue (red) points on the cube surface indicate chains with $\cramped{\N_{\c} < 0}$ ($\cramped{\N_{\c} > 0}$), while white points indicate chains with $\cramped{\N_{\c} \approx 0}$.
\Cref{fig:nematic_cubes}(a) shows a snapshot above $\TCuni$ in the paramagnetic phase, where all of the chain order parameters are nearly zero on every chain in all directions.
\Cref{fig:nematic_cubes}(b) shows a snapshot below the transition to the uniaxial phase: chains along the $c$-axis exhibit a large $\cramped{\N_{\c}\approx -1}$ (blue face of the cube), while the $a$- and $b$-axis chains exhibit weaker but opposite ordering, $\cramped{\N_{\c} \sim +0.5}$ (washed-out red faces).
This pattern clearly breaks the rotational symmetry of the cube to a tetragonal subgroup, and is consistent with a thermally disordered version of the uniaxial fracton crystal state state shown in \cref{fig:fracton-crystal}(a), which would produce a cube with one blue face ($\N_{\c}=-1$) and two red faces ($\N_{\c}=+1$).

Despite the highly unconventional structure of the ground-state manifold established in \cref{sec:fracton_crystal_spinon_excitations_subsystem_symmetry}, the highest-temperature ordering transition at $\TCuni$ is governed primarily by local single-octahedron energetics and appears fairly conventional.
The role of the accidental subsystem symmetries only becomes apparent upon further cooling, as we discuss next.

\subsection{Fluctuation-Driven Biaxiality and Dimensional Reduction}
\label{sec:biaxial}

As the temperature continues to decrease, the density of gapped excitations diminishes the system begins to approach its ground state manifold, with the uniaxial crystalline structure becoming more rigid.
As discussed in \cref{sec:fracton_crystal_spinon_excitations_subsystem_symmetry} and illustrated in \cref{fig:fracton-crystal}(d-f), the ground state manifold exhibits accidental subsystem symmetries: in each plane orthogonal to the $c$-axis, chains parallel to \emph{either} the $a$- or $b$-axis can be flipped at zero energy cost, but not both.
At the lowest energies, the minimal excitations propagating on top of the uniaxial fracton crystal state are 1D antiferro-spinons, illustrated in \cref{fig:nematic_AFspinon_chain_deconfinement}.
These excitations can in principle deconfine along either the $a$- or $b$-axis in each plane, implying an instability of the uniaxial fracton crystal ground state at low temperatures.

Indeed, returning to the numerical results in \cref{fig:nematic-thermodynamics}, a second phase transition occurs at a lower temperature $\TCbi/J\approx 0.96$, signaled by a small peak in the specific heat, which is only clearly resolvable for relatively large system sizes---the bottom right inset of \cref{fig:nematic-thermodynamics}(a) shows its weak system size dependence.
Despite the weakness of the specific heat response, the low-temperature phase is strikingly characterized by the eigenvalues of the quadrupolar order parameter, \cref{fig:nematic-thermodynamics}(c)---the two degenerate eigenvalues split below $\TCbi$, resulting in a \emph{biaxial} nematic phase, i.e. one with $Q_{a^2-b^2} \neq 0$ (\cref{eq:quadrupole_OP_2_components}).
The real-space snapshot shown in \cref{fig:nematic_cubes}(c) illustrates what is happening: within each plane orthogonal to the $c$-axis, chains along either the $a$- or $b$-axis are polarized with saturated $\vert \N_{\c}\vert \approx 1$, while chains along the third axis have become highly disordered with $\N_{\c} \approx 0$. 
This is consistent with the deconfinement of antiferro-spinons in each plane along one of the two axes---by the standard Peierls argument, the presence of these deconfined quasiparticles disorders the chains at finite temperature, yielding a large finite-temperature entropy and a significant reduction in free energy, resulting in a fluctuation-driven instability of the uniaxial fracton crystal structure.
Such a fluctuation-driven mechanism is a type of order-by-disorder~\cite{villainInsulatingSpinGlasses1979,villainOrderEffectDisorder1980,henleyOrderingDisorderGroundstate1987,henleyOrderingDueDisorder1989,wengelSpinglassAntiferromagnetCritical1996,hiziAnharmonicGroundState2009}, but instead of selecting a particular state out of a large ground state manifold, it instead drives \emph{spontaneous dimensional reduction}---the low temperature behavior is that of decoupled 1D chains which are disordered in the low-temperature limit.

The dimensional reduction to 1D chain behavior is reflected in our numerical observations at the lowest temperature. 
First, near the transition to the biaxial phase, the $A_{1u}$ order parameter declines precipitously, as seen in \cref{fig:nematic-thermodynamics}(b).
This is consistent with the deconfinement of antiferro-spinons and the disordering of the N\'eel domains: as shown in \cref{fig:nematic_AFspinon_chain_deconfinement}, the two N\'eel domains have opposite patterns of staggered monopole moments, so they tend to cancel each other in their contribution to the scalar $A_{1u}$ order parameter, which measures the staggered monopole moment.
Note that the staggered $E_u$ moments in the two domains are aligned along two different axes, therefore they do not give canceling contributions to the quadrupole tensor order parameter but rather contribute to two different components of it, giving rise to a net biaxial moment.
Second, at the lowest temperatures $T/J \lesssim 0.5$ we see clear features which can be understood as finite-size effects associate with finite-length 1D chains.
The inset of \cref{fig:nematic-thermodynamics}(a) shows that the low-temperature decay of the specific heat matches the exact solution for finite-length 1D chains (\cref{apx:1D_Chains}).
At the same temperature the order parameters shown in \cref{fig:nematic-thermodynamics}(b,c) begin to behave erratically, most strikingly the $A_{1u}$ behavior. 
\Cref{fig:nematic_cubes}(d) demonstrates that at these temperatures the disordered chains are becoming frozen into one of their two N\'eel ground states.
This occurs when the correlation length of the one-dimensional chains has exceeded the finite system size, giving random athermal contributions to the order parameters.
In the thermodynamic limit, the chains can remain disordered at any finite temperature.
Altogether our Monte Carlo data confirms that the lowest-temperature physics is that of decoupled chains, making this a realization of spontaneous dimensional reduction~\cite{mishraDirectionalOrderingFluctuations2004,batistaGeneralizedElitzursTheorem2005,XuReductionEffectiveDimensionality2005,nussinovIntermediateSymmetriesElectronic2006,taharaAntiferromagneticIsingModel2007,nussinovCompassModelsTheory2015,makutaDimensionalReductionQuantum2021} in a classical Ising model.

Returning to the snapshot shown in \cref{fig:nematic_cubes}(c), it is clear that the biaxial phase also breaks translation symmetry along the direction of the uniaxial director axis.
As discussed in \cref{sec:fracton_crystal_spinon_excitations_subsystem_symmetry}, the accidental subsystem symmetries in this system are highly constrained---for each $ab$-plane, chains along either the $a$-axis or the $b$-axis can be flipped independently, but these two sets of accidental symmetries are mutually incompatible with each other within a given plane.
Therefore, the ground state manifold is divided into sectors labeled by a choice of $a$- or $b$-axis in each plane orthogonal to the uniaxial director, so that within each such sector there is an emergent subsystem symmetry flipping chains along the chosen axes.
These accidental symmetries guarantee a massive entropy gain from deconfined quasiparticles, so it is thermodynamically favorable for the system to spontaneously choose one such sector.
One expects that thermal fluctuations will split the free energies of the different sectors at finite temperature, generating weak effective inter-plane interactions, mediated by flipping spins along the uniaxial director axis.
Because our annealed simulations are cooled through the transition at a finite rate (i.e. non-adiabatically), they do not necessarily find the sector which globally optimizes the free energy.
It therefore remains a question as to which sector is chosen in thermodynamic limit, i.e. the extent of effective inter-plane interactions induced by thermal fluctuations.

The simplest scenario one may envisage is that the system selects the uniform sector where the same axis is chosen in every plane, but that is not what is observed in \cref{fig:nematic_cubes}(c).
In our annealed Monte Carlo simulations, each independent replica finds a different pattern of plane-dependent rotational symmetry breaking. 
This is clear from the quantized plateaus reached by the sub-dominant eigenvalues of the quadrupole tensor order parameter shown in \cref{fig:nematic-thermodynamics}(c).
We can compute the allowed plateau values as follows: assume that all $c$-chains are with $\N_{\c} = -1$, and that $N\in[0,L]$ of the $ab$-planes are polarized along $a$ (with $\N_{\c} = +1$) and deconfined along $b$ (with $\N_{\c} \approx 0$), where $L$ is the number of planes.
Correspondingly, $L-N$ planes are polarized along $b$ and deconfined along $a$. 
For such a state the quadrupole tensor order parameter, \cref{eq:quadrupole_OP}, is given by
\begin{equation}
    Q \to \frac{3}{2\sqrt{6}}\mathrm{diag}\left(\frac{N}{L},1 - \frac{N}{L},-1\right).
    \label{eq:plateau_Q}
\end{equation}
Each value of $N$ gives a different set of plateau values for the two sub-dominant eigenvalues, which are marked in \cref{fig:nematic-thermodynamics}(c) with thin dashed horizontal lines. 
Different values of $N$ also give rise to slightly different magnitudes of the $E_u$ order parameter, which can be seen in \cref{fig:nematic-thermodynamics}(b).

Weak inter-plane correlations are evidenced in the distribution of $N$.
If the plane polarizations were completely independent, $N$ would follow a binomial distribution peaked at $N = L/2$.
On the right side of \cref{fig:nematic-thermodynamics}(c), we show a histogram of the plateau values of the two subleading eigenvalues taken from 250 annealed simulations, with a binomial distribution superimposed (dashed black line)
Small systematic deviations from the binomial distribution are visible: the weight near $N=L/2$ is somewhat suppressed and the distribution is slightly broadened.
This result demonstrates measurable but weak correlations between planes and some free energy preference among the different ground state sectors.
Determining the nature of the inter-plane interactions and whether they select a specific sector in the thermodynamic limit would require a careful study of system-size dependence of such histograms over significantly more slowly annealed simulations, and comparison to low-temperature series expansions~\cite{slawnyLowtemperatureExpansionLattice1979}, a topic that we leave for future study.

This concludes our study of the spin nematic $E_u \boxplus A_{1u}$ phase. 
Before closing, we briefly comment on how the physics of this phase relates to similar behavior in other spin models.
Spontaneous dimensional reduction and spin nematicity have been conclusively demonstrated in so-called compass models, which are multi-component (continuous) spin models with directional interactions hosting \emph{exact} chain-based subsystem symmetries~\cite{mishraDirectionalOrderingFluctuations2004,batistaGeneralizedElitzursTheorem2005,nussinovIntermediateSymmetriesElectronic2006,nussinovDiscreteSlidingSymmetries2005,wenzelMonteCarloSimulations2008,wenzelReexaminingDirectionalorderingTransition2010,wenzelUnveilingNatureThreeDimensional2011,nussinovCompassModelsTheory2015,franciniExactNematicMixed2025}.
For Ising systems, a somewhat similar model is the Xu-Moore or square-plaquette model, involving four-spin interactions on the square lattice with exact subsystem symmetries~\cite{XuReductionEffectiveDimensionality2005}, but the physics there involves immobile fracton-like excitations.

A model with more closely related physics is the domino model originally studied by Villain et al. in their pioneering work on order-by-disorder~\cite{villainOrderEffectDisorder1980}.
The domino model is a square lattice Ising model with alternating ferromagnetic and antiferromagnetic chains, with ferromagnetic inter-chain coupling.
Since the Hamiltonian explicitly breaks rotational symmetry, it cannot exhibit spontaneous nematic behavior, but it does host chain-flipping accidental symmetries and correspondingly has subextensive ground state degeneracy.
This system is disordered at zero temperature, yet exhibits finite magnetization upon cooling towards zero temperature, a classic example of order-by-disorder---fluctuations in the antiferro-chains generate effective interactions between the ferro-chain, making it entropically favorable for them all to polarize the same way.
Drawing insight from the discussion in this Section regarding the octochlore spin nematic phase, one can reinterpret the origin of this behavior: the disorder at zero temperature in the domino model may be viewed result of the lowest-energy excitations being deconfined spinons. 
However, these spinons need not be deconfined at \emph{finite} temperature---the ability of a spinon to traverse its chain at zero energy cost is only guaranteed to cost zero energy if the surrounding spins are frozen in their ground state configuration.
Thus, fluctuations of those off-chain spins can confine the spinons, or in other words generate a string tension, allowing the system to be ordered at finite temperature.
Such a putative string tension would be the analog of the thermally-generated pseudo-goldstone gap in systems with accidental continuous U(1) symmetries~\cite{rauPseudoGoldstoneGapsOrderbyQuantum2018,khatuaPseudoGoldstoneModesDynamical2023,hickeyUniversalTemperaturedependentPower2025}.
Indeed our thermodynamic data shown in \cref{fig:nematic-thermodynamics}(b) are potentially consistent with the presence of a weak string tension between antiferro-spinons, because the $A_{1u}$ order parameter does not vanish in the biaxial phase.
As illustrated in \cref{fig:nematic_AFspinon_chain_deconfinement}, the deconfinement of antiferro-spinons on the chains should disorder the N\'eel domains and thereby fully disorder the $A_{1u}$ order parameter.
The fact that the $A_{1u}$ order parameter remains small but finite may imply that the disordered chains retain a small but finite N\'eel order parameter. 
Given this interpretation, we would then predict that in the thermodynamic limit the disordered chains' N\'eel order parameters, and the $A_{1u}$ order parameter, should remain finite in the biaxial phase but tend towards zero at $T=0$, corresponding to the vanishing of a (fluctuation-generated string tension between antiferro-spinons.
We leave the exploration of such an interesting possibility for future work.

\section{Magnetic Octochlore Materials}
\label{sec:materials}

Having established in detail the phase diagram of the classical Ising Hamiltonian on the octochlore lattice, \cref{eq:H}, in the preceding sections, we now turn to the question of its physical realization. 
From a materials perspective, the octochlore geometry is realized by the magnetic-moment-bearing sites of anti-perovskite compounds and also in a class of magnetic trivalent rare-earth fluoride insulators. 
In this section, we survey these two material families, which  offer a concrete experimental arena in which the exotic physics harbored by our minimal model may be sought. 

\subsection{Rare Earth Anti-Perovskites}

The octochlore structure is perhaps best known in the anti-perovskite (or inverse-perovskite) structure, realized in materials with chemical formula \ce{R3AX}~\cite{PhysRevB.97.060401}. 
The R ions form the octochlore lattice as shown in \cref{fig:unit_cells}(a), with X ions sitting at the centers of octahedra and A ions in the voids between octahedra. 
Of note, anti-perovskites are typically metallic, with a transition metal ion such as Mn occupying the R site~\cite{PhysRevB.97.060401}. 
Placing rare-earth ions on the R site of an ideal insulating anti-perovskite is highly unfavorable---their strong preference for the 3+ oxidation state would require unrealistically large negative charge on the A and X ions to maintain charge neutrality.
Therefore rare-earth anti-perovskites should generally be expected to be metallic.
If the $4f$ moments remain strongly localized and unquenched (weak Kondo coupling), RKKY interaction would naturally generate spin-spin exchange in such materials, though not limited to $J_1$ and $J_{2a}$.
Many lanthanide anti-perovskite nitrides have been proposed as stable materials~\cite{zhouStabilityTopologicalBehaviors2025}.
Indeed a series of metallic rare earth nitrides \ce{R3NIn} with R a rare earth have been synthesized and characterized~\cite{KIRCHNER20031247,niewaMetalRichTernaryPerovskite2019,ALHARBI2024416544}.
One compound in particular, \ce{Ce3NIn}, was found to exhibit a transition to commensurate antiferromagnetic order with moments aligned along the crystal axes in Ising-like fashion, apparently with each octahedron in the state shown in \cref{fig:spinon_interactions}(d)~\cite{gablerMagneticStructureInverse2008}.
Another rare-earth antiperovskite compound, \ce{Ho3InC}, was recently reported to have a ground state consisting of a staggered arrangement of the configurations shown in \cref{fig:spinon_interactions}(c), suggested to be stabilized by first and second (including $J_{2b}$) neighbor interactions~\cite{luoAbsenceFragmentedSpin2026}.
In a somewhat different context, the rare earth antiperovskite compound \ce{Eu3PbO}, while not having Ising moments, exhibits a rich variety of tunable topological semimetal behaviors~\cite{hirschmannCreatingControllingDirac2022}.
While anti-perovskites may not be an ideal place to realize rare earth octochlore \emph{insulators}, they are a promising platform for \emph{metallic} frustrated magnets, which have generated much interest as platforms for the study of exotic quantum criticality~\cite{savaryNewTypeQuantum2014,zhaoQuantumcriticalPhaseFrustrated2019} and non-Fermi liquid behaviors~\cite{vojtaFrustrationQuantumCriticality2018}.

\subsection{Rare Earth Octochlore Fluorides}

\begin{figure}[t]
    \centering
    \vspace{1cm}
    \begin{overpic}[width=\linewidth]{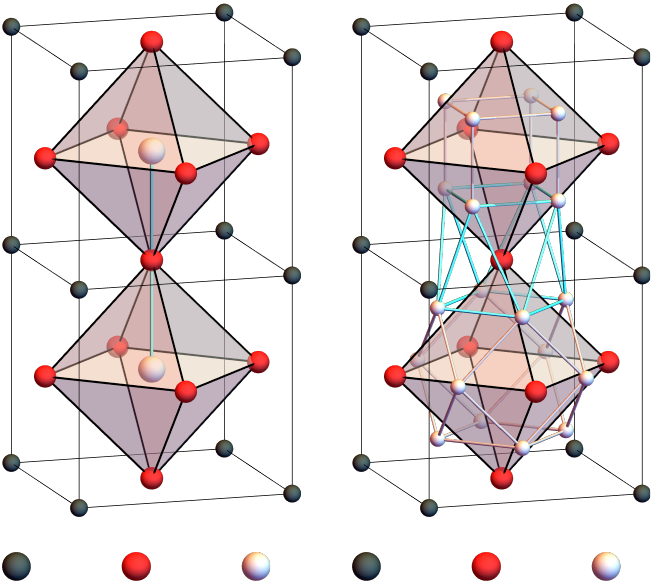}
        \put(06.5,1.2){A}
        \put(25.0,1.2){R}
        \put(43.5,1.2){X}
        \put(60.0,1.2){A}
        \put(78.5,1.2){R}
        \put(97.0,1.2){F}
        \put(00,95){\large (a)}
        \put(50,95){\large (b)}
        \put(15,95){\large {\ce{R3AX}}}
        \put(65,95){\large {\ce{AR3F10}}}
    \end{overpic}
    \caption{\textbf{Composition of octochlore materials.} (a) The anti-perovskite structure with chemical formula \ce{R3AX}. The R sites form an octochlore lattice. (b) The octochlore fluoride structure with chemical formula \ce{AR3F10} with \ce{A=K,Rb,Cs}. The R sites form an octochlore lattice. The R site sits inside of a square anti-prism cage of F ions, shown with cyan lines.}
    \label{fig:unit_cells}
\end{figure}

A promising family of insulating materials is \ce{AR3F10}~\cite{chamberlainthesis,chamberlainMagneticOrderingKDy3F102003,chamberlainMagneticOrderingRbGd3F102003,chamberlainMagneticOrderingRareearth2005,chamberlainParamagneticPropertiesCubic2006,youHydrothermalSynthesisMixed2010,McMillenCrystalChemistryFluorides2015,demortierUnusualPlanarAnisotropy2025}, with R a trivalent rare earth ion, and \ce{A=K,Rb,Cs}, which realize the octochlore structure as shown in \cref{fig:unit_cells}(b). 
Each rare earth ion sits in a square anti-prism cage of F ions with $C_{4v}$ symmetry (compare to the distorted cube cage of O ions with $D_{3d}$ symmetry surrounding each rare earth ion in pyrochlore oxides~\cite{subramanianOxidePyrochloresReview1983}).
Interestingly, and of significant importance to make experimental progress in the study of highly frustrated magnetic materials, these compounds can be grown as single crystals~\cite{McMillenCrystalChemistryFluorides2015,demortierUnusualPlanarAnisotropy2025}.
Chamberlain {\it et al.}\ synthesized the \ce{KR3F10} series with \ce{R=Dy}~\cite{chamberlainMagneticOrderingKDy3F102003}, Er, Yb ~\cite{chamberlainMagneticOrderingRareearth2005}, each of which likely exhibit a crystal field ground state doublet with predominantly Ising-like moments (KTm$_3$F$_{10}$ may, given its reported thermodynamics~\cite{chamberlainMagneticOrderingRareearth2005}, have a nearly-degenerate singlet pair).
The compounds with \ce{R=Ho}~\cite{chamberlainParamagneticPropertiesCubic2006} and Tb~\cite{demortierUnusualPlanarAnisotropy2025,chamberlainMagneticOrderingRareearth2005} have a non-magnetic singlet crystal field ground state, but the remaining Lanthanide rare earth compounds in this series remain promising candidates for interesting frustrated magnetic insulators.
It is also possible to substitute K by Rb~\cite{chamberlainMagneticOrderingRbGd3F102003,chamberlainParamagneticPropertiesCubic2006} or Cs~\cite{McMillenCrystalChemistryFluorides2015}, yielding additional series in the family. 
In particular the compound \ce{RbSm3F10} has been reported to have Ising moments with no order down to 8~mK~\cite{chamberlainParamagneticPropertiesCubic2006}.
It appears that little further work has been done to study this family as a platform for frustrated magnetism, but the ground work has been laid to explore these compounds in more detail.

Chamberlain {\it et al.}\ reports competition between dipole and exchange couplings in the \ce{KR3F10} compounds they studied~\cite{chamberlainMagneticOrderingKDy3F102003,chamberlainMagneticOrderingRareearth2005,chamberlainParamagneticPropertiesCubic2006}, which may be favorable for observing spin ice physics of dipolar origin~\cite{denhertogDipolarInteractionsOrigin2000} in some of those compounds at low temperatures~\cite{szaboFragmentedSpinIce2022}.
In a similar vein as in current efforts devoted to some rare-earth pyrochlore oxides~\cite{smithExperimentalInsightsQuantum2025,hallasExperimentalInsightsGroundState2018,rauFrustratedQuantumRareEarth2019}, it will be especially interesting to explore whether such compounds with small magnetic moments can be synthesized and whose behavior is governed primarily by exchange couplings as opposed to long-range magnetostatic dipole-dipole interactions.
While the second-neighbor $J_{2a}$ exchange may be expected to be weaker than first-neighbor $J_1$ exchange interaction, for lack of an anion ligand mediating a superexchange pathway, making the realization of the fracton CSL and frustrated chains phase potentially challenging, a positive $J_1$ with a small $J_{2a}$ could stabilize either the fragmented spin ice CSL or the spin nematic phase in the phase diagram of~\cref{fig:phase_diagram}(f).
Observation of either of these would be a significant achievement in the exploration of frustrated magnetic phenomena in three dimensions.
Given advances in crystal growth methods which have been applied intensively to grow large clean single crystals of a variety of pyrochlore compounds~\cite{powellHydrothermalCrystalGrowth2019}, and growing interest in non-oxide pyrochlores~\cite{reig-i-plessisFrustratedMagnetismFluoride2021}, this family of non-oxide octochlore compounds is a promising avenue for exploring the rich physics discussed in this work.

\section{Discussion}
\label{sec:discussion}

In this work, we have explored in detail the phase diagram of Ising moments on a three-dimensional octochlore lattice, shown in \cref{fig:phase_diagram}(a), with first-neighbor $J_1$ and second-neighbor $J_{2a}$ interactions inside each octahedron, leveraging the notion of irreducible cluster multipole moments~\cite{suzukiMultipoleExpansionMagnetic2019} for each octahedron which have played a key role in the study of pyrochlore magnetism~\cite{bentonSpinliquidPinchlineSingularities2016,taillefumierCompetingSpinLiquids2017,yanTheoryMultiplephaseCompetition2017,noculakClassicalQuantumPhases2023,lozano-gomezCompetingGaugeFields2024,chungMappingPhaseDiagram2024,lozano-gomezAtlasClassicalPyrochlore2024,franciniExactNematicMixed2025,franciniHigherrankSpinLiquids2025,chungGeometricallyFrustratedQuadrupoles2025,yogendraSymmetrySuperpositionFragmentation2024}.
Some of the physics is familiar from the pyrochlore lattice: there is a spin ice string-net classical spin liquid with quasiparticle excitations carrying monopole moments and an all-in/all-out phase whose ground states can be viewed as a fully packaged crystalline arrangement of monopoles. 
We identified three parts of the phase diagram hosting remarkable frustrated physics without clear precedent: 
a frustrated chains phase exhibiting a smooth dimensional crossover and quasicritical Kasteleyn-like  physics (\cref{sec:frustrated_chains}); a fracton cage-net CSL that hosts lineon quasiparticle excitations carrying quadrupole moments and realizing a classical X-cube model (\cref{sec:fracton_CSL}); and a spin nematic phase exhibiting spontaneous dimensional reduction (\cref{sec:spin_nematic}).
The frustrated chains and spin nematic phases exhibit subextensive degeneracies intermediate between spin liquids and ordered phases, and their low-temperature physics is driven by deconfinement of 1D spinon quasiparticles and corresponding emergent subsystem symmetries.
Our work opens numerous avenues for further research into the physics of 3D frustrated magnets with Ising moments.
Most importantly, it lays the basic foundation for experimental work studying octochlore materials such as those discussed in \cref{sec:materials}.
Before concluding, we briefly outline interesting directions for future work building on our results.

\subsection{Open Questions for the Frustrated Chains Phase}

With regard to the classical model studied here, a number of interesting questions remain.
A prominent question is the nature of the avoided Kasteleyn-like transitions at the ends of the frustrated chains phase discussed in \cref{sec:kasteleyn} where there are at least two key directions to explore. 
A first would be to characterize the nature of the confinement-deconfinement transition driven by the ``string bending energy'' described by \cref{eq:F_string_kink}, which is similar yet distinct from the Kasteleyn transition driven by string tension described by \cref{eq:F_string_Kastleyn}.
A second direction would be to study the analogous physics near the fractonic X-cube spin liquid, where we conjecture that a similar avoided crossover occurs in our phase diagram---what is the nature of the Kasteleyn-like confinement transitions out of a \textit{fracton} spin liquid described by \cref{eq:F_cage_corner}?
Furthermore, it is interesting to ask whether the zero-modes of this phase guarantee that there is no phase transition, along the lines of the generalized Elitzur's theorem of Ref.~\cite{batistaGeneralizedElitzursTheorem2005} which guarantees no phase transition for exact chain-based subsystem symmetries.

\subsection{Open Questions for the Spin Nematic Phase}

With regard to the biaxial spin nematic phase discussed in \cref{sec:spin_nematic}, we have not determined whether the direction of the secondary axes within the biaxial phase is truly random in different planes.
We found evidence for weak inter-plane correlations, but it is an interesting statistical mechanism problem to explore the consequence of these interactions in the thermodynamic limit and whether a particular ordered pattern is selected via order-by-disorder.
Furthermore, at the end of \cref{sec:spin_nematic} we commented on similarity between the physics of the biaxial phase and the domino model~\cite{villainOrderEffectDisorder1980}, and hypothesized that the antiferro-spinon excitations could be weakly confined by a temperature-dependent string tension analogous to the thermal gap of pseudo-Goldstone bosons in systems with accidental U(1) symmetries~\cite{rauPseudoGoldstoneGapsOrderbyQuantum2018,khatuaPseudoGoldstoneModesDynamical2023,hickeyUniversalTemperaturedependentPower2025}.
Exploring these issues further could deepen our understanding of order-by-disorder in classical Ising models, examples of which are sparse~\cite{villainOrderEffectDisorder1980,wengelSpinglassAntiferromagnetCritical1996,stubelFinitesizeScalingMonte2018,polgreenMonteCarloSimulation1984}.

\subsection{Spinon Bound State Condensation Mechanism}

In \cref{sec:CSLs_as_Spinon_Condensates} we  showed that the two CSLs in the current model can be viewed as condensates of different bound states of spinons arising from coupled chains.
This is similar in spirit to the concept of gauging a subsystem symmetry to generate fracton topological order~\cite{shirleyFoliatedFractonOrder2019,williamsonTypeIIFractonsCoupled2021,rayhaunHigherformSubsystemSymmetry2023,gorantlaStringMembraneNets2025,hsinSubsystemSymmetryFractionalization2025,gorantlaGaugingNexusTopological2026}.
The spinon bound state condensation approach can be generalized to build Ising CSLs on a variety of lattices constructed from intersecting chains.
While the octochlore lattice is the next-simplest after the pyrochlore, it will be very interesting to explore Ising systems on other corner-sharing-cluster lattices with more spins per cluster, giving rise to additional multipole moments, more spin liquids, and more exotic phases intermediate between spin liquids and ordered phases.

\subsection{Extensions of the Octochlore Ising Model}

One obvious path forward is to explore extensions of the minimal Ising model we have studied here with the addition of the inter-octahedra $J_{2b}$ interaction, and to develop an irrep analysis for system with interaction between the corner-sharing clusters.
Studying the extended $J_1$-$J_{2a}$-$J_{2b}$ phase diagram of the Ising model will be relevant for modeling the ground states of the rare earth anti-perovskites \ce{Ce3NIn}~\cite{gablerMagneticStructureInverse2008} and \ce{Ho3InC}~\cite{luoAbsenceFragmentedSpin2026}.
Furthermore, we expect interesting frustrated magnetism physics to occur along the $J_1$-$J_{2b}$ line---whereas the pure-$J_{2a}$ model leads to decoupled 1D chains, the pure-$J_{2b}$ model leads to decoupled 2D \emph{planar} Ising models, which are coupled together by the $J_1$ interaction, likely leading to some analog of the frustrated chains phase.
Moreover, we expect an additional exotic classical spin liquid on the $J_1$-$J_{2b}$ line, such as a spin vorticity model which realizes a 2-form U(1) spin liquid whose excitations are strings rather than particles~\cite{chung2formU1Spin2025}.
In particular, it will be interesting to examine how this spin liquid arises from a mechanism similar to the spinon bound state condensation discussed in \cref{sec:CSLs_as_Spinon_Condensates}.

\subsection{Anisotropic Octochlore Spin Hamiltonians}

Another promising  direction is to investigate the three-component classical spin model on this lattice, retaining the $S^x$ and $S^y$ spin components and incorporating all symmetry-allowed anisotropic interactions.
Such an approach would parallel the corresponding program on the pyrochlore lattice, which has provided impetus for the study of exotic phases in strongly correlated frustrated spin systems~\cite{rossQuantumExcitationsQuantum2011,huangQuantumSpinIces2014,rauFrustratedQuantumRareEarth2019}.
The phase diagram of the three-component bilinear spin model on the pyrochlore lattice including all symmetry-allowed anisotropic interactions has been fully mapped in the classical limit using the irreducible multipole approach~\cite{chungMappingPhaseDiagram2024,yanTheoryMultiplephaseCompetition2017}, and  found to harbor a variety of classical spin liquids~\cite{taillefumierCompetingSpinLiquids2017,lozano-gomezAtlasClassicalPyrochlore2024,chungMappingPhaseDiagram2024,bentonSpinliquidPinchlineSingularities2016,franciniHigherrankSpinLiquids2025,lozano-gomezCompetingGaugeFields2024,noculakClassicalQuantumPhases2023,rauFrustratedQuantumRareEarth2019} and other frustrated behaviors including spin nematicity~\cite{moessnerLowtemperaturePropertiesClassical1998,taillefumierCompetingSpinLiquids2017,franciniHigherrankSpinLiquids2025,franciniExactNematicMixed2025}.
There is little doubt that the analogous anisotropic model on the octochlore lattice will harbor analogous phases and spin liquids with quite likely even more complex behaviors.
Indeed, as have shown in this work, the octahedral unit exhibits richer physics than the tetrahedral unit owing to its increased number of spins and corresponding additional competing irreducible multipole modes.
A concrete problem of experimental relevance to the rare-earth fluoride insulating compounds discussed in \cref{sec:materials} is to determine the types of symmetry-allowed crystal field doublets which occur in this family of compounds; in the pyrochlore there exist three distinct types of doublets---pseudospin-1/2, non-Kramers, and dipolar-octupolar~\cite{huangQuantumSpinIces2014}---each with their own spin Hamiltonian dictating their low-temperature physics~\cite{rauFrustratedQuantumRareEarth2019,smithExperimentalInsightsQuantum2025}.

\subsection{U(1) X-Cube Quantum Spin Liquid}

The full anisotropic model including all spin components will contain terms such as $S_i^+S_j^-$, $S_i^+ S_j^+$, and $S_i^z S_i^+$, where $\smash{\cramped{S^{\pm} = S^x \pm i S^y}}$ act to flip spins in the $S^z$ basis.
If these are added perturbatively to the classical $S^z$ model studied in this work, they may be viewed as generating quantum dynamics (kinetic energy) for the various quasiparticle excitations which appear throughout the phase diagram. 
It is of particular interest to explore the physics of the classical X-cube cage-net CSL studied in \cref{sec:fracton_CSL} when perturbed by such terms, which perturbatively generate \textit{ring exchange}---quantum processes which tunnel between classical ground states, promoting a classical spin liquid to a quantum spin liquid~\cite{hermelePyrochlorePhotonsU12004}.
For the X-cube CSL, it is straightforward to write down a quantum ring exchange term which performs the cage-net move shown in \cref{fig:cages_and_thermodynamics}(a).
The resulting effective Hamiltonian will map to a U(1) lattice gauge theory~\cite{hermelePyrochlorePhotonsU12004,chung2formU1Spin2025}, 
specifically a spin-1/2 quantum link model~\cite{chandrasekharanQuantumLinkModels1997} realization of the U(1) X-cube lattice gauge theory~\cite{seibergExoticU1Symmetries2020}.
This should give rise to a quantum spin liquid hosting a photon-like excitation which is gapless along 1D lines in reciprocal space~\cite{seibergExoticU1Symmetries2020}, arising from the nodal line in the band structure. 
Quantum fluctuations will likely also give rise to rich sub-dimensional dynamics in both the frustrated chains and spin nematic phases, which will be of great interest to explore.

\subsection{Outlook}

Altogether, our work lays the groundwork for the exploration of highly frustrated 3D Ising systems beyond the pyrochlore lattice, advancing the theoretical methods for studying these systems and paving the road towards new experimentally-accessible materials for studying 3D frustrated magnetism.
On the theoretical front, we have extended techniques used in the continuous-spin analysis of frustrated magnets to the study of Ising systems, providing a clean demonstration of the usefulness and physical meaning of the multipolar irreps in the characterization of frustrated phases.
We uncovered a classical fracton CSL that broadens the spin ice paradigm, together with a striking set of frustrated phases intermediate between spin liquids and conventional long-range order.
On the experimental front, our work establishes that Ising-like rare-earth magnets in corner-sharing geometries beyond the pyrochlore's tetrahedra are a highly promising target for further study. 
We hope our work will inspire the further development and fruitful investigation of both rare-earth anti-perovskite and octochlore fluoride compounds, and spur on the search for other magnetic rare-earth octochlore materials which, one may enthusiastically hope, could constitute a next-generation platform for 3D frustrated magnets.

\begin{acknowledgments}
We thank Han Yan, Johannes Reuther, and Zohar Nussinov for useful discussions and thank Peter Holdsworth for some comments on the manuscript.
The work at the University of Waterloo was funded by the NSERC of Canada as well as supported by the  University of Waterloo Research Chair Program.
M.D.B. acknowledges CGS-M and CGS-D scholarships from NSERC. 
M.S. and J.R. were supported by the NSF through grant DMR-2142554.
This work was in part supported by the Deutsche Forschungsgemeinschaft under the cluster of excellence ct.qmat (EXC-2147, project number 390858490).
This research was enabled in part by computing resources provided by the Digital Research Alliance of Canada. 
\end{acknowledgments}


\appendix

\section{Lattice Conventions}
\label{apx:conventions}

The octochlore lattice is the line graph of the cubic lattice, meaning that each octochlore site sits at the midpoint of a cubic lattice edge~\cite{henleyCoulombPhaseFrustrated2010}. 
Since there are three edges per cubic lattice unit cell, it is natural to use a 3-site primitive unit cell.
On the other hand, it is convenient to use a 6-site unit cell which contains all six sites of a single octahedron, with an fcc lattice.
For the purpose of Monte Carlo simulations with periodic boundaries, on the other hand, it is convenient to use a a larger 24-site unit cell with a cubic space lattice with enlarged spacing.
Denoting the Cartesian cubic axis unit vectors $\hat{\bm{x}}$, $\hat{\bm{y}}$, $\hat{\bm{z}}$, and general vectors by 
\begin{equation}
    [abc] \equiv a \,\hat{\bm{x}} + b\,\hat{\bm{y}} + c\,\hat{\bm{z}},
\end{equation}
and $-[abc] \equiv [\bar{a} \bar{b} \bar{c}]$,
these unit cells have translation vectors
\begin{equation}
     {
     \setlength{\arraycolsep}{10pt}
     \begin{array}{*4{>{\displaystyle}c}p{4cm}}
        &
        \underline{\text{3-site}} & 
        \underline{\text{6-site}} & 
        \underline{\text{24-site}}
        \\[1ex]
        \bm{a}_1^{(n)} \,\,\,= &
        a_0 [\tfrac{1}{2}00] &
        a_0 [\tfrac{1}{2}\tfrac{1}{2}0] &
        a_0 \, [100]
        \\[1ex]
        \bm{a}_2^{(n)} \,\,\,= &
        a_0 [0\tfrac{1}{2}0] &
        a_0 [\tfrac{1}{2}0\tfrac{1}{2}] &
        a_0 \, [010]
        \\[1ex]
        \bm{a}_3^{(n)} \,\,\,= &
        a_0 [00\tfrac{1}{2}] &
        a_0 [0\tfrac{1}{2}\tfrac{1}{2}] &
        a_0 \, [001] ,
    \end{array}
    }
\end{equation}
where $a_0$ is the edge length of the enlarged 24-site cubic cell and the superscript $n\in\{3,6,24\}$  in  $\bm{a}_i^{(n)}$indicates the number of sites in the unit cell. 
The sublattice positions for the 3-site unit cell are
\begin{equation}
    \bm{c}_1^{(3)} = \frac{a_0}{4}[100],
    \,\,
    \bm{c}_2^{(3)} = \frac{a_0}{4}[010],
    \,\,
    \bm{c}_3^{(3)} = \frac{a_0}{4}[001].
\end{equation}
For the 6-site unit cell they are
\begin{equation}
    \begin{aligned}
    \bm{c}^{(6)}_1 &= \frac{a_0}{4}[100],
    \,\,
    \bm{c}^{(6)}_2 = \frac{a_0}{4}[010],
    \,\,
    \bm{c}^{(6)}_3 = \frac{a_0}{4}[001],
    \\[1ex]
    \bm{c}^{(6)}_4 &= \frac{a_0}{4}[\bar{1}00],
    \,\,
    \bm{c}^{(6)}_5 = \frac{a_0}{4}[0\bar{1}0],
    \,\,
    \bm{c}^{(6)}_6 = \frac{a_0}{4}[00\bar{1}].
    \end{aligned}
\end{equation}
For the 24-site unit cell, we define
\begin{equation}
    \bar{\bm{c}}_{\mu}^{(24)} = \bm{a}^{(6)}_{\lfloor (\mu-1) / 6 \rfloor} + \bm{c}^{(6)}_{(\mu \text{ mod } 6) +1},
\end{equation}
where $\mu \in \{1\ldots 24\}$.
We define $\bm{a}_0 \equiv [000]$ and $\lfloor\ldots\rfloor$ denotes the floor function.
This enumerates the six corners of each of the four octahedra centered at $\bm{a}_0$, $\bm{a}_1^{(6)}$, $\bm{a}_2^{(6)}$, and $\bm{a}_3^{(6)}$.

In the 3-site unit cell we define local $C_{4v}$ easy-axis vectors at each sublattice $\smash{\bm{c}_\mu^{(3)}}$ as
\begin{equation}
    \hat{\bm{z}}_1 = [100],\quad 
    \hat{\bm{z}}_2 = [010],\quad 
    \hat{\bm{z}}_3 = [001],
    \label{eq:SM_zhat_3_site}
\end{equation}
which point along the positive Cartesian directions.
For the 6-site unit cell, we define the easy-axis vectors at each sublattice $\smash{\bm{c}_\mu^{(6)}}$ as
\begin{equation}
    \begin{aligned}
    \hat{\bm{z}}_1 = [100],\quad 
    \hat{\bm{z}}_2 = [010],\quad 
    \hat{\bm{z}}_3 = [001],
    \\
    \hat{\bm{z}}_4 = [\bar{1}00],\quad 
    \hat{\bm{z}}_5 = [0\bar{1}0],\quad 
    \hat{\bm{z}}_6 = [00\bar{1}],
    \end{aligned}
    \label{eq:SM_zhat_6_site}
\end{equation}
such that the six easy-axis vectors point from the center of an octahedron outwards, as in \cref{fig:phase_diagram}(b), or from the $\A$ to $\B$ sublattices of the bipartite cubic lattice formed by the octahedron centers, see \cref{fig:phase_diagram}(a). 
For the 24-site unit cell, we use the same easy-axes as the 6-site unit cell.
In terms of these, the local moment at site $i$ points along
\begin{equation}
    \hat{\bm{\moment}}_i = S_i^z \hat{\bm{z}}_i.
\end{equation}
Note that $\hat{\bm{d}}_i$ is \emph{independent} of the unit cell choice, whereas $\hat{\bm{z}}_i$ are chosen by convention, and $S_i^z = \hat{\bm{d}}_i \cdot \hat{\bm{z}}_i$ depends on that convention.

Given a choice of $n$-site unit cell, we consider an $\cramped{L\times L \times L}$ system with periodic boundaries, for which there are $L^3$ allowed reciprocal space wavevectors $\bq$ given be
\begin{equation}
    \bq_{m_1,m_2,m_3} 
    = 
    \frac{1}{L}(m_1 \bm{b}_1^{(n)} + m_2 \bm{b}_2^{(n)}+ m_3 \bm{b}_3^{(n)}),
\end{equation}
where the $\bm{b}_\alpha^{(n)}$ are reciprocal lattice vectors defined to satisfy $\cramped{\bm{a}_\alpha^{(n)} \cdot \bm{b}_\beta^{(n)} = 2\pi \delta_{\alpha\beta}}$, and $\cramped{m_1,m_2,m_3\in\{0,\cdots L-1\}}$.
We define Fourier transforms of single-site variables $\varphi_i$ as
\begin{equation}
    \varphi_\mu(\bq) = \frac{1}{\sqrt{L^3}} \sum_{i\in \mu} \varphi_i \, \exp(-i\bq\cdot \bm{r}_i),
    \label{apx_eq:fourier-transform}
\end{equation}
where $\mu$ in the sublattice index and we assume  system with $L^3$ unit cells and periodic boundaries.
For translationally-invariant two-site variables $\varphi_{ij}$, we define the Fourier transform as 
\begin{equation}
    \varphi_{\mu\nu}(\bq) = \frac{1}{L^3}\sum_{i \in \mu} \sum_{j \in \nu} \varphi_{ij} \, e^{-i\bq \cdot (\bm{r}_i - \bm{r}_j)}.
    \label{eq:matrixFT}
\end{equation}
For specifying reciprocal space wavevectors, we use Miller indices with respect to the 24-site unit cell cubic cell,
\begin{equation}
    \bq = \frac{2\pi}{a_0} \left( h \hat{\bm{x}} + k \hat{\bm{y}} + l \hat{\bm{z}}\right) \equiv (hkl), 
\end{equation}
which are used in \cref{fig:X-cube_bands_structure_factors}.

\begin{table}[t]
    \centering
    \includegraphics[width=\columnwidth]{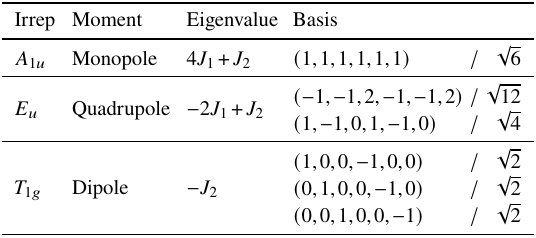}
    \caption{Eigenvalues and eigenvectors of the interaction matrix for a single octahedron, \cref{eq:interaction_matrix_octahedron}, labeled by irreducible representations of $O_h$ and corresponding multipole moment.}
    \label{tab:irrep_table}
\end{table}

\section{Irreducible Representations and Multipoles}
\label{apx:irreps_multipoles}

In \cref{sec:hamiltonian_irreps}, we described the phase diagram of the Hamiltonian \cref{eq:H} using irreducible representations of the point group, illustrated in \cref{fig:phase_diagram}(b-d); we expand further on this here.

The use of irreps to study the phase diagram of corner-sharing clusters has seen extensive use in studying nearest-neighbor anisotropic Heisenberg models on the pyrochlore lattice~\cite{yanTheoryMultiplephaseCompetition2017,yanRank2U1Spin2020,bentonSpinliquidPinchlineSingularities2016,chungMappingPhaseDiagram2024,lozano-gomezAtlasClassicalPyrochlore2024,rauFrustratedQuantumRareEarth2019}. 
In particular, spin liquids occur when multiple irreps are tuned to be degenerate in the ground state~\cite{yanRank2U1Spin2020,bentonSpinliquidPinchlineSingularities2016,chungMappingPhaseDiagram2024,lozano-gomezAtlasClassicalPyrochlore2024,rauFrustratedQuantumRareEarth2019}.
Previous works have not explored the application of these irreps for Ising models.
The octochlore lattice is an excellent platform for its application to Ising models due to having two inequivalent interactions within the octahedra and thus the resulting rich phase diagram studied in the main text. 
The octochlore Hamiltonian we study decomposes into a sum over octahedra
\begin{equation}
    H = \frac{1}{2}\sum_{\o} \sum_{\langle ij \rangle \in \o} S_i^z \mathcal{J}_{ij} S_j^z,
\end{equation}
where the single-octahedron interaction matrix is given using the sublattice ordering convention in \cref{eq:SM_zhat_6_site} by
\begin{equation}
    J_{ij} = \begin{pmatrix}
          0 & J_1 & J_1 & J_{2a} & J_1 & J_1 \\
        J_1 &   0 & J_1 & J_1 & J_{2a} & J_1 \\
        J_1 & J_1 &   0 & J_1 & J_1 & J_{2a} \\
        J_{2a} & J_1 & J_1 &   0 & J_1 & J_1 \\
        J_1 & J_{2a} & J_1 & J_1 &   0 & J_1 \\
        J_1 & J_1 & J_{2a} & J_1 & J_1 &   0 \\
    \end{pmatrix}.
    \label{eq:interaction_matrix_octahedron}
\end{equation}
The symmetries of the octahedron are described by the group $O_h$, which acts on the six Ising spins by permutations of the sublattice position. This 6-dimensional representation $P$ can be decomposed into irreducible representations, where the number of copies of irrep $\Gamma$ is given by the formula~\cite{zeeGroupTheoryNutshell2016}
\begin{equation}
    n_\Gamma = \frac{d_\Gamma}{\vert O_h \vert} \sum_{g \in O_h} \chi_\Gamma(g)^* \Tr[P(g)],
\end{equation}
where $g$ are the group elements of $O_h$, $\vert O_h \vert = 48$, $\chi_\Gamma(g)$ is the character (trace) of $g$ in irrep $\Gamma$, and $d_\Gamma$ is the dimension of the irrep. 
This results in the decomposition ${P = A_{1u} \mixed E_u \mixed T_{1g}}$. 
Since \cref{eq:interaction_matrix_octahedron} commutes with all of the $P(g)$ permutation matrices, its eigenvalues are labeled by $\Gamma$ and are each $d_\Gamma$ degenerate. 
The eigenvalues and eigenvectors are given in \cref{tab:irrep_table}, which are used to write the Hamiltonian in the form of \cref{eq:H_irreps}.

\begin{table}[t]
    \centering
    \includegraphics[width=.95\linewidth]{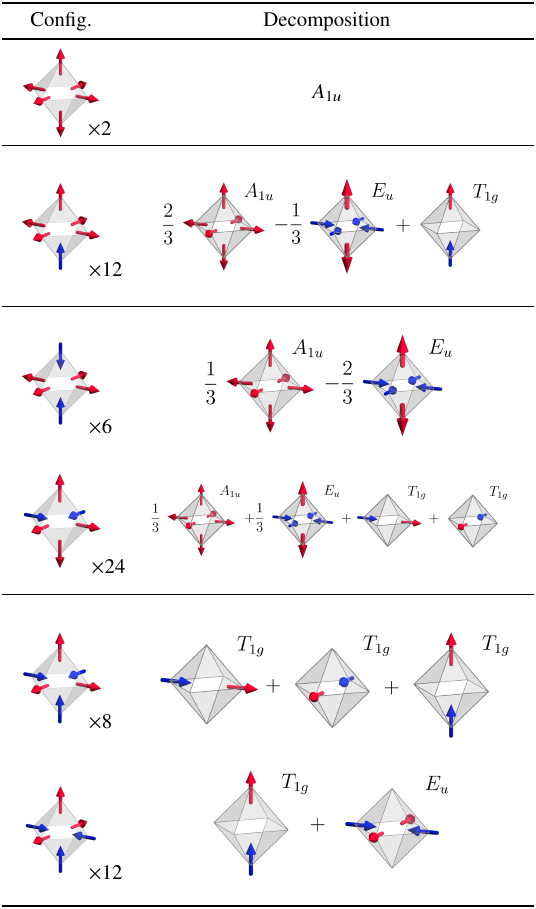}
    \caption{The decomposition of all $64$ spin configurations on an octahedron according to the symmetry group $O_h$ of an octahedron. Spin configurations are grouped into into symmetry classes under the action of $O_h$ and time reversal. The first column shows one state from each symmetry class, with the number of states in each class indicated. 
    We group the various states by their total $S^z$ value, separated by horizontal lines.  
    The only states transforming as a pure irrep are the all-in/all-out states, transforming as $A_{1u}$, and the dipolar states constructed from ferromagnetic chains, transforming as $T_{1g}$. All other states are fragmented.} 
\label{tab:64states}
\end{table}%

These irreps correspond to cluster multipole moments~\cite{suzukiMultipoleExpansionMagnetic2019} of the octahedral spin configuration, as defined in \cref{eq:irrep_multipoles} using the 6-site basis. 
To see this, let $\cramped{\bm{\moment}_i =(1/2) g \mu_{\text{B}} S_i^z \hat{\bm{z}}_i}$ denote the local dipole moment vector. 
The three-dimensional $T_{1g}$ irrep describes the net dipole moment of an octahedron, 
\begin{equation}
    \moment^\alpha_{\o} = \sum_{i \in \o} \moment_i^\alpha 
    \propto 
    \sum_{i \in \o} S_i^z \hat{z}_i^\alpha .
\end{equation}
The other two irreps are contained in the rank-2 moment tensor, 
\begin{equation}
    \mathcal{M}^{\alpha\beta}_{\o} 
    = 
    \sum_{i \in {\o}} (r_i^\alpha - r_{\o}^\alpha)\, \moment_i^\beta ,
\end{equation}
where $\bm{r}_i$ is the position of site $i$ and $\bm{r}_{\o}$ is the position of the center of the octahedron. 
The trace of this tensor gives the $A_{1u}$ irrep order parameter, 
\begin{equation}
    \Tr[\mathcal{M}_{\o}] = \sum_{i \in \o} \hat{\bm{z}}_i \cdot \bm{d}_i\propto \sum_{i\in {\o}} S_i^z,
\end{equation}
which may be interpreted as the ``monopole moment'' of the octahedron since it is a scalar maximized in the all-in or all-out configurations. 
Note that $\mathcal{M}$ has zero off-diagonal components because the $\hat{\bm{z}}_i$ point along the cubic axes, so $\smash{\hat{z}_i^\alpha \hat{z}_i^\beta = 0}$ if $\alpha \neq \beta$. 
Subtracting its trace, we obtain the quadrupole moment tensor,
\begin{equation}
    \mathcal{Q}^{\alpha\beta}_{\o} 
    = 
    \mathcal{M}^{\alpha\beta}_{\o} 
    - 
    \frac{1}{3}
    \Tr[\mathcal{M}_{\o}] 
    \delta^{\alpha\beta} 
    \propto
    \sum_{i \in \o} S_i^z \left(\hat{z}_i^\alpha\hat{z}_i^\beta - \frac{1}{3}\delta_{\alpha\beta}\right).
\end{equation}
The off-diagonal elements of this tensor are zero, leaving only two independent degrees of freedom of the trace-free diagonal, forming the two-component $E_u$ irrep. They correspond to the $3z^2 - r^2$ and $x^2 - y^2$ components of the quadrupole tensor, the first and second $E_u$ eigenvectors listed in \cref{tab:irrep_table}, respectively.
The expressions in \cref{eq:irrep_multipoles} are dimensionless versions of these multipole moments.

A few comments on the multipoles defined in \cref{eq:irrep_multipoles} are in order.
The alternating signs $(-1)^{\o}$ ensure that the \textit{ungerade} irrep monopole and quadrupole moments receive positive contributions from ``out'' spins and negative contributions from ``in'' spins. 
Note, however, that the \textit{gerade} irrep dipole moment $\bm{D}_{\o}$ points in the \emph{opposite} direction of the net magnetic moment for $\B$-sublattice octahedra.
While we could omit the sign in the definition of $\bm{D}_{\o}$ so that it always points in the direction of the physical dipole moment, including the sign allows for a more unified treatment of the phase diagram because the staggered moment is locally conserved by N\'eel domain walls.
In particular, the $A_{1u}\oplus E_u$ decoupled AFM chains point can be treated in a parallel manner to the $A_{1u}\oplus T_{1g}$ X-cube CSL and the $E_u \oplus T_{1g}$ spin ice CSL, whose excitations carry locally conserved $E_u$ quadrupole and $A_{1u}$ monopole charge, respectively.
Lastly, a terminological point---we have labeled $\phi_{\o}$ the monopole moment, which agrees with the notion that the ``hedgehog''-like all-out configuration in \cref{fig:phase_diagram}(a) looks like a positively charged magnetic monopole, and indeed behaves like one in the spin ice spin liquid, especially when long-range dipole-dipole interactions are included~\cite{castelnovoMagneticMonopolesSpin2008,castelnovoSpinIceFractionalization2012}.
In terms of the physical magnetic field produced by the dipole moments, however, this configuration sources a hexadecapolar field.

\mysubsection{Irrep Decomposition of Spin Configurations}

An octahedron, decorated with Ising spins, has $2^6=64$ different configurations.
These split into $(32\times 2)$ states and their time-reversal partners and can be further classified according to their total spin and the symmetry group of the octahedron.
We collect the spin configurations in \cref{tab:64states}, showing their decomposition into the irreducible representations $A_{1u}$ (monopole), $E_u$ (quadrupole), and $T_{1g}$ (dipole).

\section{Interaction Matrices and Structure Factors}
\label{apx:interaction_matrix_SCGA_structure_factors}

In this Appendix we provide the interaction matrices from which the band structures are computed, summarize the self-consistent Gaussian approximation which is used to compute correlation functions, and discuss the structure factors shown in \cref{fig:X-cube_bands_structure_factors}.

\mysubsection{Interaction Matrices}

Using the 6-site unit cell, the bilinear Hamiltonian \cref{eq:H} can be written in terms of adjacency matrices of the octochlore lattice,
\begin{align}
    H &= \frac{J_1}{2} \sum_{ij} S_i^z A^{(1)}_{ij} S_j^z + \frac{J_{2a}}{2} \sum_{ij} S_i^z A^{(2a)}_{ij} S_j^z 
    \nonumber
    \\
    &\equiv \frac{1}{2}\sum_{ij} S_i^z \underbrace{ [J_1 A^{(1)}_{ij} + J_{2a} A^{(2a)}_{ij}]}_{\mathcal{J}_{ij}} S_j^z,
    \label{eq:SCGA_Hamiltonian}
\end{align}
where $\cramped{A^{(n)}_{ij} = 1}$ if sites $i$ and $j$ are $n$'th-nearest-neighbors, and $0$ otherwise, and the factors of $1/2$ removes the double counting. 
We refer to the matrix $\mathcal{J}_{ij}$ as the interaction matrix.
Here, the Ising variables, normalized as $\cramped{S_i^z = \pm 1}$, refer to a dipole oriented along the local $\hat{\bm{z}}_i$ axis defined in \cref{eq:SM_zhat_3_site} and \cref{eq:SM_zhat_6_site}.
This means that $\cramped{J_{2a} > 0}$ energetically prefers \emph{ferromagnetic} alignment of the physical dipoles along the 1D chains coupled by the second neighbor interaction.

In the 3-site unit cell, the Fourier transformed adjacency matrices have the form
\begin{align}
    A^{(1)}_{\mu\nu}(\bq)
    &=
    4\begin{pmatrix}
        0 & s_x s_y & s_x s_z \\[1pt]
        s_x s_y & 0 & s_y s_z \\[1pt]
        s_x s_z & s_y s_z & 0
    \end{pmatrix},
        \nonumber 
    \\[2ex]
    A^{(2a)}_{\mu\nu}(\bq)
    &=
    -2
    \begin{pmatrix}
        c_{xx}^+ & 0 & 0 \\[1pt]
        0 & c_{yy}^+ & 0 \\[1pt]
        0 & 0 & c_{zz}^+
    \end{pmatrix} , 
    \label{eq:SM_A_mats_3site_UC}
\end{align}
where $\mu,\nu$ are sublattice indices and we define
\begin{equation}
    s_\alpha \equiv \sin\left(\frac{q_\alpha}{4}\right) \quad\text{and}\quad
    c_{\alpha\beta}^\pm \equiv \cos\left(\frac{q_\alpha \pm q_\beta}{4}\right).
\end{equation}
The negative sign in front of $A_{\mu\nu}^{(2a)}(\bq)$ is included in \cref{eq:SM_A_mats_3site_UC} because in the unit cell of 3-sites, a positive $J_{2a}$ prefers the antiferromagnetic alignment of the chains. 
With this choice, \cref{eq:SCGA_Hamiltonian} is the same Hamiltonian for either basis choice.
In the 6-site unit cell the adjacency matrices are given by
\begin{equation}
    A^{(1)}_{\mu\nu}(\bq)
    = 
    2
    \begin{pmatrix}
        0 & c_{xy}^- & c_{xz}^- & 0 & c_{xy}^+ & c_{xz}^+ \\[1pt]
        c_{xy}^- & 0 & c_{yz}^- & c_{xy}^+ & 0 & c_{yz}^+ \\[1pt]
        c_{xz}^- & c_{yz}^- & 0 & c_{xz}^+ & c_{yz}^+ & 0 \\[1pt]
        0 & c_{xy}^+ & c_{xz}^+ & 0 & c_{xy}^- & c_{xz}^- \\[1pt]
        c_{xy}^+ & 0 & c_{yz}^+ & c_{xy}^- & 0 & c_{yz}^- \\[1pt]
        c_{xz}^+ & c_{yz}^+ & 0 & c_{xz}^- & c_{yz}^- & 0
    \end{pmatrix},
\end{equation}
\begin{equation}
    A^{(2a)}_{\mu\nu}(\bq)
    = 
    2
    \begin{pmatrix}
     0 & 0 & 0 & c_{xx}^+ & 0 & 0 \\
     0 & 0 & 0 & 0 & c_{yy}^+ & 0 \\
     0 & 0 & 0 & 0 & 0 & c_{zz}^+ \\
     c_{xx}^+ & 0 & 0 & 0 & 0 & 0 \\
     0 & c_{yy}^+ & 0 & 0 & 0 & 0 \\
     0 & 0 & c_{zz}^+ & 0 & 0 & 0 
    \end{pmatrix}.
\end{equation}
For the purpose of counting the zero-energy locus of the interaction matrix (after shifting its minimum eigenvalue to zero), as reported in the last column of \cref{tab:summary_degeneracies}, one should always use the primitive 3-site unit cell, since the bands of the enlarged conventional unit cells yields are folded versions of the primitive bands.

\mysubsection{Self-Consistent Gaussian Approximation}

The self-consistent Gaussian approximation (SCGA)~\cite{garaninClassicalSpinLiquid1999} is a simple soft-spin approximation which works well for computing correlation functions so long as the system remains paramagnetic, in particular for spin liquids. 
This method replaces the hard spin length constraint $\vert S_i^z \vert = 1$ with a soft one, meaning that the $S_i^z$ are treated as real-valued variables which vary continuously, with the spin length enforced on average.
Given a bilinear spin Hamiltonian
\begin{equation}
    H = \frac{1}{2} \sum_{ij} S_i^z \mathcal{J}_{ij} S_j^z, 
\end{equation}
we add a term which energetically regulates the spin length from growing arbitrarily large,
\begin{equation}
    H_{\text{SCGA}} = H + \frac{1}{2}\sum_i \lambda_i T(S_i^z)^2,
    \label{eq:H_SCGA}
\end{equation}
where the $\lambda_i$ are to be determined self-consistently by the requirement that $\langle \vert S_i^z \vert^2 \rangle_{\lambda,T} = 1$ $\forall i$, and where the angle brackets denote a thermal expectation value at fixed $\lambda$ and temperature $T$. 
The Hamiltonian simplifies to
\begin{equation}
    \beta H_{\text{SCGA}} = \frac{1}{2} \sum_{ij} S_i^z (\Lambda_{ij} + \beta \mathcal{J}_{ij}) S_j^z
\end{equation}
where $\cramped{\beta=1/T}$ is inverse temperature, and $\cramped{\Lambda_{ij} = \lambda_i \delta_{ij}}$ with $\delta$ the Kronecker delta.
The partition function is then a simple multi-variate Gaussian, and one can directly read off the correlation functions
\begin{equation}
    \mathscr{G}(\{\lambda_i\})_{ij} = \langle S_i^z S_j^z \rangle = [\Lambda + \beta \mathcal{J}]^{-1}_{ij},
\end{equation}
and the self-consistency condition reads $\mathcal{G}_{ii} = 1$ for all $i$.
For a system in which every site $i$ is equivalent by symmetry, the problem simplifies since all $\lambda_i = \lambda$. 
The self-consistency condition can then be simplified to an average over all sites, 
\begin{equation}
    \frac{1}{N_{\text{sites}}} \Tr[\mathscr{G}(\lambda)] = 1.
\end{equation}
In practice this is solved by Fourier transformation to block-diagonalize $\mathcal{J}$, then numerically solving the self-consistency equation. 
In this work we utilize the SCGA to compute the structure factors shown in \cref{fig:X-cube_bands_structure_factors}(b,c), which show excellent agreement with the Monte Carlo calculations.

\mysubsection{Structure Factors and Neutron Scattering}

The full two-point correlation function $\mathscr{G}_{ij} = \langle S_i^z S_j^z \rangle$ can be Fourier transformed as
\begin{equation}
    \mathscr{G}_{\mu\nu}(\bq) = \langle S^z_\mu(-\bq) S^z_\nu(\bq) \rangle,
\end{equation}
using \cref{apx_eq:fourier-transform}. 
Due to the sublattice structure, this quantity is a tensor at each wavevector $\bq$, which contains the full information about the spin correlations.
Structure factors are a contraction of this tensor with a form factor to obtain one scalar value per $\bq$.
The simplest is the spin structure factor, defined as
\begin{equation}
    S(\bq) = \sum_{\mu\nu} \mathscr{G}_{\mu\nu}(\bq).
\end{equation}
Note that this structure factor depends on the choice of unit cell as it is defined in terms of the gauge-dependent quantities $S_i^z$.
Furthermore, because the form factor is $\bq$-independent, this structure factor may not fully reflect the angular dependence of the correlation tensor. 
This is the case for the X-cube fracton spin liquid discussed in \cref{sec:fracton_CSL}, which has nodal line singularities with quadrupolar angular dependence---the spin structure factor only exposes two-fold pinch points at the zone centers and exhibits smooth behavior along the nodal lines. 

Exposing the angular dependence requires the introduction of a $\bq$-dependent form factor.
Such a form factor appears naturally in neutron scattering experiments---neutrons are scattered by the divergence-free magnetic field produced by the local Ising magnetic moments $\bm{\moment}_i\propto S_i\hat{\bm{z}}_i$ of the ions in the lattice~\cite{loveseyPolarizationEffectsMagnetic1987}.
We define the dimensionless moment-moment correlation tensor,
\begin{equation}
    \mathscr{G}_{\mu\nu}^{\alpha\beta}(\bq) = \langle \hat{d}_\mu^\alpha(-\bq) \hat{d}_\nu^\beta(\bq) \rangle \equiv \hat{z}_\mu^\alpha \hat{z}_\nu^\beta \mathscr{G}_{\mu\nu}(\bq),
\end{equation}
which is independent of the gauge choices of quantization axes.
For unpolarized neutrons, the scattering cross section is proportional to
\begin{equation}
    S_{\text{unp}}(\bq) = 
    \sum_{\mu,\nu}
    \sum_{\alpha,\beta}
    \left(
    \delta_{\alpha\beta} - \hat{q}^\alpha \hat{q}^\beta
    \right) 
    \mathscr{G}_{\mu\nu}^{\alpha\beta}(\bq),
\end{equation}
where the projector arises because the magnetic field is divergence-free, so neutrons do not ``see'' the longitudinal (along $\hat{\bq}$) component of the correlations.
Polarization analysis further refines this cross-section into two channels, denoted spin-flip (SF) and non-spin-flip (NSF), referring to filtering the scattered neutrons according to whether they are flipped or not in the scattering process~\cite{loveseyPolarizationEffectsMagnetic1987}.
We consider incident neutrons initially polarized in the direction $\hat{\bm{z}}_s$.
For each scattering wavevector $\bm{q}$ such that $\bm{q}\perp\hat{\bm{z}}_s$, we define a local orthonormal frame $\{\hat{\bm{x}}_s, \hat{\bm{y}}_s, \hat{\bm{z}}_s\}$, where $\hat{\bm{x}}_s=\hat{\bm{q}}$ and $\hat{\bm{y}}_s=\hat{\bm{z}}_s \times\hat{\bm{x}}_s$~\cite{fennellMagneticCoulombPhase2009,chungProbingFlatBand2022}.
The SF and NSF cross-sections are then proportional to the following structure factors,
\begin{equation}
    \begin{aligned}
        S_{\textrm{SF}}(\bm{q}) 
        &= 
        \sum_{\mu,\nu}\sum_{\alpha,\beta}
        \hat{y}^\alpha_s \hat{y}^\beta_s 
        \mathscr{G}_{\mu\nu}^{\alpha\beta}(\bq) 
        \quad 
        (\bq\perp \hat{\bm{z}}_s),
        \\
        S_{\textrm{NSF}}(\bm{q}) 
        &= 
        \sum_{\mu,\nu}\sum_{\alpha,\beta}
        \hat{z}^\alpha_s \hat{z}^\beta_s 
        \mathscr{G}_{\mu\nu}^{\alpha\beta}(\bq)
        \quad 
        (\bq\perp \hat{\bm{z}}_s),
    \end{aligned}
\end{equation}
which satisfy
\begin{equation}
    S_{\text{unp}}(\bq) = S_{\text{SF}}(\bq) + S_{\text{NSF}}(\bq),
\end{equation}
because the vectors $\hat{\bm{x}}_s$, $\hat{\bm{y}}_s$, $\hat{\bm{z}}_s$ form an orthonormal basis at each $\bq \perp \hat{\bm{z}}_s$.

\section{Specific Heat of 1D Chains}
\label{apx:1D_Chains}

In \cref{fig:nematic-thermodynamics}(a), we fit the low-temperature specific heat of the spin nematic phase to the exact solution for decoupled 1D chains. This solution is easily derived from the transfer matrix formulation of the partition function. For a periodic 1D chain with Hamiltonian 
\begin{equation}
    H_{\text{1D-chain}} = -\tilde{J}\sum_{i=1}^L S_i^z S_{i+1}^z
\end{equation}
where $S_i^z = \pm 1$ and $S_{L+1}\equiv S_1$, the partition function at inverse temperature $\beta = 1/T$ is 
\begin{equation}
    Z_L(\beta) = \sum_{\{S_i^z = \pm 1\}} \prod_{i=1}^L \exp(+\beta \tilde{J} S_i^z S_{i+1}^z) \equiv \Tr[\mathcal{T}^L]
\end{equation}
where $\mathcal{T}$ is the transfer matrix, 
\begin{equation}
    \mathcal{T} = \begin{pmatrix}
        e^{+\beta \tilde{J}} & e^{-\beta \tilde{J}} \\
        e^{-\beta \tilde{J}} & e^{+\beta \tilde{J}}
    \end{pmatrix}.
\end{equation}
The heat capacity for the 1D chain is then 
\begin{equation}
    C_V^{(\text{1D},L)}(T) =
    \frac{\partial\langle H\rangle }{\partial T}
    = 
    \frac{\partial}{\partial T}
    \left(
    -
    \frac{\partial }{\partial \beta}\, \ln Z_L
    \right).
\end{equation}
In the thermodynamic limit $L \to \infty$, the partition function is approximated by the largest eigenvalue of the transfer matrix, so
\begin{equation}
    Z_{L\to \infty} \propto \cosh(\beta \tilde{J})^L .
\end{equation}
From this we obtain the exact expression for the specific heat in the thermodynamic limit
\begin{equation}
    C_V^{\text{(1D)}}/L \,\xrightarrow{L\to\infty} \,(\beta \tilde{J})^2 \mathrm{sech}^2(\beta\tilde{J})
\end{equation}
which is precisely the form of a Schottky anomaly for a 2-level system with gap $2\tilde{J}$, i.e. a smooth crossover associated to thermal depopulation of domain walls (spinons) with energy $2\tilde{J}$.

We show in \cref{fig:nematic-thermodynamics}(a) a comparison of the low-temperature heat capacity per spin (with $3L^3$ total spins) of the 3D system to that of $L^2$ decoupled length-$L$ 1D chains with $\tilde{J} = J_{2a}$,
\begin{equation}
    \frac{C_V^{(\text{3D})}(T)}{3L^3} \approx \frac{L^2 C_V^{(\text{1D},L)}(T)}{3L^3}.
\end{equation}
At the point in the phase diagram where the data in \cref{fig:nematic-thermodynamics}(a) is taken, with $2J_1=-J_{2a}>0$, we have $J_{2a} = 2J/\sqrt{5}$, with $J$ defined in \cref{eq:theta_J1_J2a} and used in the temperature scale of \cref{fig:nematic-thermodynamics}(a).

\section{Cluster Algorithm for Fracton CSL}
\label{apx:cage_net_cluster_algorithm}

Otsuka developed a highly efficient cluster algorithm for the simulation of nearest-neighbor pyrochlore spin ice~\cite{otsukaClusterAlgorithmMonte2014}, which can be straightforwardly adapted for octochlore spin ice~\cite{szaboFragmentedSpinIce2022}. 
Here we explain how the algorithm works, how it is adapted to work for octochlore spin ices, and how it is amended for the X-cube fracton spin studied in \cref{sec:fracton_CSL}.

\begin{figure}[t]
    \centering
    \vspace{4ex}
    \begin{minipage}[b]{.9\linewidth}
        \begin{overpic}[width=\linewidth]{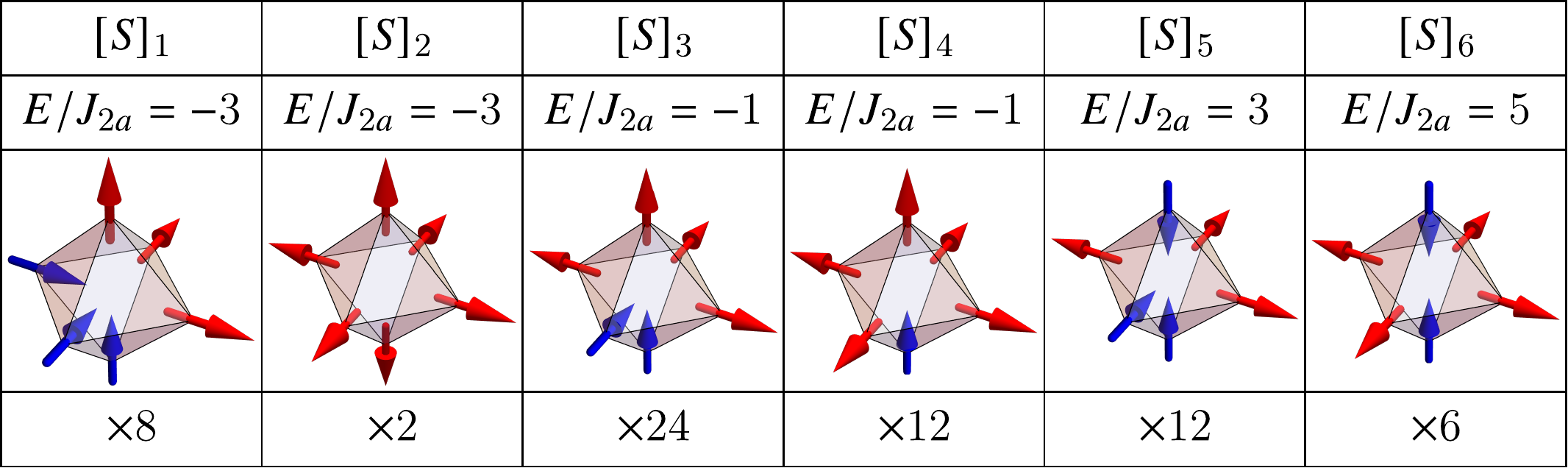}
            \put(0,31){(a)}
        \end{overpic}
        \\[4ex]
        \begin{overpic}[width=\linewidth]{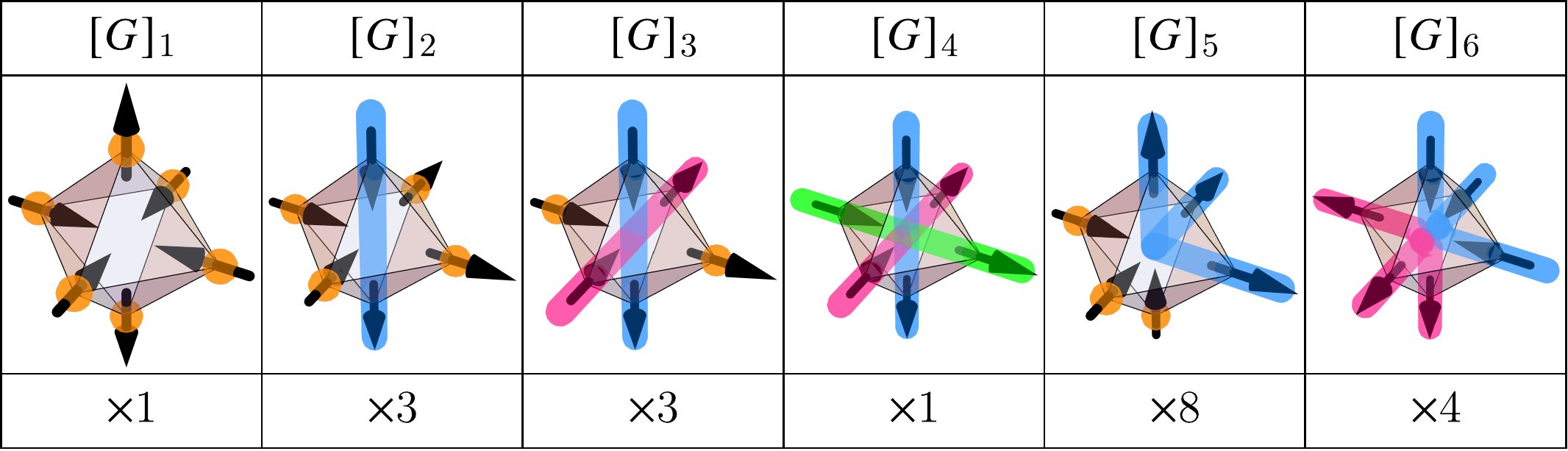}
            \put(0,31){(b)}
        \end{overpic}
    \end{minipage}
    \\[3ex]
    \begin{overpic}[width=\linewidth]{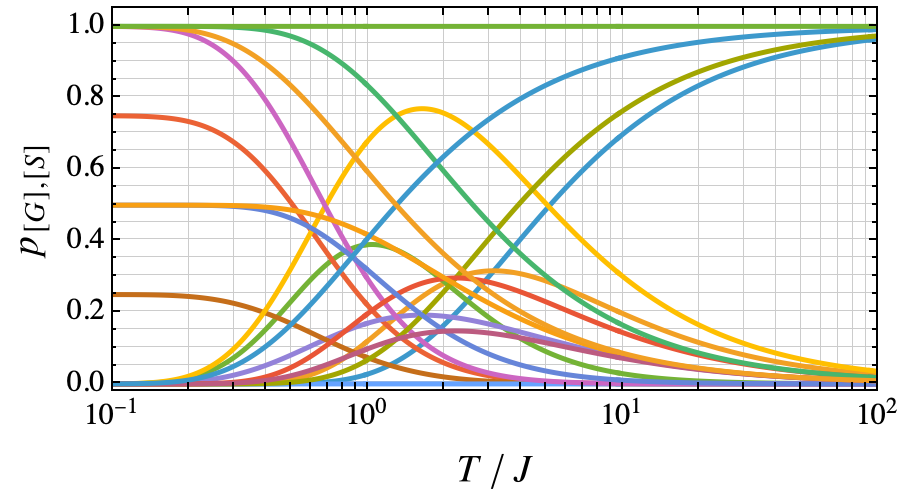}
        \put(0,54){(c)}
        \put(14,20.5){\scalebox{.7}{$p_{6,1}$}}
        \put(14,30.7){\scalebox{.7}{$p_{3,3}$}}
        \put(14,34.3){\scalebox{.7}{$p_{5,3}$}}
        \put(14,44){\scalebox{.7}{$p_{4,1}$}}
        \put(25,46){\rotatebox{-50}{\scalebox{.7}{$p_{6,2}$}}}
        \put(34,46){\rotatebox{-46}{\scalebox{.7}{$p_{2,5}$}}}
        \put(41,48){\rotatebox{-48}{\scalebox{.7}{$p_{5,4}$}}}
        \put(53,50.5){\scalebox{.7}{$p_{1,6}$}}
        \put(53,44){\rotatebox{29}{\scalebox{.7}{$p_{1,5}$}}}
        \put(63,42){\rotatebox{29}{\scalebox{.7}{$p_{1,3}=p_{1,4}$}}}
        \put(71,40){\rotatebox{29}{\scalebox{.7}{$p_{1,1}=p_{1,2}$}}}
        \put(75,22){\rotatebox{-27}{\scalebox{.7}{$p_{5,2}$}}}
    \end{overpic}\\
    \caption{
    \textbf{Cluster algorithm for the X-cube CSL.}
    (a)~Symmetry classes of single-octahedron spin configurations along with their energies at the X-cube point (where $J_{2a}/\vert J_1\vert=2$ and $J= {\sqrt{5}J_{2a}}$), and their multiplicities below (64 total states), 
    (b)~Symmetry classes of graphs for the X-cube point along with their multiplicities. 
    For each graph, we also show a spin configurations compatible with that graph, i.e. with $\Delta_{S,G}=1$.
    (c)~Probabilities of graph class assignments as a function of temperature, \cref{eq:SM_total_graph_probability}. Curves are labeled according to $p_{n,m}\equiv p([G]_n\vert [S]_m)$.
    }
    \label{fig:SM_X-cube_cluster_algo}
\end{figure}

\mysubsection{Graph Decomposition Algorithm}

The algorithm works by a graph decomposition of the partition function, see Ref.~\cite{otsukaClusterAlgorithmMonte2014} for details. 
The result is a decomposition of any spin configuration into clusters of spins connected by graphs, each of which may be flipped independently with 50$\%$ probability. 
Each octahedron is assigned one of a number of graphs $G$ such that the Boltzmann weight for a spin configuration $S$ can be expanded as a sum over graphs,
\begin{equation}
    \exp[-\beta (E([S]) - \epsilon_0)]  \equiv w([S]) = \sum_{G} \Delta_{S,G} \,w([G]).
    \label{eq:SM_boltzmann_decomp}
\end{equation}
Here, $[S]$ denotes an equivalence class of single-octahedron spin configurations that are related by a point group transformation, and $[G]$ denotes an equivalence class of graphs. 
For convenience, we subtract the single-octahedron ground state energy $\epsilon_0$, so that the minimum-energy configurations have $w([S])=1$.
The $w([G])$ are the \emph{graph weights}, and the $\Delta_{S,G}$ are the \emph{compatibility coefficients}, defined as
\begin{equation}
    \Delta_{S,G} = \begin{cases}
        1 &\quad \text{if $G$ is compatible with $S$,} \\
        0 &\quad \text{otherwise}.
    \end{cases}
\end{equation} 
Each graph equivalence class $[G]$ must contain all symmetry-equivalent graphs, but may contain multiple graphs which are not related by symmetry.  
To obtain the graph weights, \cref{eq:SM_boltzmann_decomp} is rewritten as 
\begin{equation}
    w([S]) = \sum_{[G]} \Gamma_{[S],[G]}\,w([G]),
    \label{eq:SM_boltzmann_expansion_matrix}
\end{equation}
where we have defined the \emph{reduced compatibility matrix}
\begin{equation}
    \Gamma_{[S],[G]} = \sum_{G \in [G]} \Delta_{S,G},
\end{equation}
which counts how many graphs in a given class $[G]$ are compatible with a given spin configuration $S$. 
This quantity depends only on the symmetry class $[S]$ since each $[G]$ contains all symmetry-equivalent graphs.
In this form, we can see that solving for $w([G])$ can be done by inverting the matrix $\Gamma_{[S],[G]}$, assuming that the number of spin classes $[S]$ is the same as the number of graph classes $[G]$,
\begin{equation}
    w([G]) = \sum_{[S]} \Gamma^{-1}_{[G],[S]}\, w([S]).
    \label{eq:SM_graph_weights}
\end{equation}
The probabilities for assigning graph $G$ to spin configuration $S$ is then given by~\cite{otsukaClusterAlgorithmMonte2014}
\begin{equation}
    p(G\vert S) = \Delta_{S,G} \, w([G])/w([S]).
    \label{eq:assignment_prob}
\end{equation}
Note that $\sum_G p(G\vert S) = 1$ using \cref{eq:SM_boltzmann_expansion_matrix}, i.e. the total probability of assigning a graph to a spin configuration is unity.
The algorithm works by fixing a spin configuration, applying a random graph $G$ to every octahedron with probability given by \cref{eq:assignment_prob}, identifying the resulting clusters obtained by gluing graphs on neighboring octahedra, and flipping each cluster with 50$\%$ probability. See Ref.~\cite{otsukaClusterAlgorithmMonte2014} for further details. It is useful to define the total probability of assigning any graph in class $[G]$ to a spin configuration $S$,
\begin{equation}
    p([G]\vert [S]) = \sum_{G\in [G] }p(G\vert S) \,=\, \Gamma_{[S],[G]} \frac{w([G])}{w([S])},
    \label{eq:SM_total_graph_probability}
\end{equation}
which only depends on the symmetry class $[S]$ since we sum over all symmetry-equivalent graphs $G$.
In practice, for each octahedron we pick a graph class according to $p([G]\vert[S])$ and then pick one of the compatible $G \in [G]$ with equal probability.

\mysubsection{Cluster Algorithm for Spin Ice CSL}

The graph types $[G]$ must be chosen in a way that is compatible with the ground state constraints. 
In the case of spin ice, the ground state satisfies a zero-divergence constraint and is therefore a string condensate (where a string connects an in-spin to an out-spin), with no restriction on the geometry of the strings. 
Thus the graphs should be chosen to pair up in with out spins on each octahedron and their weights $w([G])$ depend only on the number of ``loose ends'' of the strings, see Ref.~\cite{szaboFragmentedSpinIce2022} for further details.

\mysubsection{Cluster Algorithm for X-cube CSL}

At the X-cube point, $J_{2a}=-2J_1>0$, we distinguish the six symmetry-classes of spin configurations and their energies in \cref{fig:SM_X-cube_cluster_algo}(a). Defining
\begin{equation}
    z = \exp(-2\beta J_{2a}) ,
\end{equation}
the Boltzmann weights are
\begin{equation}
    \begin{aligned}
    w([S]_1) &= 1, \quad 
    w([S]_3)  = z,  \quad 
    w([S]_5)  = z^3, 
    \\
    w([S]_2) &= 1, \quad
    w([S]_4)  = z, \quad
    w([S]_6)  = z^4. 
    \end{aligned}
    \label{eq:SM_Xcube_spin_weights}
\end{equation} 
Whereas spin ice realizes a string-net liquid, according to our discussion in \cref{sec:cage_net_tunneling}, the X-cube CSL realizes a cage-net liquid where the flippable motifs are cages like the ones shown in \cref{fig:cages_and_thermodynamics}(a,b).
Accordingly, we choose graphs as shown in \cref{fig:SM_X-cube_cluster_algo}(b), which contain either straight segments ($[G]_1$ through $[G]_4$) or corners ($[G]_5$ and $[G]_6$).
Below each graph, we give the number of symmetry-equivalent graphs in the equivalence class.
Gluing such graphs together produces cage nets like those shown in \cref{fig:cages_and_thermodynamics}(a,b).
Graphs $[G]_1$, $[G]_2$, $[G]_3$, and $[G]_5$ have disconnected spins, indicated in orange, which produce cage nets with open ends. 
The compatibility coefficients $\Delta_{S,G}$ depend only on the graph class, and are given by
\begin{equation}
    \Delta_{S,{G\in [G]_n}} = 
    \begin{cases}
    \begin{array}{ll}
        \begin{cases}
            1 \text{ if $S_1^z = - S_2^z$} \\
            0 \text{ otherwise}
        \end{cases}
        &
        n=2,3,4 , 
        \\[.5ex]
        \begin{cases}
            1 \text{ if $S_1^z = S_2^z=S_3^z$} \\
            0 \text{ otherwise}
        \end{cases}
        &
        n=5,6, 
        \\[.5ex]
        \qquad\quad 1 
        &
        n=1 ,
    \end{array}
    \end{cases}
    \label{eq:SM_Xcube_Deltas}
\end{equation}
where the $S_i^z$ are given in the 6-site basis, and refer only to spins which are connected by the graph.
In words: 
the first case refers to the straight segment graphs $n=2,3,4$.
These graphs can only connect pairs of spins with one spin ``in'' and the other ``out''.
The second case refers to the corner graphs  $n=5,6$, and these graphs can only connect a trio of spins with all three spins ``in'' or all three spins ``out''.
The third case refers to the free graph $n=1$, which is compatible with any spin configuration.
The reduced compatibility matrix $\Gamma_{[S],[G]}$ is given by 
\begin{equation}
    \Gamma_{[S],[G]} 
    = 
    \left(
    {
    \setlength{\arraycolsep}{2pt}
    \begin{array}{c|cccccc}
    & [G]_1
    & [G]_2
    & [G]_3
    & [G]_4
    & [G]_5
    & [G]_6
    \\[2pt]
    \hline
    \\[-8pt]
    {[S]}_1 & 1 & 3 & 3 & 1 & 2 & 1 \\[6pt]
    {[S]}_2 & 1 & 0 & 0 & 0 & 8 & 4 \\[6pt]
    {[S]}_3 & 1 & 2 & 1 & 0 & 2 & 0 \\[6pt]
    {[S]}_4 & 1 & 1 & 0 & 0 & 4 & 0 \\[6pt]
    {[S]}_5 & 1 & 1 & 0 & 0 & 0 & 0 \\[6pt]
    {[S]}_6 & 1 & 0 & 0 & 0 & 0 & 0 
    \end{array}
    }
    \right).
\end{equation}
Inverting this matrix, we obtain the graph weights, \cref{eq:SM_graph_weights}, 
\begin{equation}
    \begin{aligned}
    w([G]_1) &= z^4, \\
    w([G]_2) &= z^3-z^4 ,\\
    w([G]_3) &= (z-3z^3+2z^4)/2 ,\\
    w([G]_4) &= (3 - 6z + 6z^3 - 3z^4)/4 ,\\
    w([G]_5) &= (z - z^3)/4 ,\\
    w([G]_6) &= (1-2z+2z^3-z^4)/4 .
    \end{aligned}
    \label{eq:SM_Xcube_graph_weights}
\end{equation}
Combining \cref{eq:SM_Xcube_spin_weights,eq:SM_Xcube_Deltas,eq:SM_Xcube_graph_weights},  we can compute the corresponding probabilities using \cref{eq:assignment_prob}.
In \cref{fig:SM_X-cube_cluster_algo}(c) we plot the resulting graph class assignment probabilities, \cref{eq:SM_total_graph_probability}.
Note that in a ground state consisting of only $[S]_1$ and $[S]_2$ octahedra, in the limit $T\to 0$ only the fully connected graphs $[G]_4$ and $[G]_6$ have non-zero probabilities, and the algorithm decomposes each spin configuration into a fully packed cage net covering, analogous to the fully packed loop coverings for Coulomb liquids~\cite{jaubertAnalysisFullyPacked2011}.

\section{Monte Carlo Simulation Details}
\label{apx:monte_carlo}

In this appendix, we provide details of the Monte Carlo simulations.
Except where noted, these are performed using the 24-site unit cell, for a system of $L\times L\times L$ unit cells with periodic boundaries.
Some figures instead label system sizes by the equivalent 3-site linear dimension $2L$ for consistency with the rest of the manuscript; we indicate the correspondence where relevant.

\mysubsection{Thermodynamics}
 
In our Monte Carlo simulations we compute the specific heat from the variance of the energy,
\begin{equation}
    C_V = \frac{1}{nL^3}\,\frac{\langle E^2\rangle - \langle E\rangle^2}{T^2},
\label{eq:apx-cv}
\end{equation}
where $n$ is the number of spins in the simulated unit cell (\cref{apx:conventions}), and uncertainties are obtained by jackknife binning.
The entropy per spin is then obtained by integrating $C_V/T$~\cite{chung2formU1Spin2025},
\begin{equation}
    S(T) = S(\infty) - \int_T^{T_{\max}}\frac{C_V}{T}\,\mathrm{d}T - \int_{T_{\max}}^{\infty}\frac{C_V}{T}\,\mathrm{d}T,
\label{eq:apx-entropy}
\end{equation}
where $S(\infty)=\ln 2$ and $T_{\max}$ is the highest temperature simulated.
Since the simulations extend only up to $T_{\max}$, we split the integral into a numerically sampled part and a high-temperature tail.
The first integral is evaluated by numerical quadrature using the $C_V$ data in \cref{fig:cages_and_thermodynamics}.
For the second, we truncate the high-temperature series expansion of the specific heat,
\begin{equation}
    C_V \approx \frac{a}{T^2} + \frac{b}{T^3},
\label{eq:apx-cv-hightemp}
\end{equation}
extract $a$ and $b$ by fitting the high-temperature tail of the measured $C_V$, and integrate analytically,
\begin{equation}
    \int_{T_{\max}}^{\infty}\frac{C_V}{T}\,\mathrm{d}T \approx \frac{a}{2T_{\max}^2} + \frac{b}{3T_{\max}^3}.
\label{eq:apx-tail}
\end{equation}

\begin{figure}[t]
    \centering
    \includegraphics[width=1.0\linewidth]{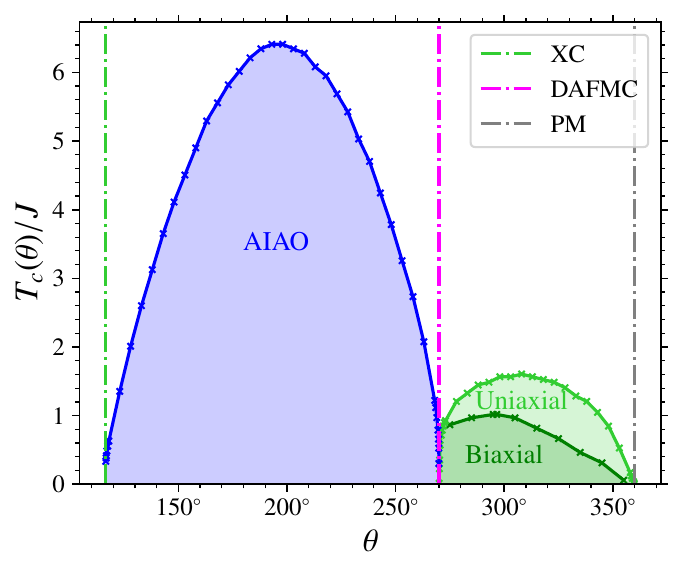}
    \caption{\textbf{Finite-temperature phase boundaries.} Critical temperatures $T_c(\theta)/J$ for the intra-octahedral second-nearest-neighbor octochlore Ising phase diagram. 
    The couplings are parametrized as $J_1 = J\cos\theta$ and $J_{2a} = J\sin\theta$, normalized to the unit circle. 
    Monte Carlo estimates of $T_c$ are shown for the ordered phases [all-in/all-out (AIAO), uniaxial and biaxial spin-nematic], with vertical dashed lines indicating special points of interest: the X-cube fracton CSL (XC), decoupled antiferromagnetic chains (DAFMC) and the trivial paramagnet (PM). 
    These values serve as the input for the polar representation of the phase diagram presented in~\cref{fig:phase_diagram}(f) on the non-linear radial scale: $r(T/J) = 1-\tanh[(T/10J)^{1/2}]\in[0,1]$.}
    \label{fig:TcValues}
\end{figure}

\mysubsection{Finite-Temperature Phase Diagram}

In \cref{fig:phase_diagram}(f), we report critical temperatures $T_c$ for a range of angles $\theta$ on the unit-circle parametrization [\cref{eq:theta_J1_J2a}] of the intra-octahedral phase diagram.
These sweeps were performed on an $L=24$ system using the 6-spin unit cell, with parallel tempering over 100 replicas.
Each simulation comprised $10^4$ thermalization sweeps followed by $10^5$ measurements, with measurements taken every 10 sweeps and parallel-tempering swaps attempted every 5 sweeps.
The critical temperatures $T_c^{\mathrm{AIAO}}$ and $T_c^{\mathrm{uni}}$ were identified from the location of the $C_V$ peak.
The biaxial transition produces no diverging $C_V$ anomaly [\cref{fig:nematic-thermodynamics}(a)]; we therefore identify $T_c^{\mathrm{bi}}$ as the temperature at which the two degenerate subleading eigenvalues of the quadrupole tensor split apart [\cref{fig:nematic-thermodynamics}(c)].
The resulting $T_c(\theta)$ are plotted on a linear scale in \cref{fig:TcValues}, and reproduced on the non-linear radial scale $r(T/J)=1-\tanh[(T/10J)^{1/2}]$ in \cref{fig:phase_diagram}(f).

\mysubsection{Frustrated Chain Phase}

In \cref{fig:frustrated_chains_arc} we present specific-heat curves for angles $\theta\in\{\theta_{\mathrm{SI}},\dots,\theta_{\mathrm{XC}}\}$ spanning the frustrated chain phase, where $\theta_{\mathrm{SI}}=45^\circ$ ($J_1=J_{2a}$) and $\theta_{\mathrm{XC}}$ ($J_1/J_{2a}=-1/2$) are the two spin-liquid endpoints.
At these endpoints we use the dedicated cluster algorithms of \cref{apx:cage_net_cluster_algorithm}, combined with simulated annealing to cool $L=8$ systems from $T/J=10^2$ to $T/J=10^{-2}$ while making $10^6$ measurements at each temperature.
All other angles were simulated with an $L=8$ system using parallel tempering with single-spin-flip updates.
For angles near the spin-ice point ($\theta\le54^\circ$), we supplement the single-spin flips with the short spin-ice loop moves of Ref.~\cite{melkoMonteCarloStudies2004}.
All replicas were initialized in a random ferromagnetic configuration and thermalized for $10^4$ sweeps, with parallel-tempering swaps every 3 sweeps, before recording $5\times10^5$ measurements separated by 5 sweeps.

\mysubsection{X-cube Spin Liquid}

In \cref{fig:cages_and_thermodynamics}(c) and \cref{fig:X-cube_bands_structure_factors}(b,c), we present thermodynamic data for the X-cube CSL at $J_{2a}=-2J_1>0$, discussed in \cref{sec:fracton_CSL}.
These use the cluster algorithm of \cref{apx:cage_net_cluster_algorithm} on $L=16,20,24$ systems with the 24-site unit cell (\cref{apx:conventions}).
Starting from a random configuration, we used simulated annealing to cool the system through 40 temperatures spanning $T_{\min}<T<T_{\max}$, distributed as: linearly spaced over $0.35<T/J_{2a}<1.46$; linearly spaced over $1.5<T/J_{2a}<2.5$; and logarithmically spaced over $10^{0.4}<T/J_{2a}<10^2$, with $J_{2a}=J/\sqrt5$.
These ranges were chosen to resolve the $C_V$ peak.
At each temperature we ran 1000 independent simulations, each with $10^4$ thermalization sweeps followed by 5000 measurements sampled every 5 sweeps, for a total of $5\times10^6$ measurements per temperature.

To evaluate the entropy [\cref{eq:apx-entropy}], we fit the high-temperature tail of the $L=24$ specific heat for $T\ge50J_{2a}$ to \cref{eq:apx-cv-hightemp}, obtaining
\begin{equation}
a = 7.91\pm0.02, \qquad b = -31.0\pm0.3,
\label{eq:apx-ab-fit}
\end{equation}
which gives a high-temperature entropy contribution [\cref{eq:apx-tail}] of
\begin{equation}
\int_{T_{\max}}^{\infty}\frac{C_V}{T}\,\mathrm{d}T = (4.06\pm0.01)\times10^{-4}.
\label{eq:apx-tail-value}
\end{equation}
The entropy $S(T)$ was then obtained by numerically integrating $C_V(T)/T$ as in \cref{eq:apx-entropy}.
Error bars on the full integral were estimated by resampling: we used the jackknife uncertainties on $C_V$ to generate $N_{\mathrm{samp}}=10^3$ synthetic data sets, repeated the integration for each, and took the standard deviation of the resulting entropy curves as the uncertainty.

\mysubsection{Spin Nematic}

\Cref{fig:nematic-thermodynamics} presents thermodynamic data at the point diametrically opposite the fractonic X-cube model in \cref{fig:phase_diagram}(f) labeled ``Fully Packed Fractons''.
Three sets of simulations underlie the figure, each targeting a different feature; all use the 24-site unit cell.
Following the convention above, the $L=20$ simulations in this section correspond to the label $L=40$ in \cref{fig:nematic-thermodynamics,fig:nematic_cubes}.

To expose the spontaneous breaking of rotational symmetry, we performed 250 independent simulated-annealing runs on an $L=20$ system (40 $ab$-planes), each cooled from a random configuration at $T/J=2.0$ to $T/J=0.2$, equilibrating for $10^4$ sweeps and then recording $10^5$ measurements (every 5 sweeps) at each temperature.
These runs produce the faint ensemble of eigenvalue curves in \cref{fig:nematic-thermodynamics}(c): as discussed in \cref{sec:biaxial}, once the $ab$-planes order [\cref{eq:plateau_Q}] the two subleading eigenvalues of $Q$ lock onto plateaus set by the number $N\in[0,40]$ of planes polarized along each axis, and each seed selects a different $N$.
Collecting these plateau values across the 250 seeds at $T/J\approx0.46$ gives the histogram on the right axis of \cref{fig:nematic-thermodynamics}(c), where the integer levels $N$ are marked.
Although the plateau distribution is representative of the biaxial phase as a whole, at higher temperatures within the phase the eigenvalue trajectories have not yet settled cleanly onto their plateaus owing to finite-size effects; we therefore sample at $T/J\approx0.46$, the low-temperature end of the biaxial phase just above the 1D-chain freezing regime, where the plateaus are most sharply resolved.
The histogram is compared to the binomial distribution expected if the plane polarizations were statistically independent.
The observed distribution deviates measurably from binomial, indicating that the planes are weakly correlated rather than perfectly decoupled---the sense in which the low-temperature ``decoupled 1D chains'' picture is a very good first approximation but not exact, as discussed further in \cref{sec:biaxial}.
The temperature $T/J\approx0.46$ also marks the high-temperature edge of the shaded region in \cref{fig:nematic-thermodynamics}, below which the correlation length of the 1D chains exceeds the system size and the chains freeze into one of their two N\'eel states, producing the erratic finite-size order-parameter values in \cref{fig:nematic-thermodynamics}(b,c) [cf.\ \cref{fig:nematic_cubes}(d)].
Within this shaded region the specific heat matches the exact finite-length 1D Ising result (\cref{apx:1D_Chains}), shown for the 250 seeds in the inset of \cref{fig:nematic-thermodynamics}(a); the close agreement confirms that the thermodynamics is well described by decoupled 1D chains at leading order, even though the histogram reveals residual inter-plane correlations.

To resolve the two transitions we ran parallel tempering at $L=16,20,24$ (labels $32,40,48$), producing the scaling insets of \cref{fig:nematic-thermodynamics}(a).
At $\TCuni$, $N_T=64$ replicas were initialized randomly, thermalized for $10^4$ sweeps, and sampled $10^6$ times (every 5 sweeps).
The biaxial transition requires more care: cooling through $\TCuni$ breaks cubic symmetry to tetragonal, so independent replicas would select different director axes and dilute a fixed-component order parameter.
We therefore fix the sector by initializing all $N_T=64$ replicas (for each $L$) from a single configuration annealed through $\TCuni$, equilibrating for $10^4$ sweeps with swaps every 3 sweeps before $10^6$ measurements (every 10 sweeps).

The solid curves in \cref{fig:nematic-thermodynamics}(a--c) come from a single $L=20$ parallel-tempering run over $0.1\le T/J\le2.5$ with $N_T=96$ replicas, initialized from an annealed ground state [as in \cref{fig:nematic_cubes}], equilibrated for $10^4$ sweeps with swaps every 3 sweeps, and sampled $10^6$ times (every 3 sweeps).
This run supplies the specific-heat and order-parameter curves in \cref{fig:nematic-thermodynamics}(a--c) and guides the eye through the 250-seed ensemble in \cref{fig:nematic-thermodynamics}(c).


%

\end{document}